\documentclass[fleqn,usenatbib]{mnras}

\usepackage{newtxtext,newtxmath}
\usepackage[T1]{fontenc}

\DeclareRobustCommand{\VAN}[3]{#2}
\let\VANthebibliography\thebibliography
\def\thebibliography{\DeclareRobustCommand{\VAN}[3]{##3}\VANthebibliography}

\usepackage{graphicx}	% Including figure files
\usepackage{amsmath}	% Advanced maths commands

\title[Disequilibrium in dwarf galaxies]{Dynamical disequilibrium in dwarf galaxies: rethinking gas dynamics, rotation curves, and dark matter inference}

\author[D. Dado et al.]{Diego Dado,$^{1,2}$\thanks{E-mail: diego.dado2@durham.ac.uk}  % putting on a new line inserts an ugly space in render
Kyle A. Oman,$^{1,2}$
Katherine E. Harborne,$^{1,2}$
Francesca Fragkoudi,$^{1}$
\newauthor
Joop Schaye,$^{3}$
Matthieu Schaller,$^{3,4}$
Alejandro Ben\'{i}tez-Llambay,$^{5}$
Evgenii Chaikin,$^{3}$
\newauthor
Carlos S. Frenk,$^{1}$
Filip Hu\v{s}ko,$^{3}$
Sylvia Ploeckinger$^{6}$
and Alexander J. Richings$^{7,8}$
\\
$^{1}$ Institute for Computational Cosmology, Department of Physics, Durham University, South Road, Durham DH1 3LE, UK\\
$^{2}$ Centre for Extragalactic Astronomy, Department of Physics, Durham University, South Road, Durham DH1 3LE, UK\\
$^{3}$ Leiden Observatory, Leiden University, PO Box 9513, 2300 RA Leiden, the Netherlands\\
$^{4}$ Lorentz Institute for Theoretical Physics, Leiden University, PO Box 9506, 2300 RA Leiden, the Netherlands\\
$^{5}$ Dipartimento di Fisica G. Occhialini, Universit\`{a} degli Studi di Milano Bicocca, Piazza della Scienza, 3 I-20126 Milano MI, Italy\\
$^{6}$ Department of Astrophysics, University of Vienna, T\"{u}rkenschanzstrasse 17, 1180 Vienna, Austria\\
$^{7}$ Centre for Data Science, Artificial Intelligence and Modelling, University of Hull, Cottingham Road, Hull, HU6 7RX, UK\\
$^{8}$ E. A. Milne Centre for Astrophysics, University of Hull, Cottingham Road, Hull, HU6 7RX, UK\\
}

\date{Accepted XXX. Received YYY; in original form ZZZ}

\pubyear{2025}

\begin{document}
\label{firstpage}
\pagerange{\pageref{firstpage}--\pageref{lastpage}}
\maketitle

% Abstract of the paper
\begin{abstract}
We quantify departures from hydrodynamical and centrifugal equilibrium in the gas discs of low-mass ($10^{10.75}<M_\mathrm{200c}/\mathrm{M}_\odot<10^{11}$) galaxies from the COLIBRE cosmological hydrodynamical simulations. We evaluate the full Eulerian acceleration balance in the midplane and show that disequilibrium is widespread: equilibrium-based circular velocity estimates typically have errors of $\geq 10$~per~cent ($\approx 75$~per~cent of midplane gas by mass). Disequilibrium is strongest and the largest associated errors occur in the inner few kiloparsecs that are crucial for constraining the dark matter density profile. Correcting the circular velocity to account for pressure and convective terms does not reliably improve its recovery in strongly perturbed systems where time-dependent forces dominate the residual acceleration budget. Stellar feedback, self-gravitating gas clumps and AGN energy injection account for most strong local perturbations, and large-scale gravitational asymmetries act as a scaffold for disequilibrium. We classify gas discs into coherent, perturbed, and slow/erratic rotators and show that this classification correlates with galaxy properties like mass, morphology and tracers of recent feedback. A majority of galaxies in our sample would be unsuitable for standard rotation curve analysis. Much of the observed diversity in the shapes of dwarf galaxy rotation curves may stem from non-equilibrium gas motions rather than diversity in mass profiles -- resolving the discrepancy is then first and foremost a problem in gas dynamics.
\end{abstract}


%%%%%%%%%%%%%%%%%%%%%%%%%%%%%%%%%%%%%%%%%%%%%%%%%%

%%%%%%%%%%%%%%%%% BODY OF PAPER %%%%%%%%%%%%%%%%%%
\section{Introduction} \label{sec:Introduction}
Despite the remarkable successes of the standard model of cosmology ($\Lambda$CDM; dark energy, $\Lambda$, and cold dark matter, CDM) in reproducing cosmic microwave background (CMB) anisotropies \citep[e.g.][]{Planck2016}, the clustering signal in large-scale structure \citep[e.g.][]{Eisenstein2005}, and gravitational lensing observations \citep[e.g.][]{Heymans2013}, significant challenges remain on galactic scales -- particularly for low-mass galaxies.

Dwarf galaxies -- broadly defined as systems with stellar masses $M_\star < 10^9 \, \mathrm{M}_\odot$ -- fall in a crucial regime for testing both galaxy formation physics and the nature of dark matter. As the most dark-matter-dominated galaxies known, they (potentially) offer clean laboratories for studying gravitational dynamics, and thus dark matter, in low baryon-fraction environments \citep[see e.g.][]{Sales2023}. At the same time, their shallow potential wells render them highly susceptible to feedback processes, introducing complex baryonic effects that complicate both simulations and observations \citep[e.g., baryon-induced core creation, BICC,][]{Pontzen2012, Llambay2019}.

A longstanding issue in this context is the cusp-core problem: the discrepancy between the steep inner density profiles ($\rho \propto r^{-1}$) predicted by dark-matter-only N-body simulations \citep{Navarro1997} and the flatter profiles inferred from observed galaxy rotation curves, once baryonic contributions are accounted for \citep{Flores1994,Moore1994,deBlok2010}. This is typically evaluated by fitting a power law to the inferred dark matter density in the innermost resolved region, and comparing its slope to simulated haloes at similar radii \citep[e.g,][]{deBlok2001,deBlok_Bosma2002, Swaters2003, Oh2011DM_L, Robles2017}. A mismatch suggests tension with $\Lambda$CDM predictions.

Recent advances have shifted the focus from inner slopes to more global discrepancies. Cosmological parameters are now tightly constrained on large scales \citep[e.g.][]{Planck2016}, and dark matter halo scaling relations -- such as the dependence of halo structural properties on cosmology -- are well established \citep[e.g.][]{Correa2015c, Ludlow2016, DiemerJoyce2019}. This enables robust predictions of halo circular velocity profiles once a single parameter (e.g., the maximum circular velocity, $V_{\rm max}$) is known.

Following this logic, \citet{Oman2015} reformulated the cusp–core problem in terms of an inner mass deficit: a shortfall in the circular velocity at a fixed inner radius \citep[e.g. 2 kpc, see also][for a later, more flexible definition]{Santos2020} relative to that expected from the dark matter halo alone. Some galaxies exhibit rotation velocities far below this baseline, even when sharing the same $V_{\rm max}$ as systems without such deficits. The phenomenon spans a wide range of galaxy masses, from dwarfs to large discs, and varies substantially at fixed galaxy properties (e.g., the baryonic surface mass density $\Sigma_{\rm bar}$, or the baryonic mass contribution in the inner regions $\eta_{\mathrm{bar}} = (V_{\rm bar,fid}/V_{\rm fid})^2$).

In extreme cases, the inferred deficits exceed the total baryonic mass, challenging the hypothesis that they arise from BICC \citep{Oman2015}. While such processes -- driven by bursty feedback and gravitational coupling \citep{Navarro_Eke_Frenk1996, Mashchenko2008, Governato2012, Pontzen2012} -- can flatten central density profiles, simulations cannot reproduce the most extreme cores inferred from observations and indicate that these baryonic processes are limited to a narrow range of halo masses \citep{DiCintio2014, Chan2015}. Moreover, not all state-of-the-art simulations predict cores: cosmological hydrodynamical simulations like EAGLE \citep[][see also \citealp{Llambay2019}]{Schaye2014, Schaller2015} and Illustris \citep{Vogelsberger2014, Chua2019} reproduce key galaxy statistics without featuring them. 

In addition to a solution based on modifications due to baryonic feedback, alternative explanations fall into two main categories, as outlined by \citet{Oman2015}: (i) dark matter is not cold and collisionless; or (ii) mass distributions are incorrectly inferred from observations. 

Departures from the central assumption of the cold dark matter framework -- that dark matter behaves as a cold, collisionless fluid -- can alter the internal kinematics of dwarf galaxies. Self-interacting dark matter (SIDM) models have been proposed \citep{Spergel2000}, in which scattering interactions among DM particles are introduced. For interaction cross-sections per unit mass of order $\sigma/m = 1~\mathrm{cm^2~g^{-1}}$, such interactions can thermalize the central regions of DM haloes, producing the constant-density cores observed in dwarfs while retaining the large-scale successes of CDM \citep[see e.g.][]{Rocha2013, Elbert2015, Ren2019}. SIDM has also been invoked to explain the broad diversity of dwarf galaxy rotation curves, however, additional effects due to galaxy formation physics complicate this picture: introducing baryon-DM coupling and fine-tuning seems inevitable \citep{Creasey2017, Kaplinghat2020, Burger2022, Correa2025}.

Concerns about residual systematics in kinematic modeling and observational mass inference have also been a long-standing topic of discussion in the field. Emission line-widths can be affected by ‘beam smearing’ \citep[see][and references therein]{Swaters2009}, or centering, alignment, and seeing in the case of H$\alpha$ slit spectroscopy \citep{Swaters2003, Spekkens2005}. Mass models can be affected by ambiguities in the mass-to-light ratios of galaxies, or the rotation velocity-inclination degeneracy ($V_{\rm rot}-i$). These effects can lead to biased inferences of the underlying dark matter distribution, creating the illusion of a core in intrinsically cuspy haloes \citep[e.g.][]{vandenBosch2001, Dutton2005}. However, analyses based on high-resolution 21-cm radio observations of dwarf galaxies -- particularly from the THINGS \citep{Walter2008} and LITTLE THINGS \citep{Hunter2012} surveys, and the SPARC compilation \citep{Lelli2016} -- continue to indicate the presence of some central cores, where concerns around many of these systematics are greatly alleviated. The mass-to-light ratios for galaxies in the THINGS and LITTLE THINGS surveys were estimated using empirical relations based on stellar optical colours \citep{Bell2001, Bruzual2003}, which have been extensively validated through independent methods \citep[see e.g.][]{Oh2011}. In the SPARC compilation, stellar masses are derived from near-infrared (NIR) luminosities at 3.6$\mu$m -- a wavelength widely considered one of the most reliable tracers of stellar mass \citep{Bell2001}, as the corresponding mass-to-light ratio depends only weakly on age and metallicity across a broad range of star formation histories.

Nevertheless, the possibility that the observed cores are merely artifacts of systematic uncertainties in observation or modelling has not been conclusively ruled out. Dark matter haloes have long been predicted to be triaxial in shape, as indicated by some of the earliest CDM simulations \citep{Davis1985, Frenk88}. Such triaxiality gives rise to an aspherical gravitational potential, with the strongest deviations from spherical symmetry occurring in the central regions \citep{Hayashi2007}. A disc galaxy embedded within such a halo -- typically aligned with the plane defined by the halo’s intermediate and major axes \citep{Hayashi2006} -- thus experiences a non-axisymmetric potential. This induces non-circular motions in the disc, as gas orbits become elongated, producing primarily bisymmetric ($m = 2$ harmonic) variations in azimuthal velocity as a function of projection angle \citep[e.g.][]{Marasco2018, Oman2019}. Even mild asphericity in the potential can generate appreciable non-circular motions \citep[e.g.][]{Hayashi2006}. Although harmonic decompositions of observed dwarf galaxy velocity fields generally indicate small non-circular motion amplitudes \citep{Trachternach2008}, these are likely underestimated in standard model fitting procedures \citep{Chemin2020}. In many cases, velocity variations arising from non-circular motions are absorbed by degenerate model parameters \citep[e.g.][]{Roper2023}, potentially leading to substantial errors in rotation curve measurements -- that is, significant discrepancies between the inferred and true rotational velocities at a given radius.

Despite the promising efforts to resolve the diversity of rotation curve shapes tension in studies targeting the effects of non-circular motions \citep{Marasco2018, Oman2019, Roper2023} and/or pressure gradients \citep[e.g.][]{Valenzuela2007, Chemin2016, Read2016, Pina2025}, they have yet to provide conclusive evidence that the tension has been fully resolved (\citealt{Santos2020}; see also \citealt{Sales2023} for a recent review). This realisation has motivated a closer examination of the hierarchy of assumptions required to reduce the complex dynamics of gaseous discs to a one-dimensional rotation curve -- i.e. an azimuthally averaged rotational velocity profile, $V_{\rm rot}(r)$ -- as a tracer of the underlying gravitational potential.

At the heart of this reduction lies a critical physical assumption: that the gas is in a quasi-steady state, or equivalently, that the system’s timescale of intrinsic velocity evolution are long compared to dynamical timescales -- or, in general, $\partial_t \boldsymbol{v}$ is negligible compared to other terms. This (quasi-)equilibrium condition is invoked -- either explicitly or implicitly -- when equating the observed rotation velocity of the gas to the circular velocity induced by the gravitational potential. Mathematically, this equivalence follows from Euler’s equation for fluid motion. Hence, any correction that relies solely on time-independent terms, such as accounting for thermal or turbulent pressure gradients (e.g. asymmetric drift) or non-centrifugal convective motions (e.g. higher-order harmonics), remains inherently limited by neglecting the time-dependent evolution of the velocity field.

Recent simulation studies have shown that gas in dwarf galaxies often violates these equilibrium assumptions, owing to a range of perturbative processes. Bursty feedback, accretion flows, misaligned angular momentum, and environmental interactions can generate warps, asymmetries, and large-scale non-circular motions \citep[e.g.][]{ElBadry2018, DowningOman2023, Jahn2023, Rey2024, Sands2024}. These effects challenge the validity of conventional 3D or 2D velocity-field modelling pipelines, even when sophisticated corrections to the rotation curve are applied, as such methods implicitly assume steady-state dynamics.

Following this motivation, we present a novel, fully local, Eulerian analysis of midplane gas dynamics in a large sample of dwarf galaxies drawn from the high-resolution COLIBRE cosmological hydrodynamical simulations \citep{Schaye2025, Chaikin2025}. COLIBRE directly and self-consistently models the multiphase interstellar medium (ISM), including dust grains coupled to the chemical network. It reproduces key galaxy statistics at $z = 0$, such as the observed stellar mass function, sizes, star formation rates, H\textsc{i} and H$_2$ masses, gas and stellar metallicities, and dust abundances \citep{Schaye2025}. These successes, and in particular the self-consistent treatment of the cold and dense phases of the ISM and the super-sampling of dark matter particles \citep[hence enhanced resolution of the gravitational potential, mitigating spurious baryon–DM energy transfer][]{Ludlow2019a, Ludlow2021, Ludlow2023}, make COLIBRE particularly well suited to identifying where steady-state assumptions break down, and to tracing the physical origins of such departures from equilibrium.

In this work, we quantify when and where traditional dynamical modelling assumptions fail by directly computing gravitational, pressure-gradient, and convective accelerations in the simulation midplane. Our framework also provides objective diagnostics for identifying dynamically perturbed systems and isolating the physical processes responsible for deviations from equilibrium. 

In the sections that follow, we describe the simulation setup and analysis methodology (Sec.~\ref{sec:Methods}); present results on dynamical diagnostics and the accuracy of mass modeling across our sample (Sec.~\ref{sec:ModelsResults}); identify the primary astrophysical processes driving these systems out of dynamical equilibrium (Sec.~\ref{sec:Origins_of_diseq}); and investigate some broad differences between perturbed and unperturbed systems (Sec.~\ref{sec:Classification}). We conclude by discussing the implications of our findings for interpreting galaxy rotation curves and testing dark matter models (Sec.~\ref{sec:Conclusions}).

\section{Methods} \label{sec:Methods}

\subsection{COLIBRE simulations} \label{sec:Methods:Colibre}
The COLIBRE (COLd ISM and Better REsolution) simulation suite\footnote{COLIBRE is a project of the Virgo Consortium for cosmological supercomputer simulations. Further information on the project, including publications, team members, and simulation visualisations, is available at \url{https://colibre.strw.leidenuniv.nl/}.} consists of cosmological hydrodynamical runs performed with the public \textsc{Swift} code \citep{Schaller2024}, version 2025.04, extended with a bespoke subgrid physics implementation. The simulations adopt the flat $\Lambda$CDM cosmology from the "DES Y3 + All Ext." dataset \citep{Abbott2022}, consistent with the fiducial FLAMINGO simulations \citep{Schaye2023}, with parameters: $h = 0.681$, $\Omega_\mathrm{m} = 0.306$, $\Omega_\mathrm{b} = 0.0486$, $\Omega_\Lambda = 0.6939$, $n_s = 0.967$, $\sigma_8 = 0.807$, $A_s = 2.099 \times 10^{-9}$, and $\sum m_\nu c^2 \approx 0.06~\mathrm{eV}$.

COLIBRE uses the \textsc{sphenix} hydrodynamics scheme \citep{Borrow2022}, a smoothed particle hydrodynamics (SPH) formulation optimized for galaxy formation. This method employs a density–energy approach with artificial viscosity and conduction terms, using a quartic spline kernel with smoothing length $\eta = 1.2348$, corresponding to approximately 65 weighted neighbours. 

Key subgrid physics modules in COLIBRE enhance the realism of the interstellar and circumgalactic media, allowing gas to cool to temperatures as low as $\sim$10 K. Radiative cooling, heating, and species abundances (atoms, ions, and molecules) are computed using \textsc{hybrid-chimes} \citep{Richings2014a, Richings2014b, Ploeckinger2025}, which evolves the non-equilibrium abundances of electrons and nine hydrogen and helium species on the fly, and treats an additional 147 metal species using quasi-equilibrium chemistry. Dust physics is modeled self-consistently for three grain species and two size bins, coupled to the chemistry and cooling network \citep{Trayford2025}. Star formation follows a Schmidt law based on gravitational instability, without imposing an effective ISM equation of state \citep{Nobels2024}.

Stellar mass loss includes metal injection from asymptotic giant branch (AGB) stars, core-collapse supernovae (CCSNe), and type Ia supernovae (SNIa) with updated yields and a calibrated SNIa delay-time distribution (Correa at al., in preparation). Pre-supernova feedback from massive stars -- including stellar winds, radiation pressure, and H\textsc{ii} region photoionization -- is explicitly modelled \citep{Llambay2025}. CCSN and SNIa feedback follow a stochastic thermal scheme \citep{DallaVecchia2012}, supplemented by a lower-energy kinetic mode that enhances turbulence in the case of CCSN \citep{Chaikin2023}. The simulations analysed here employ the fiducial purely thermal active galactic nucleus (AGN) feedback model \citep[][]{BoothSchaye2009}, updated to improve feedback sampling for low-mass black holes. We have explicitly checked that similar runs using the hybrid AGN model of \citet{Husko2025b} -- which tracks black hole spin and combines thermally driven winds with kinetic jet feedback -- yield results consistent with the thermal AGN model in all parts of our analysis.

COLIBRE employs four times more dark matter than baryonic particles, improving sampling of the gravitational potential and suppressing spurious CDM–baryon energy transfer \citep{Ludlow2019a, Ludlow2021, Ludlow2023}. The absence of an imposed ISM equation of state allows gas structure to develop on sub-kiloparsec scales, yielding a higher effective resolution than in previous simulations with comparable baryonic particle masses. Together, these features enable a fully multiphase ISM and improved modelling of the gravitational potential, representing a major resolution improvement over earlier galaxy formation simulations (see \citealt{Schaye2025, Chaikin2025} for full model and calibration details, respectively).

Haloes and galaxies are identified through a three-step process: (i) a friends-of-friends (FoF) algorithm applied to dark matter particles with a linking length of 0.2 times the mean interparticle spacing; (ii) the hydrodynamics-optimised \textsc{hbt-herons} subhalo finder \citep{ForouharMoreno2025}, a descendant of \textsc{hbt+} \citep{Han2018}, which identifies bound structures and constructs merger trees; and (iii) the \textsc{soap} tool \citep{McGibbon2025}, which computes galaxy and halo properties within fixed and overdensity-based apertures, adopting $M_{200\mathrm{c}}$ and $R_{200\mathrm{c}}$ as the halo mass and radius definitions\footnote{The subscript $200\mathrm{c}$ indicates a quantity evaluated within a spherical aperture with a mean enclosed density of $200$ times the critical density for closure, $\rho_\mathrm{c}=3H^2/8\pi G$.}. Throughout the analysis, we refer to individual galaxies via their \textsc{hbt-herons} track ID, which uniquely links each $z=0$ halo to its main progenitor across simulation snapshots and thus provides a stable and physically meaningful identifier for each system.

For this work, we use a COLIBRE run with a comoving box size of $25~\mathrm{cMpc}$ at m5 resolution, corresponding to baryonic and dark matter particle masses of $\approx2\times10^5~\mathrm{M_\odot}$ and $\approx3\times10^5~\mathrm{M_\odot}$, respectively. The Plummer-equivalent gravitational softening length is $0.35~\mathrm{kpc}$ (physical) for $z<1.57$.

\subsection{Gas discs of dwarf galaxies: sample and definitions} \label{sec:Methods:GasDiscs}

We analyse central haloes, defined as those with no more massive structure identified within their FoF group, with $10.75\leq\log_{10}(M_{200\mathrm{c}}/\mathrm{M_\odot}) \leq 11$ at $z = 0$. This selection yields 122 galaxies. Their corresponding stellar masses, $M_\star \sim 10^8$–$10^9~\mathrm{M_\odot}$ (with only a few slight exceptions), place them in the regime of bright dwarf galaxies while ensuring sufficient particle resolution ($>10^5$ total particles per system, and typically $\gtrsim10^3$ stellar or gas particles). No preselection is applied based on gas content, morphology, or dynamical state to avoid biasing correlations with the (dis)equilibrium state of the gas. 

Our sample spans diverse morphologies and gas fractions, with neutral hydrogen (H\textsc{i}) masses ranging from $10^7$ to $10^9\,\mathrm{M_\odot}$ and H\textsc{i} fractions $f_{\rm H\textsc{i}} = M_{\rm H\textsc{i}}/M_\star$ between (roughly) 0.1 and 10. This diversity is essential for testing the robustness of dynamical modeling assumptions across different galaxy types.

\subsubsection{Disc extent}

The radial extent of each gas disc is defined using the cumulative H\textsc{i} mass profile\footnote{While not unique to COLIBRE, this approach is greatly facilitated by the self-consistent treatment of H\textsc{i} fractions directly within the simulation outputs, avoiding the need for post-processing schemes commonly required in other simulations \citep[see, e.g., the discussion in sec.~3.3 of][]{Oman2019}.}, $M_{\rm H\textsc{i}}(<r)$. We identify the radius $R_{\rm ext,H\textsc{i}}$ at which this profile flattens, i.e., where the logarithmic slope satisfies
\[
\frac{\mathrm{d} \log_{10} [M_{\rm H\textsc{i}}(<r)/\mathrm{M_\odot}]}{\mathrm{d}r} = \lambda,
\]
with $\lambda = 0.01~\mathrm{dex}~\mathrm{kpc}^{-1}$ chosen empirically to ensure robustness across the sample. For a rotationally supported disc this is usually similar to the radius enclosing 90~per~cent of the H\textsc{i} mass in a galaxy \citep[$R_{90,\rm H\textsc{i}}$, see e.g.][]{DowningOman2023, Sands2024}; but unlike an enclosed mass threshold, it is a more reliable tracer of the outer edge of a galaxy's gas in disturbed or irregular morphologies.

\subsubsection{Warped disc frame}

To accommodate warped gas morphologies, we construct a coordinate system locally aligned with the radially-varying angular momentum vector of the gas:
\[
\hat{L}_{\rm gas}(r) = \frac{\sum_i m_i \, \boldsymbol{x}_i \times \boldsymbol{v}_i}{\left| \sum_i m_i \, \boldsymbol{x}_i \times \boldsymbol{v}_i \right|},
\]
where the sums run over gas particles in a spherical shell at radius $r$. Here, $m_i$ is the mass of particle $i$, $\boldsymbol{x}_i$ is its position vector relative to the coordinate origin, and $\boldsymbol{v}_i$ is its peculiar velocity vector in the global frame. The profiles are evaluated in radial bins of fixed width $\Delta r = 0.2~\mathrm{kpc}$, which are identical to those used for the cumulative H\textsc{i} mass profiles described above. Interpolating $\hat{L}_{\rm gas}(r)$ as a continuous function of $r$ -- in this work achieved via a broken third-order polynomial, `cubic spline' -- yields a smoothly varying local $z$-axis that defines the warped disc midplane.

The coordinate origin is fixed at the potential minimum as identified by \textsc{hbt-herons}. The velocity centroid $\boldsymbol{v}_{\rm cen}$ is computed as the mass-weighted average velocity of dark matter particles within a sphere of radius $r_{\rm vel\,cen} = 2\,r_{\rm conv}$, where $r_{\rm conv}$ is the convergence radius following \citet{Power2003}. This choice ensures the frame is centered on a physically meaningful, dynamically stable origin and velocity frame, minimizing artificial harmonic modes (e.g., spurious $m=1$ modes) in the potential and velocity fields caused by miscentering.

Sampling coordinates in this warped frame are
\[
\left\{ \boldsymbol{x} \right\}_{\rm local} = \left\{ (r_i, \phi_j, 0) \right\}_{\rm local},
\]
where the local spherical coordinates are radius $r_i$ and azimuthal angle $\phi_j$ -- we note that setting $\theta_{\rm local} = 0$ for all sampling points is equivalent to midplane sampling, and that in this special case the spherical and cylindrical ($r, \phi$) sets coincide. Specifically, $r_i$ ranges from $0.1\,\mathrm{kpc}$ to $R_{\rm ext,H\textsc{i}}$ in steps of $\Delta r = 0.2~\mathrm{kpc}$, and $\phi_j$ spans $0^\circ$ to $358^\circ$ in $\Delta \phi = 2^\circ$ increments. The azimuthal direction $\phi_{\rm local}$ is defined such that the local tangential velocity satisfies $\langle v_\phi \rangle_\phi(r_i) \geq 0$. We use the symbol $\langle A \rangle_\phi$ to denote the azimuthal average of any quantity $A$.

\subsection{Dynamics of gas discs: fluid models \& circular velocity reconstruction} \label{sec:Methods:FluidModels}

As introduced in Sec.~\ref{sec:Introduction}, this work aims to critically assess the validity of commonly adopted assumptions when modeling the kinematics of galactic gas discs.

The evolution of the velocity field in a collisional, inviscid fluid system -- such as galactic gas -- is governed by Euler’s acceleration equation:
\begin{equation}
\label{eq:EulerAcceleration}
\frac{\partial \boldsymbol{v}}{\partial t} + (\boldsymbol{v} \cdot \boldsymbol{\nabla}) \boldsymbol{v} + \frac{\boldsymbol{\nabla} P_{\rm eff}}{\rho} = - \boldsymbol{\nabla} \Phi,
\end{equation}
where $\boldsymbol{v}$ is the fluid velocity, $\rho$ the density, $\Phi$ the gravitational potential, and $P_{\rm eff}$ the effective pressure. The latter includes contributions from both the thermal pressure, $P_{\rm th} = (\gamma - 1)u\rho$, where $u$ is the internal energy per unit mass and $\gamma$ the ratio of specific heats, and the turbulent pressure at the resolution limit, $P_{\rm turb} = \rho \sigma_{v}^2/3 = \rho \sigma_{v, \rm 1D}^2$, where $\sigma_{v}$ is the weighted velocity dispersion inside the SPH kernel. Equation~(\ref{eq:EulerAcceleration}) can be viewed as a Jeans-equation analogue for gaseous systems, relating observable gas kinematics to the underlying gravitational potential generated by baryonic and dark matter, and is always a statement of local force balance, regardless of whether the system is in equilibrium.

In observational analyses, the partial time-derivative term $\partial_t \boldsymbol{v}$ is often neglected under the assumption of quasi-steady dynamics. This omission is not justified by observational limitations (i.e. the fact that galaxies are observed at a single epoch, rendering the term unmeasurable) but rather by the implicit physical assertion that the system’s timescales of intrinsic velocity evolution are long compared to dynamical timescales. When this assumption breaks down, the equilibrium approximation fails, and $\partial_t \boldsymbol{v}$ becomes physically significant. Measuring its contribution therefore provides a direct probe of the degree of disequilibrium in the gas, and quantifies the limitations inherent in steady-state kinematic and mass modelling pipelines.

Under the steady-state limit ($\partial_t \boldsymbol{v} \approx 0$), the radial component of Equation~(\ref{eq:EulerAcceleration}) defines what we refer to as the `equilibrium model'\footnote{In this framework, any residual difference between the left- and right-hand sides of Equation~(\ref{eq:EquilibriumModel}) directly quantifies the neglected time-dependent term in Equation~(\ref{eq:EulerAcceleration}):
\[
\partial_t v_r = - \left\{\left[(\boldsymbol{v}\cdot \boldsymbol{\nabla}) \boldsymbol{v}\right]_r + \frac{1}{\rho} \frac{\partial P_{\rm eff}}{\partial r} - g_r \right\},
\]
which represents the local radial acceleration of the fluid at fixed spatial coordinates, and thus provides a physically meaningful measure of departures from equilibrium.}:
\begin{equation}
\label{eq:EquilibriumModel}
\left[(\boldsymbol{v}\cdot \boldsymbol{\nabla}) \boldsymbol{v}\right]_r + \frac{1}{\rho} \frac{\partial P_{\rm eff}}{\partial r} = -\frac{\partial \Phi}{\partial r} \equiv g_r.
\end{equation}

Further simplifications arise by assuming an idealized, axisymmetric disc with negligible radial and vertical motions, and where the pressure gradient is either small or corrected separately (e.g. via `asymmetric drift corrections'; see \citealt{Valenzuela2007}, appendix~A, for a helpful discussion of this technique and related terminology). Under these conditions, the radial convective term reduces to the centrifugal acceleration, yielding the familiar relation:
\begin{equation}
\label{eq:NaiveModel}
v_\phi = \sqrt{r\left| \frac{\partial \Phi}{\partial r} \right|} \equiv v_c,
\end{equation}
where $v_\phi$ is the local azimuthal (rotational) velocity and $v_c$ the corresponding local circular velocity\footnote{The commonly used expression $v_c(r)=\sqrt{GM(<r)/r}$ only holds under spherical symmetry. In general, the fundamental dynamical variable is the radial gravitational acceleration $g_r(r,\phi)$, which is well defined for any potential, even in systems that do not admit closed circular orbits. The `circular velocity' used throughout this work therefore refers to the local force-defined quantity $v_c(r,\phi) \equiv \sqrt{r\,|g_r(r, \phi)|}$, representing the centrifugal speed required to balance the instantaneous radial gravitational force in the disc midplane, and can vary as a function of $\phi$.}.

The evaluation of Equations~(\ref{eq:EulerAcceleration})–(\ref{eq:NaiveModel}) requires a change of formalism for the hydrodynamics: in a particle-based (Lagrangian) representation, the intrinsic acceleration $\partial_t \boldsymbol{v}$ and the convective acceleration $(\boldsymbol{v}\cdot \boldsymbol{\nabla})\boldsymbol{v}$ are coupled within the total (material) derivative $\mathrm{d}\boldsymbol{v}/\mathrm{d}t$. In the following section (Sec.~\ref{sec:Methods:SPHSampling}), we describe the SPH-based continuous-field reconstruction developed to evaluate all terms in Equation~(\ref{eq:EulerAcceleration}) independently in Eulerian fashion, whilst ensuring full consistency with the simulation’s hydrodynamics conventions.

\subsection{SPH-based continuous field sampling} \label{sec:Methods:SPHSampling}

Hydrodynamical quantities and their derivatives at warped disc grid points are computed using a post-processing scheme consistent with the \textsc{sphenix} SPH formalism, using the same smoothing kernel and neighbour criteria as in the COLIBRE simulation (see Sec.~\ref{sec:Methods:Colibre}). Formally, our procedure involves:

\begin{enumerate}
    \item Smoothing length estimation: 
    At each sampling point $\boldsymbol{x}$, we solve for the smoothing length $h(\boldsymbol{x})$ via the implicit SPH number density equation
    \[
    n(\boldsymbol{x}) = \sum_i W(\boldsymbol{x} - \boldsymbol{x}_i, h(\boldsymbol{x})) = \left(\frac{\eta}{h(\boldsymbol{x})}\right)^3,
    \]
    using Newton–Raphson iteration to relative tolerance $10^{-4}$. Here, $W$ is the smoothing kernel, $\eta$ is a dimensionless kernel parameter fixing the weighted neighbour number (defined in Sec.~\ref{sec:Methods:Colibre}), and the sum goes over all the particle neighbours of the point, $i$, with position vectors $\boldsymbol{x}_i$.

    \item Density at sampling points and interpolation:
    The density at an arbitrary sampling point $\boldsymbol{x}$ is computed via the traditional SPH equation, 
    \[
    \rho(\boldsymbol{x}) = \sum_i m_i W(\boldsymbol{x} - \boldsymbol{x}_i, h(\boldsymbol{x})),
    \]
    where $m_i$ is the mass of the neighbour particle $i$ and $h(\boldsymbol{x})$ is the smoothing length evaluated at $\boldsymbol{x}$.
    Scalar, vector, and tensor fields are then interpolated at sampling points $\boldsymbol{x}$ via
    \[
    A^a(\boldsymbol{x}) = \frac{\sum_i m_i A_i^a \, W(\boldsymbol{x} - \boldsymbol{x}_i, h(\boldsymbol{x}))}{\rho(\boldsymbol{x})},
    \]
    where $A^a_i$ is the field value of particle $i$. The notation $A^a$ indicates any component of an $n$-dimensional field.

    \item First-order derivative operator and evaluation:
    The interpolation scheme described above allows a self-consistent first-order derivative operator for any $n$-dimensional field $A$. This operator is given by 
    \[
    \partial_b A^a(\boldsymbol{x}) = \frac{\sum_i m_i (A_i^a - A^a(\boldsymbol{x})) \partial_b W(\boldsymbol{x} - \boldsymbol{x}_i, h(\boldsymbol{x}))}{\rho(\boldsymbol{x})},
    \]
    where $\partial_b$ denotes the partial derivative of a quantity with respect to the $b$ component of the sampling position vector, i.e. $\partial/\partial \boldsymbol{x}^b$.

\end{enumerate}

Gravitational acceleration is computed as
\[
\boldsymbol{g}(\boldsymbol{x}) \equiv -\boldsymbol{\nabla} \Phi(\boldsymbol{x}) = \sum_j \frac{G~m_j~(\boldsymbol{x}_j - \boldsymbol{x})}{\bigg(\sqrt{|\boldsymbol{x}_j - \boldsymbol{x}|^2 + \epsilon_{\rm soft}^2}\bigg)^3},
\]
i.e. via direct $N$-body summation over all particles within $R_{200\mathrm{c}}$ (both baryonic and dark, indexed by $j$), employing a softened Newtonian potential consistent with the COLIBRE redshift-dependent softening length, $\epsilon_{\rm soft}$, and the adopted gravitational constant, $G$. Since we focus our analysis on the scale of the disc ($\approx 0.2R_\mathrm{200c}$) this truncation of the particle distribution when evaluating $\boldsymbol{g}(\boldsymbol{x})$ is an acceptable approximation.

All calculations are performed in the global coordinate frame and projected into the warped local frame only at the final step, ensuring that radial, azimuthal and vertical components of fields and their derivatives remain physically meaningful. For illustrative purposes, Appendix~\ref{app:MorphologySampling} presents a representative example from a galaxy in our sample, illustrating the warped disc frame construction alongside the output of our SPH-based continuous field sampling procedure. These visualisations provide context for the numerical procedures outlined here and in Sec.~\ref{sec:Methods:GasDiscs}.

This continuous-field reconstruction extends particle-based analyses into smooth, spatially resolved descriptions of the gas dynamics. Technically, it corresponds to a Shepard \citep{Shepard1968} SPH–gather scheme employing the same kernel definitions and neighbour criteria as the simulation itself, thereby ensuring consistency with its hydrodynamical framework. An important advantage of this scheme is that each sampling point attains an effective spatial resolution determined self-consistently by the local fluid density. This guarantees that reconstructed quantities vary smoothly across the disc, mitigating the shot noise typical of scatter-type interpolations, where individual particle contributions dominate the local estimate. The Shepard normalization further enforces that the effective kernel sum satisfies $\sum_i \tilde{W} = 1$, eliminating one of the main sources of bias in standard SPH interpolation -- deviations from unity kernel normalization -- which can otherwise lead to systematic under- or overestimation of field amplitudes in sparsely sampled regions.

A caveat of standard Shepard formalisms is that they are not strictly conservative in mass or momentum \citep{Reinhardt2019}. However, since the sampling points are not interpreted as finite-volume fluid cells, but as infinitesimal tracers defining a continuous warped midplane, the lack of strict conservation poses no conceptual inconsistency for the subsequent equilibrium analysis. Our reconstruction does not attempt to reproduce the exact numerical forces used during the simulation’s time integration, but rather to recover the instantaneous, continuous field structure implied by the discrete SPH representation.

When evaluating the hydrodynamical Equations~(\ref{eq:EulerAcceleration})–(\ref{eq:EquilibriumModel}), we adopt several deliberate departures from the native SPH formulation used during the simulation. In particular, we exclude (i) artificial viscosity terms, which are numerical constructs introduced for entropy dissipation and shock capturing \citep{Borrow2022}, and (ii) the contribution from the large-scale gravitational field of the full simulation box. Our focus is instead on the physical force balance of the gas under the assumption of an inviscid, collisional fluid.

Artificial viscosity in SPH serves a necessary numerical role but introduces several well-known side effects: (i) spurious shear stresses, (ii) unphysical angular-momentum transport, and (iii) damping of turbulent motions at the resolution scale. In our reconstruction, the latter effect is at least partially accounted for by the inclusion of the turbulent pressure term, $P_{\rm turb} = \rho\sigma_{v,\mathrm{1D}}^2$, which captures the local subgrid velocity dispersion and its contribution to pressure support. Consequently, neglecting the explicit artificial-viscosity term does not significantly bias the inferred quasi-static balance of forces, since its dominant dynamical impact -- turbulent damping -- is represented implicitly through $P_{\rm turb}$.

Framing the reconstruction in this way ensures that the recovered fields address the fundamental equilibrium problem central to rotation-curve analyses: whether pressure gradients and convective accelerations balance the internal gravitational potential, and where astrophysical processes such as feedback or accretion drive measurable departures from this balance.

\subsection{Feedback tracing and filtering} \label{sec:Methods:FeedbackTracing}

Since stellar and AGN feedback are primary candidates for driving disequilibrium in the interstellar medium, we construct a filtering scheme to identify regions of the disc that have been recently affected by these processes. This enables a direct spatial and temporal association between localised feedback sites and the gas dynamics in those sites.

For each warped midplane sampling point, we select gas and stellar particles within a spherical distance $r \leq 1~\mathrm{kpc}$. This scale encompasses the local environment dynamically sensitive to recent feedback while minimising contamination from unrelated structures. A subsequent temporal filter retains only:
\begin{enumerate}
\item gas particles flagged as having experienced feedback from an AGN or from a supernova (CCSN/SNIa) within a lookback time $t_{\rm lb} \leq t_{\rm cutoff}$; and
\item young stars with age $\tau_\star \leq t_{\rm cutoff}$.
\end{enumerate}
We test multiple cutoff times (e.g., $t_{\rm cutoff}=10$, 20, 50, and 100 Myr) to verify that the inferred spatial correlations and disequilibrium metrics are robust to this selection.

Feedback strength is quantified via energy-equivalent proxies:
\begin{itemize}
\item Thermal feedback: for purely thermal energy injection (e.g. CCSN or SNIa events), the deposited energy is taken as proportional to the local temperature increase, $\Delta E \propto k_B \Delta T$, where $k_B$ is Boltzmann’s constant.
\item Kinetic feedback: for momentum-driven injection, we compute $\Delta E = \tfrac{1}{2} m (\Delta v)^2$, where $m$ is the particle mass and $\Delta v$ the velocity kick imparted at the last feedback event.
\item AGN feedback: the simulation directly records the thermal or kinetic energy received per event, and this quantity is adopted without further conversion.
\item Generic stellar feedback: additional stellar processes such as winds, radiation pressure, and ionising radiation are traced via the mass of young stars formed. We have checked that the spatial distribution of active H\textsc{ii} regions (tracked in the COLIBRE model) closely follows that of recently formed stars, justifying this proxy.
\end{itemize}

Feedback proxies are accumulated across the warped disc and, where indicated, normalised by the corresponding local disc area to yield surface energy-density measures of feedback impact. This filtering procedure isolates regions where the gas has been recently perturbed by feedback and thus provides a physically motivated framework for linking energetic injection to local deviations from force equilibrium. The motivation for, limitations of, and implications of this analysis are discussed in detail in Sec.~\ref{sec:Origins_of_diseq:Budget}.

\section{Disequilibrium in Dwarf Galaxies} \label{sec:ModelsResults}

In rotation curve studies, it is standard practice to assume that the azimuthal velocity of gas directly traces the local gravitational potential. While this assumption is often motivated by the apparent rotational support, regularity of emission lines, and morphological coherence of gas discs, such observational signatures alone establish neither dynamical equilibrium, nor purely circular motion.

As outlined in Sec.~\ref{sec:Introduction}, a range of effects can influence mass inferences from gas kinematics. Some stem from projection and observational artifacts -- such as beam smearing, miscentering, or inclination uncertainties -- while others arise from modeling degeneracies, particularly those between geometric assumptions and non-circular motions. Yet, almost all inference methods adopt, either explicitly or implicitly, the premise that the gas is in hydrodynamical equilibrium. This assumption is seldom tested in a rigorous, spatially resolved manner in either observations or simulations. 

\subsection{Local performance of fluid models}\label{sec:ModelsResults:OneGal}

\begin{figure*}
	\includegraphics[width = \textwidth]{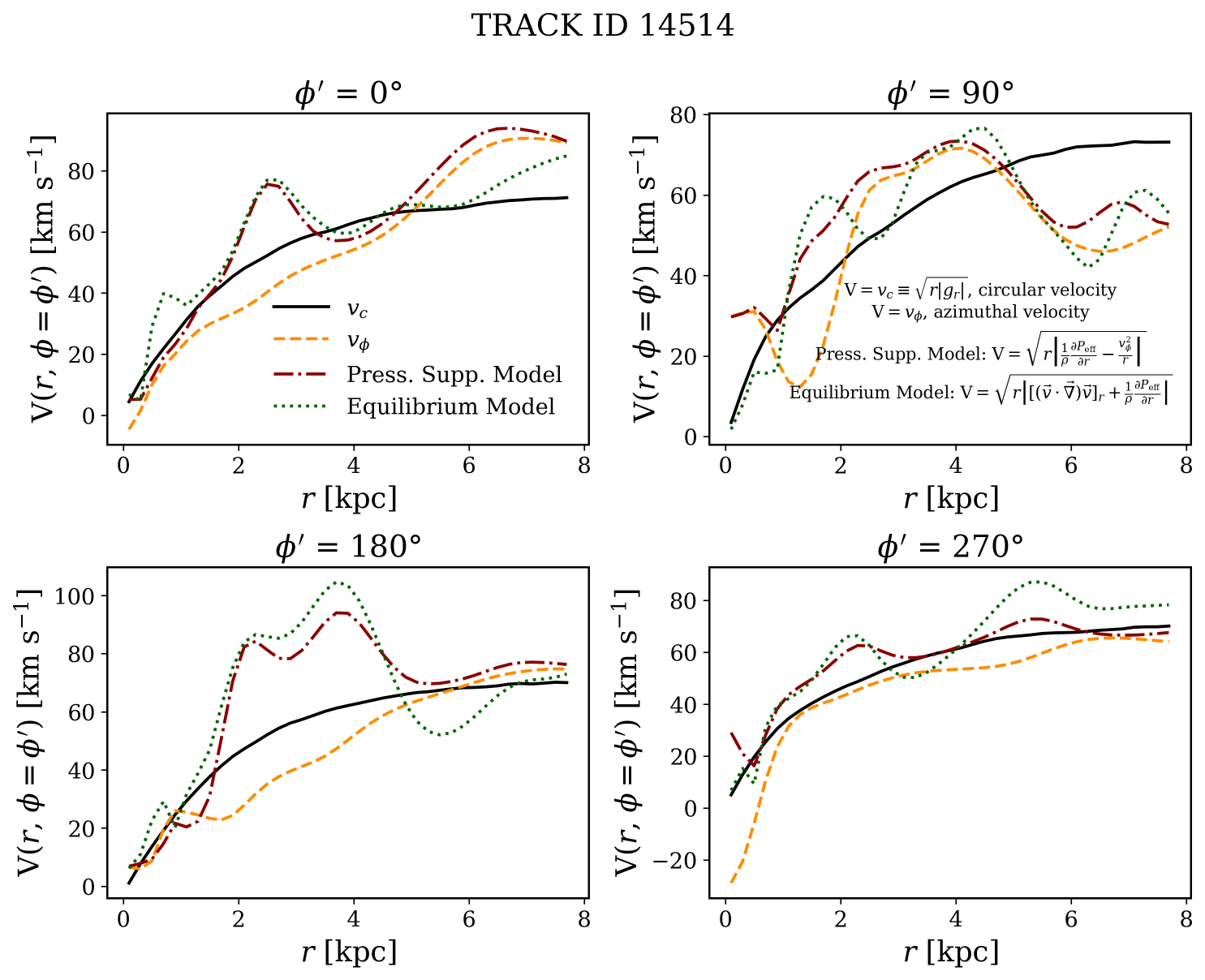}
    \caption{Comparison of different fluid dynamical models in recovering the true circular velocity profile for a representative dwarf galaxy. Each panel shows radial profiles of reconstructed circular velocity along a fixed azimuthal direction ($\phi' = {0, 90, 180, 270}^\circ$). The black solid line shows the true $v_c$ from the gravitational potential; lines that are coloured and non solid show reconstructions from different models (legend). More physically complete models do not always outperform the na\"{i}ve $v_\phi = v_c$ assumption.}
    \label{fig:RC Reconstruction}
\end{figure*}

A central aim of this work is to test fluid models in a local, physically self-consistent way. We evaluate whether the steady-state Euler equation (Equation~(\ref{eq:EquilibriumModel})) is satisfied in simulated gas discs, and how well different models recover the true gravitational potential.

Fig.~\ref{fig:RC Reconstruction} presents a case study of one simulated galaxy (\textsc{hbt-herons} track ID $= 14514$, stellar mass $M_\star \approx 2.80 \times 10^8~\mathrm{M_\odot}$), showing reconstructed circular velocity profiles along four azimuthal slices. The black solid line is the true circular velocity from the potential (i.e. the local force-defined speed $v_c(r,\phi)=\sqrt{r\,|g_r(r,\phi)|}$); broken, coloured curves correspond to different models as indicated in the legend.
Two main points stand out.

First, all reconstructions exhibit significant azimuthal variability at fixed radius, indicating strong model-dependent sensitivity to local dynamics and asymmetries. Such non‑axisymmetric structures are well documented in empirical gas discs' hydrodynamical fields (e.g., harmonic decomposition of moment maps revealing spiral patterns and/or n-symmetric velocity fields; see \citealt{Schoenmakers1997, Trachternach2008, Sellwood2010}). However, the behaviours observed here underscore a deeper implication: azimuthal averaging, which is  commonly used in rotation curve analyses (both observational and theoretical), can obscure critical localised discrepancies. These deviations are physical signatures of disequilibrium physics, driven by ongoing astrophysical processes (as we will discuss in Sec.~\ref{sec:Origins_of_diseq}).

Second, and perhaps more strikingly, the `Equilibrium model' (dark green/dotted) -- despite being the most comprehensive physical model -- is frequently outperformed by the simplistic $v_\phi = v_c$ assumption (orange/dashed); see, for example, the increased absolute deviation of the equilibrium reconstruction from the true circular velocity in the $\phi' = 180^\circ$ slice. The simpler `Pressure support model' (brown/dash-dotted), which attempts a correction to $v_\phi$ under the assumption that hydrostatic pressure gradients are the dominant non-equilibrium effect, fares similarly -- in this galaxy this assumption is justified across most of the disc, although this does not hold for all galaxies. This counterintuitive outcome echoes the findings of \citet{Sands2024} who studied disequilibrium in gas discs from the FIRE-3 simulation suite \citep{Hopkins2023}. They argue that the poor performance of the equilibrium model can be traced to the unbalanced nature of Eulerian terms. Specifically, radial components of the convective derivative, $(\boldsymbol{v}\cdot\nabla)\boldsymbol{v}$, often arise from the same non-equilibrium processes that generate time variations in the velocity field, $\partial_t \boldsymbol{v}$. This coupling undermines the validity of equilibrium reconstructions. That we find similar trends in COLIBRE reinforces a key point: adding correction terms for non‑circular motions or pressure support does not guarantee more accurate mass inferences. The limiting factor is whether the gas is actually in hydrodynamical equilibrium.

In summary, the standard equality $v_\phi = v_c$ breaks down in a systematic and physically meaningful way. Even in the most complete time-independent fluid models, disequilibrium produces frequent, significant errors in circular velocity estimates. This challenges the foundational assumptions of rotation curve analysis and suggests that purely analytic corrections are limited by unobservable, time‑dependent processes.

\subsection{Population-wide radial trends in disequilibrium physics}
\label{sec:ModelsResults:PopRadTrends}

\begin{figure*}
\centering
\includegraphics[width=1\columnwidth,height=0.33\textheight]{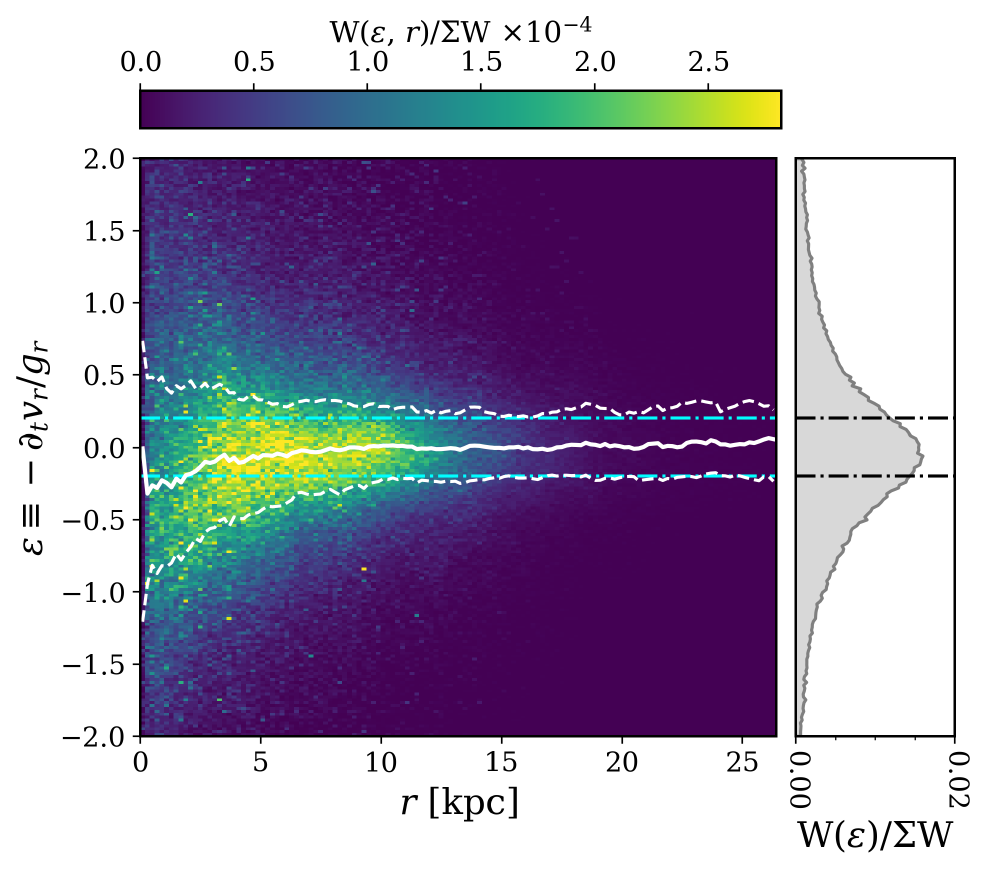}
\includegraphics[width=1\columnwidth,height=0.33\textheight]{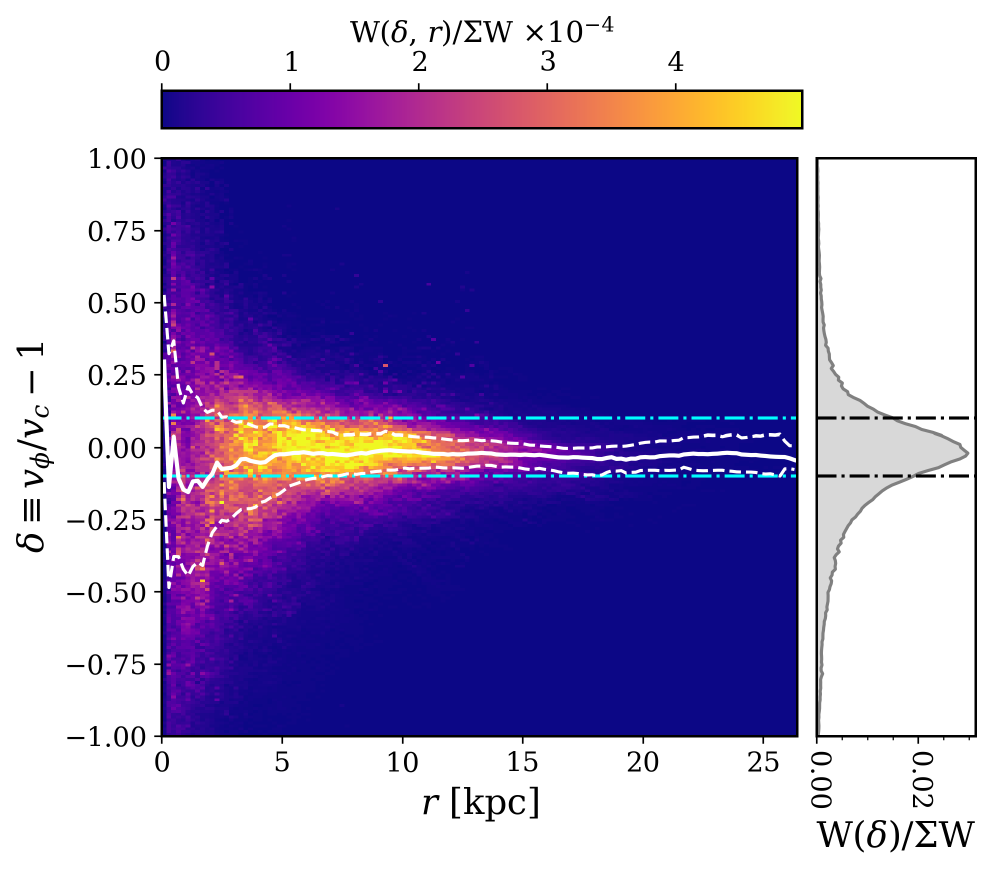}
\caption{Radial distributions of the $\varepsilon(r, \phi)$ and $\delta(r, \phi)$ parameters (Equation~(\ref{eq:Equilibrium Parameters})) weighted by gas mass across the full galaxy sample, respectively $\mathrm{W}(\varepsilon(r, \phi), r)$ and $\mathrm{W}(\delta(r, \phi), r)$. Each panel displays both the joint and marginal distributions of the respective parameter. \textit{Left panel}: Stacked 2D histogram of $\varepsilon$ vs. galactocentric radius, $r$, where the colour scale indicates the joint weighted density, $\mathrm{W}(\varepsilon, r)/\Sigma \mathrm{W}$, summed across all galaxies. The vertical subpanel to the right shows the normalised marginal (1D) distribution, $\mathrm{W}(\varepsilon)/\Sigma \mathrm{W}$.  \textit{Right panel}: Same as left panel, but for the $\delta$ parameter. Cyan dash-dotted lines in the main panel and black dash-dotted lines in the side panel denote the parameter range associated with $\leq 10$~per~cent error in circular velocity inference based on the corresponding dynamical model. In the main subpanel, white solid and white dashed lines show the median and central $50$~per~cent envelopes of the respective conditional distributions, $\mathrm{W}(\varepsilon|r)$ and $\mathrm{W}(\delta|r)$.}
\label{fig: Radial eps-delta profiles}
\end{figure*}

To test whether the results from the galaxy shown in the previous section are representative of the broader dwarf galaxy population, we extend our analysis to the full simulation sample. We introduce two diagnostic parameters that quantify deviations from hydrodynamical equilibrium and centrifugal balance, respectively. These metrics also correspond to the fractional error in inferred circular velocity when adopting either the full equilibrium model (Equation~(\ref{eq:EquilibriumModel})) or the na\"{i}ve assumption that the rotational velocity exactly traces the mass distribution (Equation~(\ref{eq:NaiveModel})).

The choice to express both discrepancies in fractional terms arises from the large variety of galaxies in the sample, which span a wide range of circular velocity profiles (i.e., rising at different rates and up to slightly different $V_{\rm max}$). Hence, we define:
\begin{subequations}
\label{eq:Equilibrium Parameters}
\begin{equation}
\varepsilon \equiv \left(\frac{[(\boldsymbol{v} \cdot \boldsymbol{\nabla}) \boldsymbol{v}]_r + \frac{1}{\rho} \frac{\partial P_{\rm eff}}{\partial r}}{g_r}\right) - 1 = -\frac{\partial_tv_r}{g_r},
\label{eq:Eq4a}
\end{equation}
\begin{equation}
    \delta \equiv \frac{v_\phi}{v_c} - 1.
     \label{eq:Eq4b}
\end{equation}
\end{subequations}
While $\delta$ has a familiar physical interpretation -- indicating the extent to which rotation supports a gas parcel against gravity --  $\varepsilon$ is better interpreted as a diagnostic of net radial imbalance. It compares the total time-independent accelerations experienced by a gas parcel to the local gravitational acceleration. As such, $\varepsilon \ne 0$ implies that the local fluid is under- ($\varepsilon < 0$) or over-supported ($\varepsilon > 0$), and would develop a net radial inflow or outflow if evolved in an Eulerian framework (assuming an initial $v_r \approx 0$). 

Fig.~\ref{fig: Radial eps-delta profiles} shows the radial distribution of both diagnostic parameters across the galaxy sample. The left panel corresponds to the mass-weighted density of $\varepsilon(r, \phi)$ -- i.e. $\mathrm{W}(\varepsilon, r)/\Sigma \mathrm{W}$ -- while the right panel shows the same for $\delta(r, \phi)$. In each case, the main subpanel presents a stacked 2D histogram of parameter values as a function of galactocentric radius, with colour indicating the joint distribution normalised by the total counts. Each galaxy here contributes 180 evenly-spaced azimuthal measurements per radial bin up to the radius marking the extent of its gaseous disc -- $R_{\rm ext, H\textsc{i}}$, as defined in Sec.~\ref{sec:Methods:GasDiscs}, enabling robust population-level statistics. The vertical subpanels display the normalised 1D distribution of the corresponding parameter. Cyan (main panels) and black (side panels) dash-dotted lines indicate the range of parameter values associated with $\leq 10$~per~cent error in circular velocity inference under the relevant fluid model (i.e. $|\varepsilon| \lesssim 0.2$, $|\delta| \leq 0.1$). In the main subpanel, white solid and white dashed lines indicate the median and central $50^{\rm th}$ percentile of the respective conditional distributions, $\mathrm{W}(\varepsilon|r)$ and $\mathrm{W}(\delta|r)$. Several results emerge.

First, both the $\varepsilon$ and $\delta$ exhibit broad distributions at all radii, demonstrating that disequilibrium is a pervasive feature of dwarf galaxy gas discs. Crucially, this is not a subtle effect -- typical deviations on the order of $10$~per~cent from radial force balance and centrifugal support are common, with a substantial fraction of measurements showing even larger discrepancies. This is more clearly seen in the marginalised 1D distributions shown in the side panels, where the parameter ranges corresponding to $\leq 10$~per~cent circular velocity inference error approximately coincide with the full width at half maximum (FWHM) of the distribution functions. This overlap indicates that a significant portion of the gas disc lies outside the regime where circular velocity estimates are accurate to better than $10$~per~cent.

As evident in the main panels, the vast majority of the samples contributing to these broad distributions originate from the inner disc regions ($r \lesssim 5$ kpc). This finding runs counter to expectations based on dynamical timescales, which predict more rapid relaxation in the central regions; instead, we find clear evidence for persistent perturbations. This is particularly relevant for core–cusp studies, since this radial range provides the strongest leverage for constraining dark matter density profiles. This is not the only hint that the centres of dwarf galaxies are continuously perturbed out of equilibrium. In fact, a recent investigation of radial trends in the radial acceleration scaling relation (RAR; \citealt{McGaugh2016}) using the SPARC sample \citep{Lelli2016} shows that the inner regions -- defined as $r < 2R_d$, where $R_d$ is the stellar disc scale radius -- are much less tight (in terms of intrinsic scatter) than the outer disc (see Fig.~S1 in \citealt{Ren2019}). This trend can be interpreted as empirical evidence that inner regions of galaxies are indeed the most dynamically disturbed.

Second, the equilibrium-based model (quantified by $\varepsilon$) does not consistently outperform the na\"{i}ve centrifugal model (quantified by $\delta$). Across the population, many regions -- particularly in the inner disc -- fall outside the parameter space where equilibrium-based corrections yield more accurate circular velocities than the azimuthal velocity, reinforcing the discussion in Sec.~\ref{sec:ModelsResults:OneGal}. This is reflected in Fig.~\ref{fig: Radial eps-delta profiles}, where the mass-weighted peak of the $\varepsilon$ distribution reaches a value $\max \big(\mathrm{W}(\varepsilon)/\Sigma \mathrm{W}\big) \approx 0.015$, compared to the peak of the $\delta$ distribution $\max \big(\mathrm{W}(\delta)/\Sigma \mathrm{W}\big) \approx 0.030$.

Moreover, the $\delta$ parameter reveals a population-wide tendency to underestimate the true circular velocity in the inner disc ($\delta < 0$), implying a systematic bias toward inferring smaller inner velocities, hence `cored' profiles.
As the white solid and dashed lines indicate, this bias arises preferentially within $r \lesssim 5$~kpc, where the median radial trend of $\delta$ departs from zero and the conditional distributions become increasingly skewed toward negative values. Although the median fractional offsets remain relatively modest -- of order 10~per~cent at $r \approx 2$~kpc -- the growing asymmetry of the distributions implies that a substantial fraction of regions systematically underestimate the true circular velocity; crucially, this behaviour is mirrored, not mitigated, when adopting the equilibrium-based model.
However, the $\delta$ distribution is broad in the inner kiloparsecs, with a minority of regions exhibiting the opposite trend ($\delta > 0$). This bidirectional scatter suggests that some of the observed diversity in dwarf galaxy rotation curves could arise from disequilibrium physics rather than intrinsic variations in the underlying dark matter distributions. In other words, the rotational structure in the inner regions of gas discs is a biased and high-variance tracer of the underlying gravitational potential even before projection effects, and the limited ability of kinematic models to accurately recover azimuthal speeds, come into play.

Together, these results point to a deeper limitation of analytic fluid models in the context of mass inference for dwarf galaxies. While efforts to correct for non-circular motions or pressure support -- such as asymmetric drift prescriptions -- have become increasingly sophisticated, our analysis suggests that these models are fundamentally constrained by the intrinsic disequilibrium of the gas. The inner kiloparsecs -- the most critical regime for dark matter constraints -- are also the least likely to satisfy steady-state assumptions.

If gas kinematics in dwarf galaxies is not a reliable tracer of the gravitational potential, mismatches between observed rotation curves and $\Lambda$CDM predictions may reflect the breakdown of the equilibrium assumption, not a failure of the cosmological model itself. This can reframe the long-standing core-cusp and diversity of rotation curves tensions into a mismatch between observed quantities and the theoretical constructs that they are used to infer. 

\section{Origins of Disequilibrium}
\label{sec:Origins_of_diseq}

Sec.~\ref{sec:ModelsResults} showed that hydrodynamical disequilibrium is a widespread and persistent feature of the gas discs in our low-mass galaxy sample, spanning a diverse set of simulated systems. The amplitude and ubiquity of these deviations from steady-state dynamics have immediate implications for dynamical mass inference. They challenge the robustness of classical rotation curve analyses and offer a plausible route to explain the diversity of dwarf galaxy rotation curves.

This naturally prompts a central question:
`If disequilibrium is so prevalent and has such strong implications, what causes it?'
Implicit in this is a methodological concern: are these departures genuine astrophysical phenomena, driven by internal feedback, gravitational instabilities, etc., or are they artifacts of numerical limitations or rare stochastic events?

In this section we dissect the physical origins of disequilibrium in our simulations, using spatially resolved diagnostics to connect local equilibrium violations with specific dynamical drivers. Our aim is to separate plausible physical mechanisms from numerical or incidental causes, while quantifying and characterising their respective dynamical signatures.

\subsection{Disequilibrium drivers}
\label{sec:Origins_of_diseq:Drivers}

We consider four broad classes of mechanisms that are expected to disrupt Eulerian steady-state velocity fields:
(i) impulsive momentum and energy injection from stellar feedback;
(ii) overdense, self-gravitating clumps;
(iii) energy injection from AGN activity; and
(iv) disc-scale gravitational asymmetries.

We stress that this taxonomy is not derived from prior studies of hydrodynamical disequilibrium per se, but rather from physical intuition and analytic expectations about the stability of galactic discs: any process capable of driving non-circular motions, injecting turbulence, or introducing rapidly evolving unrelaxed components should, in principle, generate measurable deviations from equilibrium. While the first three processes are often causally and temporally linked -- e.g. feedback can seed clump formation, which in turn can trigger gas inflow, which then may feed the AGN -- they are nonetheless localised in space and time, and can therefore be identified through spatial tracer fields. By contrast, gravitational asymmetries are inherently non-local in their dynamical influence and are treated separately in Sec.~\ref{sec:Origins_of_diseq:Grav}. We also stress that this is not a comprehensive list of all possible drivers, but a selection of the most important (as we will discuss in Sec.~\ref{sec:Origins_of_diseq:Budget}).

\begin{figure*}
    \centering
    \includegraphics[width=\textwidth, height = 0.23\textheight]{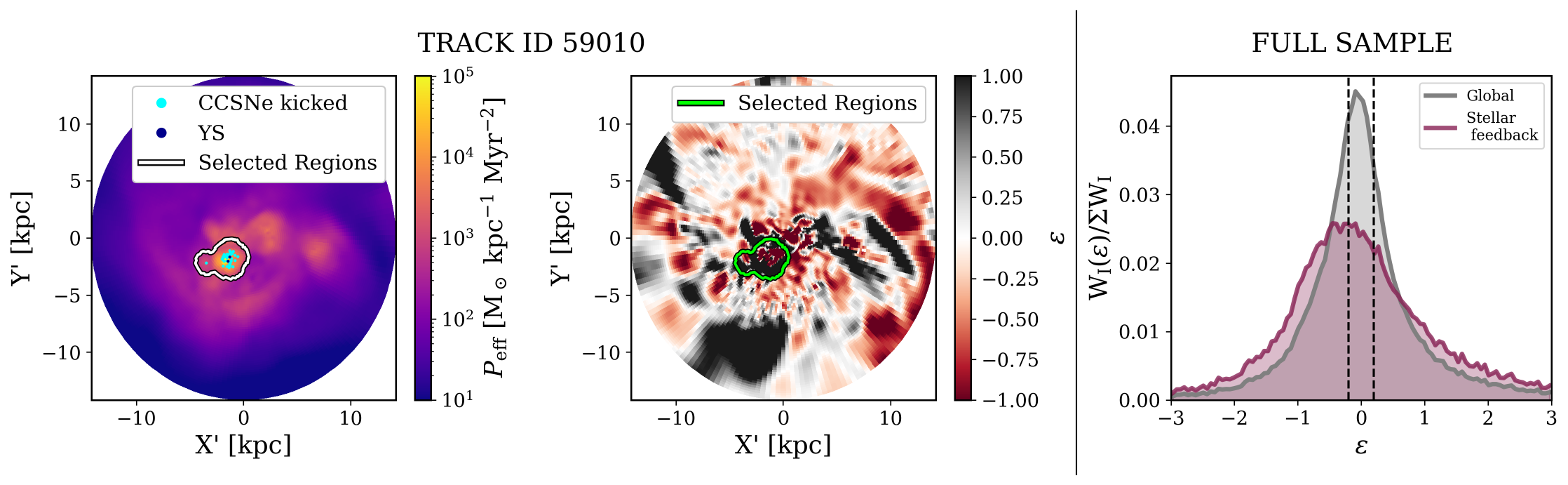}
    \includegraphics[width=\textwidth, height = 0.23\textheight]{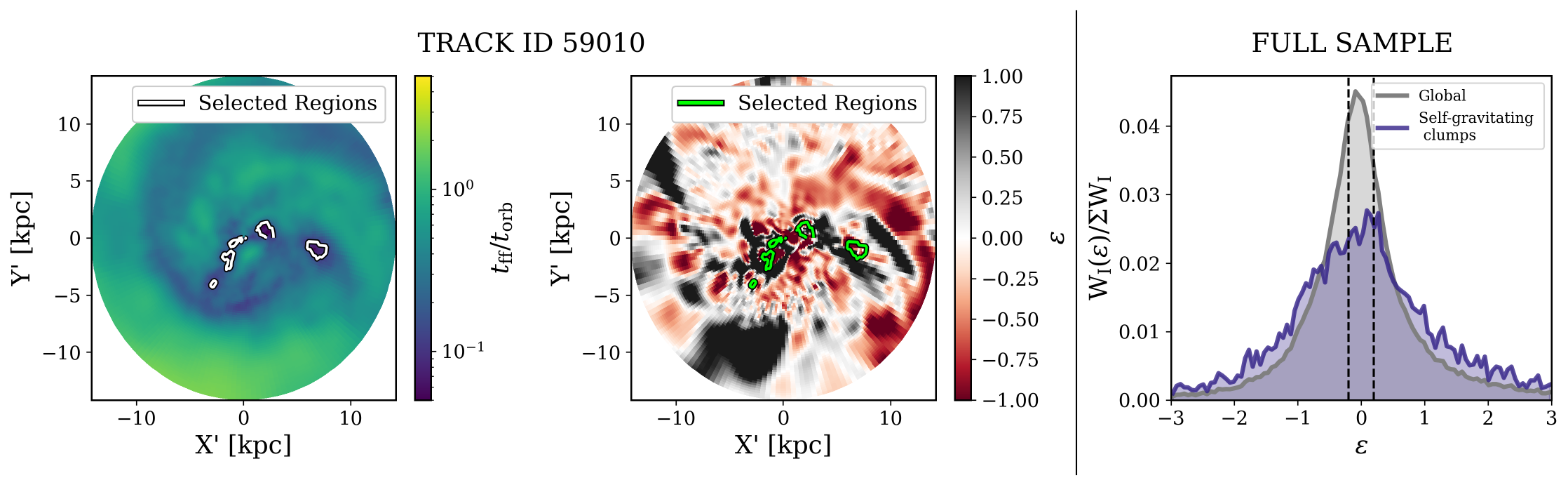}
    \includegraphics[width=\textwidth, height = 0.23\textheight]{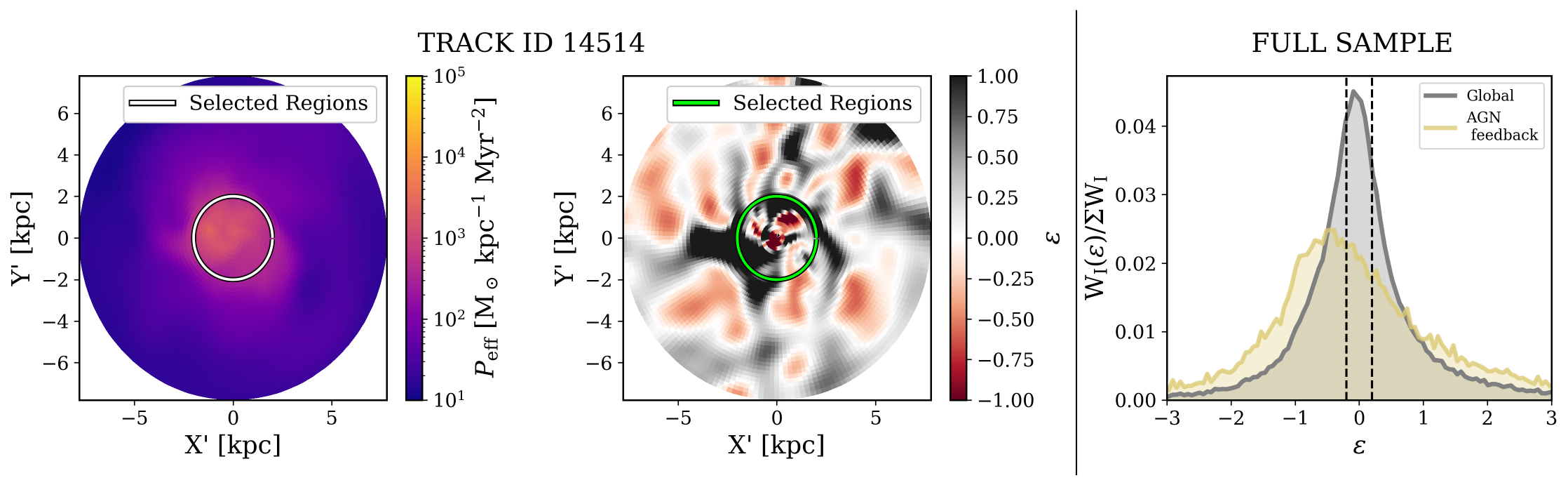}
    \caption{A visual summary of three different mechanisms inducing local disequilibrium in the interstellar medium considered in this study, along with their associated tracers. The figure is organized in rows, each corresponding to one disequilibrium-driving mechanism: stellar feedback (top), self-gravitating gas clumps (middle), and AGN activity (bottom). For the first two rows (stellar feedback and clumps), the panels in the left and middle columns show the same representative galaxy, as indicated by the annotated track IDs, whereas the bottom row illustrates a different system; the right column in each row always shows the corresponding sample-wide $\varepsilon$ distribution. \textit{Left column}: 2D Cartesian maps of a representative deprojected warped disc midplane $(X', Y')$, colour-coded by the associated tracer field -- effective pressure for both stellar and AGN feedback, and the free-fall to orbital time ratio for self-gravitating clumps. Blue and cyan dots in the upper-left panel indicate the positions of young stars (ages $\tau_\star \leq 100~\mathrm{Myr}$) and gas particles recently affected by CCSNe ($t_{\rm lb} \leq 50~\mathrm{Myr}$), respectively. Solid white contours enclose regions exceeding tracer thresholds, or satisfying proximity criteria. \textit{Middle column}: The same maps, colour-coded by the disequilibrium parameter $\varepsilon$, with the same contours as in the left panels (now in lime) overlaid to highlight the spatial correlation between tracers and disequilibrium. \textit{Right column}: Normalised 1D distributions of $\varepsilon$, weighted by gas mass, computed across the full dwarf galaxy sample for each process (coloured), compared to the global distribution (grey). The vertical dashed lines indicate the quasi-equilibrium range defined by $|\varepsilon| \lesssim 0.20$. The spatial alignment of tracer-selected regions with extreme $\varepsilon$ values, together with their skewed and/or broadened $\varepsilon$ distributions relative to the global one, supports their role as drivers of hydrodynamical disequilibrium.}
    \label{fig:Dis_Drivers}
\end{figure*}

Fig.~\ref{fig:Dis_Drivers} provides a compact summary of the spatial and statistical signatures of the first three mechanisms. Each row corresponds to one driver and one representative galaxy (unless otherwise noted), with: first column, showing a tracer-field map of a representative warped and deprojected disc midplane, $(X', Y')$; second column, the corresponding disequilibrium parameter ($\varepsilon$) map; and third column, the gas-mass-weighted $\varepsilon$ distribution for tracer-selected regions (coloured), compared to the global distribution (grey), both normalised by their total gas-mass-weight and computed across the full dwarf galaxy sample. The tracer definitions are rooted in theoretical expectations: feedback regions are identified by effective (both thermal and turbulent) overpressure near young stars and recent CCSNe, or around an active AGN; self-gravitating clumps are traced by regions with short free-fall times relative to the local orbital period. In all three cases, white (or lime) contours delineate regions exceeding tracer thresholds for overdense gas clumps, or proximity criteria for feedback. The visual and statistical correspondence between tracer-selected zones and high-$|\varepsilon|$ gas suggests that these mechanisms imprint distinct, separable signatures on the ISM kinematics. We examine each in turn, beginning with stellar feedback.

\subsubsection{Stellar feedback}
\label{sec:Origins_of_diseq:Feedback}

Stellar feedback is a well-motivated source of ISM disequilibrium, particularly in shallow gravitational potentials where momentum and energy injection from massive stars can rival the disc binding energy \citep[e.g.][]{Dekel1986, Stinson2006, Hopkins2012, KimOstriker2017}. Energy deposition from SNe, stellar winds, and radiation pressure drives expanding shells, turbulence, and large-scale pressure gradients, each of which can displace gas from local force balance for timescales exceeding the local dynamical time \citep{Martizzi2015, Orr2020}. Observations support this picture: nearby dwarfs often exhibit asymmetric H\textsc{i} profiles \citep[e.g.][]{Walter2008, Stilp2013} and vertically extended warm gas layers \citep{Boomsma2008, Bacchini2020}, with significant dispersive motions consistent with feedback-driven disturbances. A striking observational example of extreme disruption is the supergiant H\textsc{i} shell, or ‘superbubble', in IC2574 \citep{Brinks2003}, which also coincides with one of the most pronounced apparent dark matter cores known in a dwarf galaxy.

In our analysis, feedback-affected gas is defined as material lying within a projected radius of $1~\mathrm{kpc}$ in the warped disc midplane around either young stars (ages $\tau_\star \lesssim 100~\mathrm{Myr}$) or gas particles that have recently received kinetic injection from CCSNe ($t_{\rm lb} \leq 50~\mathrm{Myr}$). Particle positions are first assigned to their nearest point on the deprojected disc surface, and the selection is then made using a two-dimensional circular mask of radius $1~\mathrm{kpc}$ (see Sec.~\ref{sec:Methods:FeedbackTracing} for details). This choice reflects both the geometry of the discs and the physical dispersion of feedback-driven material: although CCSNe impart an isotropic target kick of $v_{\rm kick} = 50~\mathrm{km~s^{-1}}$ in the rest frame of the star particle \citep{Chaikin2023, Schaye2025}, the residual velocity of affected gas in the rotating disc frame is typically smaller ($\sim 20~\mathrm{km~s^{-1}}$), implying characteristic displacements of $\sim 1~\mathrm{kpc}$ over $\sim 50~\mathrm{Myr}$ (ignoring other forces). Furthermore, H\textsc{i}-dominated dwarf galaxy discs have expected vertical scale heights of $\sim 1~\mathrm{kpc}$ \citep{Banerjee2011}, making this spatial cut a physically motivated and conservative choice for identifying feedback-influenced regions in a population-wide analysis. 

We focus on kinetically kicked gas because thermally heated ejecta leave the disc too rapidly to provide meaningful information: once heated, this material rises rapidly out of the midplane (typically to several kiloparsecs) and ceases to intersect the regions where we evaluate $\varepsilon$. It therefore leaves no coherent or spatially localisable imprint within the disc, whereas the kinetic CCSNe mode produces gas that remains dynamically coupled to the disc for tens of megayears and can be robustly identified in our spatially resolved, population-wide analysis.

The top row of Fig.~\ref{fig:Dis_Drivers} shows, via a representative system (\textsc{hbt-herons} track ID $= 59010$, stellar mass $M_\star \approx 1.00 \times 10^8~\mathrm{M_\odot}$, star formation rate $\mathrm{SFR} \approx 0.013~\mathrm{M_\odot}\mathrm{yr}^{-1}$), that stellar feedback-selected regions (white contours, left panel) coincide with zones of enhanced effective pressure, often exceeding the disc median by factors of several and frequently associated with coherent superbubble structures. The corresponding $\varepsilon$ map (middle panel) indicates that these regions host some of the strongest disequilibria in the disc, with $|\varepsilon|$ commonly exceeding unity.\footnote{Since $\varepsilon$ measures the ratio of the neglected time-dependent acceleration to the radial gravitational force, $\varepsilon = -\partial_t v_r/g_r$, values with $|\varepsilon| > 1$ imply that the intrinsic local velocity-evolution timescale is comparable to or shorter than the dynamical timescale set by gravity. In such regimes the dynamics cannot be treated as small perturbations about equilibrium, rather they signify strong departures from the assumptions underlying standard disc models.}

Statistically (right panel), the stellar feedback-affected gas exhibits a strongly skewed $\varepsilon$ distribution with a pronounced tail towards large negative values, indicating a time-dependent, Eulerian, radially inward acceleration that offsets the excess outward force imparted by feedback. Physical intuition for this behaviour can be developed by first considering the rest frame of the young star launching the feedback. In this frame, a supernova injects an approximately isotropic impulse, producing a transient increase in the local radial velocity that decays as the blast wave evolves \citep{Sedov1946, Taylor1950, VonNeumann1964}; i.e. $\partial_t v_r' < 0$, where $v_r'$ is the radial component measured away from the star. When transformed to the galactocentric frame, one might expect the inward- and outward-directed components of this symmetric impulse to contribute equally to $\partial_t v_r$ at a fixed location in the disc. However, the strong radial and vertical density gradients in galactic gas discs make it easier for feedback to displace gas outward or out of the plane of the disc than inward. As a result, the Eulerian $\partial_t v_r$ in the galactocentric frame is systematically biased negative when averaged across galaxies and mass-weighted, despite the underlying isotropy of the impulse.

Moreover, the stellar feedback-selected distribution exhibits a steep rise toward the peak starting at $\varepsilon \sim -1$, which aligns with a secondary bump in the global (full sample, disc-wide) distribution, suggesting that stellar feedback makes a large fractional contribution to the mass of gas in strong disequilibrium. A quantitative breakdown of the (dis-)equilibrium budget, together with a discussion of the limitations of our tracer-based method, is deferred to Sec.~\ref{sec:Origins_of_diseq:Budget}.

Elevated $\varepsilon$ is not confined to stellar feedback-affected gas. Even in actively star-forming systems -- such as the representative galaxy in the top row of Fig.~\ref{fig:Dis_Drivers} -- substantial disequilibrium is present well outside the feedback contours, indicating the action of additional drivers. In the following subsections, we examine the role of self-gravitating clumps, AGN activity, and large-scale gravitational asymmetries in seeding and sustaining non-equilibrium dynamics.

\subsubsection{Overdense and self-gravitating gas clumps}
\label{sec:Origins_of_diseq:Clumps}

Local gravitational collapse within discs provides a natural route to hydrodynamical disequilibrium. Dense, self-gravitating clumps -- often arising from gravitational instability in the cold ISM \citep[e.g.][]{Noguchi1999, Dekel2009, Genzel2011,Goldbaum2015} -- develop internal dynamics that strongly deviate from the background shear flow.

We identify these regions using a timescale diagnostic: the ratio of the local free-fall time to the orbital time, \( t_{\rm ff} / t_{\rm orb} \). When this ratio falls below $\sim 0.1$ (a calibration value discussed in Sec.~\ref{sec:Origins_of_diseq:Budget}), collapse proceeds much more rapidly than the local orbital response time, signaling that gas self-gravity dominates over the ambient potential. This criterion, by definition, isolates dense knots and filaments whose dynamics are internally driven, rather than passively tracing the global disc rotation.

The middle row of Fig.~\ref{fig:Dis_Drivers} illustrates their spatial and dynamical imprint. In the left panel, the clump tracer field delineates spatially localised knots and filaments, frequently embedded within spiral arms or ring-like overdensities. The corresponding $\varepsilon$ map (middle) reveals that these same sites are often among the most extreme disequilibrium regions, with $\varepsilon$ commonly positive and exceeding unity. The right panel quantifies this: the clump-tagged distribution is markedly broader than the global disc-wide one, with a mild shift of the peak toward $\varepsilon \gtrsim 0$ and extended wings on both sides, reflecting a mix of accelerated collapse ($\varepsilon < 0$) and migration ($\varepsilon > 0$). This contrasts with the distribution characteristic of feedback zones, typically biased more strongly on the negative side (Sec.~\ref{sec:Origins_of_diseq:Feedback}). In idealised scenarios, a clump drives a locally symmetric, converging flow in its own rest frame, such that $\partial_t v_r' < 0$, where $v_r'$ is the radial component measured away from the clump \citep[e.g.,][]{Penston1969}. In the Eulerian galactocentric frame, because the ambient gas density decreases with radius, flux from the denser inward side exceed those from the outward side, biasing $\partial_t v_r$ towards the positive end. This naturally produces the mild positive shift in the clump-selected distribution. At the same time, many dense clumps either form within, or evolve into, regions where young stars subsequently inject feedback; sampling these mixed environments reintroduces the strongly negative tail characteristic of feedback-driven gas. The joint effect is a broad, wing-dominated distribution with a slight offset toward $\varepsilon > 0$, as seen in Fig.~\ref{fig:Dis_Drivers}.

While some clumps form or reside within feedback-active environments, a substantial fraction lie in quiescent or outer disc regions lacking obvious stellar-driven perturbations. This is evident in Fig.~\ref{fig:Dis_Drivers}: because the same representative galaxy appears in the first two rows, the stellar-feedback (top-row) and clump-selected (middle-row) contours can be directly compared. Many overdense regions fall well outside the feedback-affected zones, yet still exhibit extreme $|\varepsilon|$, demonstrating that clump-driven dynamics generate strong disequilibria independently of stellar feedback. Observational analogues -- such as the massive clumps in $z\sim2$ discs \citep[e.g.][]{Elmegreen2007, Genzel2011}, or in the nearby WLM dwarf galaxy \citep{Albers2019, Mondal2021} -- likewise show that such structures can persist without contemporaneous feedback. We therefore treat self-gravitating clumps as a separate, dynamically distinct driver of local disequilibrium.

\subsubsection{AGN feedback}
\label{sec:Origins_of_diseq:AGN}

\begin{figure}
    \includegraphics[width=\columnwidth]{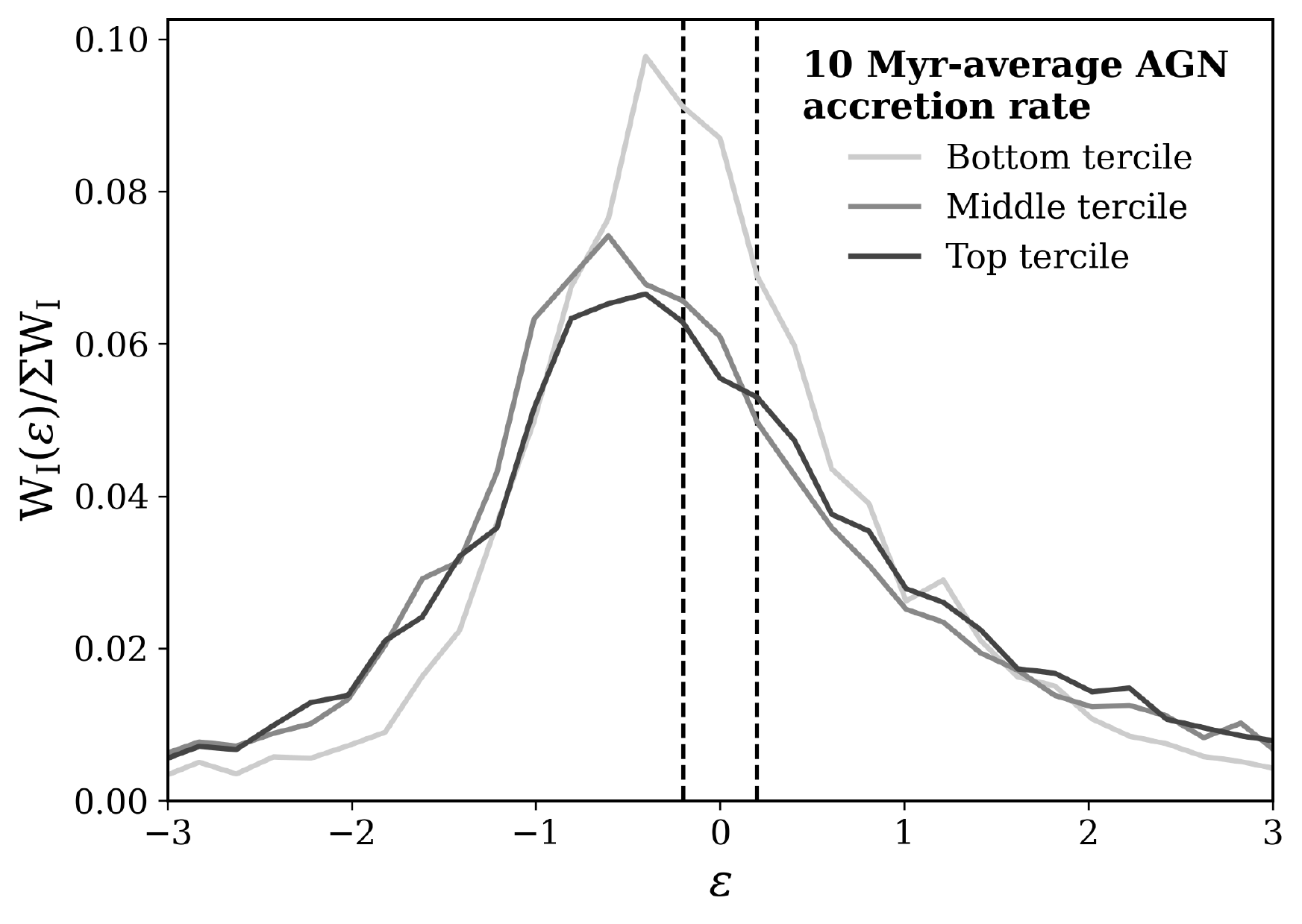}
    \caption{
    Comparison of the gaseous disequilibrium structure for galaxies binned by their $10~\mathrm{Myr}$–averaged AGN accretion rate.
    The three curves show the self-normalised, mass-weighted distribution of the disequilibrium parameter, $\varepsilon$, i.e. $\mathrm{W}_{\mathrm{I}}(\varepsilon)/\Sigma \mathrm{W}_{\mathrm{I}}$, for systems in the bottom (light grey), middle (grey), and top (dark grey) terciles of the accretion-rate distribution.
    The vertical dashed lines mark $|\varepsilon|=0.2$, as in Fig.~\ref{fig:Dis_Drivers}.
    The systematic shift of the distributions -- with higher-accretion systems favouring slightly larger $\varepsilon$ magnitudes -- under identical geometric selection (a $r = 2~\mathrm{kpc}$ mask in the deprojected midplane), supports that AGN activity is an independent, direct driver of part of the observed non-equilibrium behaviour.
    }
    \label{fig:act_vs_pass_agn}
\end{figure}

Energy and momentum injection from accreting massive black holes provide a third physical channel capable of generating local departures from hydrodynamical equilibrium. Although AGN in dwarf galaxies have historically been difficult to identify -- owing to weak emission signatures, uncertain bolometric corrections, and confusion with stellar processes \citep[e.g.][]{Reines2013, Baldassare2017} -- both theory and observations now indicate that low-mass systems can indeed host active nuclei, and that even modest AGN episodes can imprint measurable disturbances on the surrounding interstellar medium \citep[e.g.][]{ManzanoKing2020, Koudmani2025} affecting galaxy evolution \citep[e.g.][]{Koudmani2022, ArjonaGalvez2024}.

To understand how AGN activity correlates with disequilibrium in the gas surrounding it, we examine the distribution of the disequilibrium parameter within fixed central apertures for galaxies binned by their 10 Myr–averaged AGN accretion rate (Fig.~\ref{fig:act_vs_pass_agn}). Across the bottom, middle, and top terciles, the self-normalised, mass-weighted distributions of $\varepsilon$ exhibit a clear, monotonic shift: galaxies with more strongly accreting AGN show a pronounced enhancement in the mass fraction of strongly non-equilibrium gas, with substantially broader high-$|\varepsilon|$ tails. The middle tercile already has much less gas in the low-$|\varepsilon|$ regime (especially for $|\varepsilon| \lesssim 0.5$) than the bottom tercile, suggesting a threshold-like response rather than a smooth continuum. The top tercile roughly follows the middle-tercile distribution.

This behaviour motivates our working definition of AGN-affected regions: the central gas within a fixed $r = 2~\mathrm{kpc}$ aperture for galaxies in the middle and top accretion rate terciles, while the bottom tercile provides the appropriate passive comparison set under identical geometric selection. The inner-kiloparsec scale is motivated by both the feedback implementation and expectations for low-mass black holes: in the COLIBRE `Thermal' AGN model energy is deposited only into the nearest SPH neighbours, producing a compact over-pressurised region whose expansion is limited by the relatively modest energetics of dwarf galaxy AGN. As a result, the driven pressure perturbations are generally confined to the central region of the disc, and the $2\,\mathrm{kpc}$ aperture comfortably encompasses their expected extent in the midplane. As for the other channels, discussion of the arbitrariness of these selection choices is deferred to Sec.~\ref{sec:Origins_of_diseq:Budget}.

With this selection in place, we can characterise the spatial and dynamical impact of AGN activity using the representative example in Fig.~\ref{fig:Dis_Drivers}. The tracer-field map (left) reveals a compact, centrally concentrated region of elevated effective pressure -- often bubble-like in appearance, though generally less overpressurised than typical stellar feedback-driven bubbles. The corresponding $\varepsilon$ map (middle) indicates that these AGN-adjacent zones host strong disequilibria, with $|\varepsilon|$ frequently exceeding unity. Statistically (right), AGN-selected gas displays slightly heavier high-$|\varepsilon|$ tails than either stellar feedback- or clump-selected gas, and a distinctly skewed distribution with a sharp rise towards the peak at negative values ($\varepsilon_{\rm peak} \approx -0.6$). This reflects the centrally concentrated and impulsive nature of AGN energy deposition.

Two subtleties are worth noting. First, AGN-selected regions (i.e. $r \leq 2~\mathrm{kpc}$) can overlap with stellar-feedback-selected zones in systems experiencing central star formation, which partly explains the qualitative similarity of their distributions and why the bottom tercile by $10 ~\mathrm{Myr}$-averaged AGN accretion rate still includes a substantial amount of non-equilibrium gas. Second, strong AGN events are rare in dwarfs, and thus they contribute a smaller total fraction of the disequilibrium budget than stellar feedback (we explore this further in Sec.~\ref{sec:Origins_of_diseq:Budget}). Nevertheless, AGN-driven disturbances form a distinct and robust source of disequilbrium, and therefore merit treatment as an independent driver. This claim is further reinforced by the representative system shown in Fig.~\ref{fig:Dis_Drivers} (\textsc{hbt-herons} track ID 14514; $M_\star \approx 2.8 \times 10^8~\mathrm{M_\odot}$; $\mathrm{SFR} = 0.00~\mathrm{M_\odot}\,\mathrm{yr}^{-1}$), in which AGN activity is the only local source of perturbation that we identify: its disequilibrium structure cannot be attributed to stellar processes or self-gravitating gas.

\subsubsection{Gravitational asymmetries}
\label{sec:Origins_of_diseq:Grav}

\begin{figure*}
    \centering
    \includegraphics[width=0.65\textwidth]{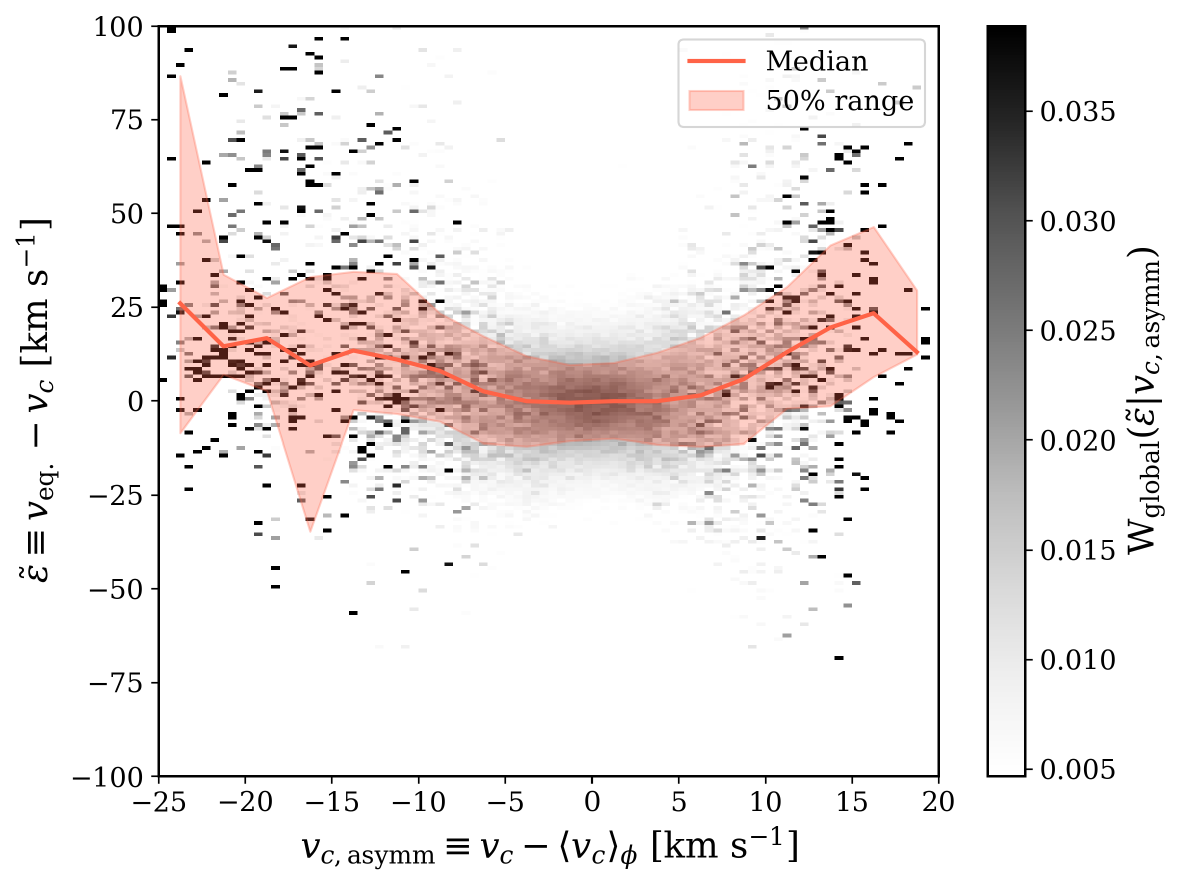}
    \caption{Correlation between gravitational asymmetries and departures from hydrodynamical equilibrium. The horizontal axis shows the non-axisymmetric component of the local circular velocity, \(v_{c,\,\mathrm{asymm}}\), defined as the local deviation from the azimuthally averaged circular speed (Equation~(\ref{eq:asymmetric potential})). The vertical axis shows the `dimensional equilibrium deviation' parameter, \(\tilde\varepsilon\), which measures the offset (in km s$^{-1}$) between the actual orbital speed and that predicted from the steady-state Euler balance (Equation~(\ref{eq:epsilon_mod})). Colours give the mass fraction in each \(\tilde\varepsilon\) bin, normalised within each column (i.e. at fixed \(v_{c,\,\mathrm{asymm}}\)), so each vertical slice represents a probability distribution. A solid and coloured (light-red) line indicates the mass-weighted median within each slice, with the shaded region showing the $50^\mathrm{th}$ percentile range, highlighting the trend and scatter.
    A clear pattern emerges: stronger gravitational asymmetries are associated with systematically larger equilibrium deviations, with increased scatter at high \(|v_{c,\,\mathrm{asymm}}|\).}
    \label{fig:Dis_Grav_conn}
\end{figure*}

The three mechanisms discussed so far -- stellar feedback, self-gravitating clumps, and AGN activity -- often operate within a broader structural context: large-scale gravitational asymmetries in the underlying potential. These asymmetries can arise from a variety of sources, including triaxial dark matter haloes \citep[e.g.][]{Hayashi2006}, lopsided stellar mass distributions \citep[e.g.][]{Rix1995, Zaritsky2013}, perturbations from orbiting subhaloes \citep{Kazantzidis2008}, massive gas clumps, or more violent events such as minor mergers \citep[e.g.][]{Peschken2020}. In practice, gravitational asymmetries are difficult to disentangle from the other drivers because they can be both cause and consequence: a clump may deepen the potential in an off-centre fashion, generating streaming motions and torques; a minor merger can simultaneously deposit fresh gas, trigger a starburst, feed the AGN and distort the potential.

To probe these asymmetries directly, we examine the non-axisymmetric component of the local circular velocity field, defined as
\begin{equation}
\label{eq:asymmetric potential}
    v_{c, ~\rm asymm}(r, \phi) \equiv v_c(r, \phi) - \langle v_c \rangle_\phi(r).
\end{equation}
We then compare these asymmetries to the dimensional equilibrium deviation parameter, \(\tilde\varepsilon\), defined by re-writing the steady-state Euler equation (Equation~(\ref{eq:EquilibriumModel})), as the difference between the true local circular velocity and the velocity reconstructed from the left-hand side, i.e.,
\begin{equation}
\label{eq:epsilon_mod}
   \tilde\varepsilon \equiv \sqrt{r \bigg|\left[(\boldsymbol{v}\cdot \boldsymbol{\nabla}) \boldsymbol{v}\right]_r + \frac{1}{\rho} \frac{\partial P_{\mathrm{eff}}}{\partial r}\bigg|} - v_c.
\end{equation}
This quantity can be interpreted as the offset between the actual and equilibrium-predicted orbital speeds -- e.g., the deviation between the dotted dark green and solid black curves in Fig.~\ref{fig:RC Reconstruction}.

Fig.~\ref{fig:Dis_Grav_conn} shows the conditional distribution of \(\tilde\varepsilon\) at fixed \(v_{c,\,\mathrm{asymm}}\). The colour scale indicates the fraction of mass in each \(\tilde\varepsilon\) bin relative to all gas at the same asymmetry amplitude. A solid coloured (light-red) line shows the mass-weighted median of \(\tilde\varepsilon\) in each slice, with the shaded region marking the $50^\mathrm{th}$ percentile range.

The trend is clear: as \(|v_{c,\,\mathrm{asymm}}|\) increases from a few to roughly 20 km s$^{-1}$, the \(\tilde\varepsilon\) distribution systematically shifts away from zero, with a slightly broadening scatter -- as indicated by the ‘bend’ of the solid coloured (light-red) line and the widening shaded areas. Large positive \(\tilde\varepsilon\) values are common for $|v_{c,\,\mathrm{asymm}}| \gtrsim 5 ~\mathrm{km}\mathrm{s}^{-1}$, indicating over-support relative to equilibrium. At low asymmetry amplitudes ($\lesssim 5~\mathrm{km}\,\mathrm{s}^{-1}$), the distribution flattens, consistent with a threshold below which the disc is largely unresponsive to the perturbation -- either because velocity dispersion exceeds the asymmetry (i.e. a hot disc) or because other processes dominate the gas dynamics, effectively hiding or breaking any underlying correlation. In fact, although not visible in the figure, in this regime we find substantial low-weight scatter, largely attributable to the mechanisms discussed in the previous subsections. Furthermore, the increasing scatter at high \(|v_{c,\,\mathrm{asymm}}|\) likely reflects the combined influence of secondary processes such as triggered star formation, inflows, or shock-driven turbulence.

In this sense, gravitational asymmetries are not a sharply localised driver like feedback or clumps; rather, they are a persistent background field that both modulates and is modulated by other processes. They provide a structural scaffold for disequilibrium, shaping the global kinematic state of the disc while enabling local perturbations to couple more effectively to the gas.

\subsection{Accounting for the disequilibrium budget}
\label{sec:Origins_of_diseq:Budget}

\begin{figure*}
    \centering
    \includegraphics[width=0.8\textwidth]{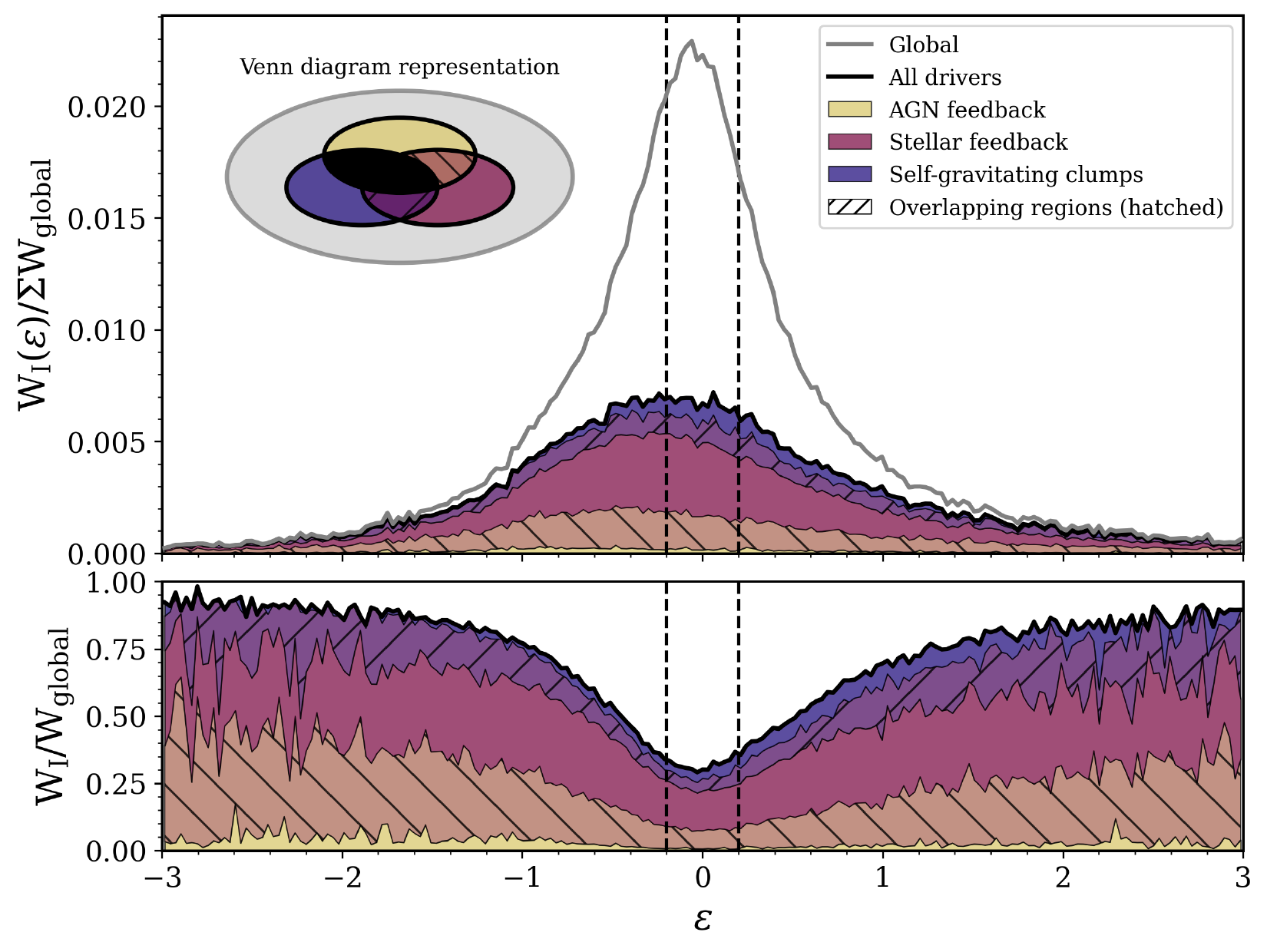}
    \caption{
    Cumulative accounting of the disequilibrium budget in simulated dwarf galaxy discs.  
    \textit{Top panel}: Gas-mass-weighted distributions of \(\varepsilon\) for the three physical drivers considered — AGN energy injection (yellow), stellar feedback (magenta), and self-gravitating clumps (purple) — plotted cumulatively on top of one another. The grey solid line shows the global distribution for the full sample. Overlaps between drivers are explicitly visualised: regions where AGN and stellar feedback coincide are hatched with `\textbackslash\textbackslash' lines; overlaps between stellar feedback and self-gravitating clumps are hatched with `//' lines. Colours in these overlaps follow a blending of the corresponding driver colours. For clarity, we omit the two remaining possible overlaps (AGN + clumps, and all three drivers simultaneously) as they contribute $\ll 1$~per~cent of the total mass-weight.  
    \textit{Bottom panel}: Fractional contribution of each process to the global mass-weight in each \(\varepsilon\) bin (same colour and hatch coding as above). This panel quantifies how each mechanism dominates, shares with, or yields to others across the disequilibrium spectrum. Vertical dashed lines in both panels mark the quasi-equilibrium range, \(|\varepsilon| \leq 0.20\).
    \textit{Inset}: The Venn diagram provides a schematic representation of how gas parcels tagged by the different physical drivers are subsets of the global gas population, illustrating their partial overlap rather than implying distinct or self-normalised distributions. To recover the distribution associated with a given driver, one should consider only the coloured regions uniquely associated with that driver and its direct combinations, rather than the cumulative stacked profile.
    }
    \label{fig:Dis_Summary}
\end{figure*}

Throughout Sec.~\ref{sec:Origins_of_diseq:Drivers}, we have examined how three localised physical mechanisms correlate with hydrodynamical disequilibrium. We also emphasised the structural role of gravitational asymmetries, which act as a persistent, non-local driver modulating these localised processes. We now turn to a more quantitative question. How much of the total disequilibrium budget across our sample can be attributed to the specific mechanisms identified in Sec.~\ref{sec:Origins_of_diseq:Drivers}?

Fig.~\ref{fig:Dis_Summary} addresses this by decomposing the global $\varepsilon$ distribution into the contributions from the three drivers.  
In the top panel, the coloured shaded distributions are stacked cumulatively, so that the total vertical extent at any given $\varepsilon$ shows the fraction of gas mass linked to at least one driver (without double counting). Hatched and colour-blended regions mark overlaps between two drivers, making their joint contribution immediately visible. For clarity, the two rarest overlaps (together $\ll 1$~per~cent of the total mass) are omitted. The solid black and grey distributions indicate respectively the sum of the drivers (i.e. including the omitted overlaps) and the global background. 
The bottom panel re-expresses the same data as fractions of the global distribution in each $\varepsilon$ bin, making the relative importance of different drivers in each regime much easier to read at a glance.

We stress that the coloured curves are not mutually exclusive: a given gas element may be tagged by multiple criteria. This makes the overlap regions physically informative, as they highlight gas where multiple mechanisms act in concert -- for example, the stellar/AGN feedback combination, which traces powerful central outflows, and the stellar feedback/self-gravitating clumps overlap, which often marks dense, star-forming regions undergoing active disruption.

Looking first at the top panel, most of the mass not tagged by any driver lies in the low-$|\varepsilon|$ regime, particularly for $|\varepsilon| < 0.5$. In the bottom panel, this pattern becomes striking: the combined three-driver contribution falls from a near-saturation of $\approx 85$~per~cent for $|\varepsilon|>1$ to only $\approx 30$~per~cent at $\varepsilon\approx0$. This behaviour provides both a sanity check and strong evidence that our mechanisms are genuinely associated with disequilibrium. In an idealised scenario, one might expect a clean step from $\sim100$~per~cent recovery for $|\varepsilon| \sim 1$ to $0$~per~cent in quasi-equilibrium, $|\varepsilon| \sim 0.1$. The persistence of a small contribution near $\varepsilon=0$ and lacking some mass for $|\varepsilon| > 0.2$ is unsurprising: the simple apertures that we use to select gas around disequilibrium drivers cannot perfectly capture the irregular morphology of the out-of-equilibrium gas. The contribution due to large-scale gravitational asymmetries is also not shown because labelling gas as affected by such asymmetries is ill defined.

The bottom panel also reveals a mild asymmetry in driver dominance. On the strongly negative side ($\varepsilon < -0.8$), feedback dominates the budget, accounting -- with overlaps -- for a steady $80$-$85$~per~cent contribution. This reflects the prevalence of expanding SNe shells and wind-driven flows in regions with large negative $\partial_t v_r$. Interestingly, the majority of the mass in this extremely negative regime comes from the AGN/feedback overlap, indicating that when the central regions host both young stars and accreting super massive black holes (SMBH), they are very strongly out of equilibrium. Whilst both AGN and self-gravitating clumps -- when ignoring overlaps -- are small contributors to the disequilibrium budget across the entire range, typically of order $1$-$5$~per~cent in total driver-tagged mass, on the positive end of the distribution (i.e., $\varepsilon \gtrsim 0.2$), self-gravitating clumps have an elevated contribution, explaining $\approx 25$~per~cent of the mass when overlaps are included, and a steady $\approx 10$~per~cent contribution individually. Overall, the impression is that stellar feedback is the dominant perturbation mechanism driving dwarf galaxies out of steady-state evolution, as expected via analytic and energetic arguments. However, the large overlaps in Fig.~\ref{fig:Dis_Summary} and, crucially, the local structure of disequilibrium in individual galaxies (Sec.~\ref{sec:Origins_of_diseq:Drivers}) signify that disequilibrium in dwarf galaxies cannot be regarded as a consequence of just stellar processes.

A key caveat is the sensitivity of our results to the exact tagging criteria. Our fiducial thresholds are physically motivated: the stellar and AGN feedback mask sizes, maximum time since energy injection, and star particle age cap reflect typical SN bubble expansion timescales and the duration of generic feedback episodes; the threshold for self-gravitating clumps are anchored to clearly outlying values in their respective tracer field. From these starting points, we iteratively tune the cuts: tightening a cut such that it reduces the mass tagged but leaves the $\varepsilon$ distribution unchanged suggests the cut is too strict, excluding relevant disequilibrium; loosening a cut such that abruptly increases the mass fraction at low-$|\varepsilon|$ indicates over-inclusiveness, contaminating the sample with recovered or ambient gas. The final selections are therefore both physically grounded and empirically calibrated.

In integrated terms, roughly $75$~per~cent of the total gas mass in our sample lies outside the quasi-equilibrium range (i.e., $|\varepsilon|>0.2$). Of this `non-equilibrium' component, the three drivers recover $\approx 85$~per~cent for $|\varepsilon|>1$, but only $\approx 30$~per~cent in the near-equilibrium regime ($|\varepsilon| \leq 0.2$) -- a scale-dependent recovery fraction that matches the visual impression from Fig.~\ref{fig:Dis_Summary}. The remaining untagged mass, particularly at $0.2 < |\varepsilon| < 1$, most likely reflects transient processes and our omission in Fig.~\ref{fig:Dis_Summary} of the asymmetric potential as a driver (Sec.~\ref{sec:Origins_of_diseq:Grav}), as well as mild turbulent, thermal, or gravitational fluctuations below our selection thresholds. We quantify the sensitivity of these recovery fractions to the adopted tagging apertures and thresholds in Appendix~\ref{app:varThresh}, where we show that the qualitative trends and relative driver contributions are robust to factor-of-two variations in the selection criteria.

To conclude, the cumulative-overlap decomposition shows that the disequilibrium budget is dominated by identifiable, physically interpretable mechanisms, often acting in concert and modulated by large-scale gravitational asymmetries. The small unaccounted fraction implies that any additional processes are either subdominant, short-lived, or subtle. By linking distribution shape and overlap structure with physical interpretation, Fig.~\ref{fig:Dis_Summary} offers a framework for both simulation diagnostics and direct observational comparison, particularly for radio interferometric and IFU surveys, and gas kinematics studies in low-mass galaxies. Moreover, the physically grounded origins of disequilibrium in our simulated sample reinforce the strong implications that non-equilibrium physics carries for rotation curve-based mass inference.

\section{Classification of Disequilibrium Galaxies} \label{sec:Classification}

Determining which galaxies yield rotation curves suitable for dynamical mass modelling remains a longstanding problem. Common observational quality cuts -- symmetric large-scale velocity gradients, high ordered-to-random motion ratios ($V_{\rm rot}/\sigma$), or thin, axisymmetric disc morphologies -- are widely employed \citep[e.g.][]{Trachternach2008, Oh2015, Pina2025}, yet none guarantees hydrodynamical equilibrium, nor do they constitute a uniform, physically grounded standard across environments. 

In low-mass systems, the prevalence of disequilibrium (Sec.~\ref{sec:Origins_of_diseq}), its disproportionate impact on the inner disc, and the inability of more complete fluid models to outperform the na\"{i}ve centrifugal balance assumption (Sec.~\ref{sec:ModelsResults}) motivate a classification scheme tied directly to the underlying dynamics. Our starting point is to identify, within the simulation, those systems whose rotation curves trace the circular velocity with minimal bias -- i.e. systems close to centrifugal balance.

\subsection{The centrifugal–hydrodynamical equilibrium link}
\label{sec:Classification:centrif_hydrod_link}

\begin{figure*}
    \centering
    \includegraphics[width=\textwidth]{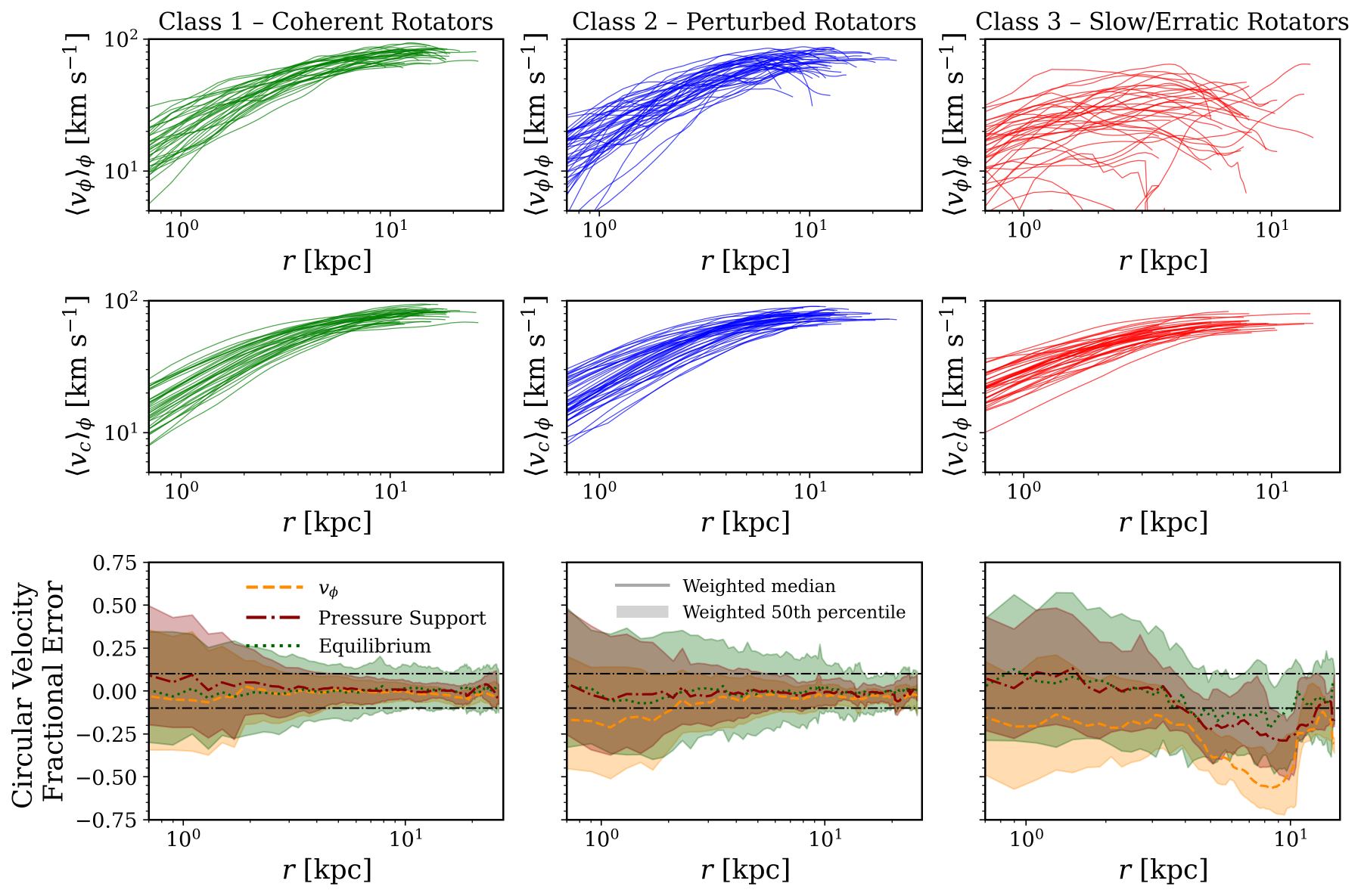}
    \caption{
    Visual summary of dynamical state classification of simulated gaseous discs. Each column corresponds to one of the three $k$-means classes (Appendix \ref{app:classAlg}), colour-coded consistently across rows: Class 1 (green), Class 2 (blue), and Class 3 (red).
    \textit{Top row}: Azimuthally averaged rotational velocity profiles, $\langle v_\phi\rangle_\phi(r)$, for all galaxies in each class.
    \textit{Middle row}: Corresponding azimuthally averaged circular velocity profiles, $\langle v_c \rangle_\phi(r)$. 
    \textit{Bottom row}: Fractional error between the model-dependent reconstructed circular velocity, $v_{c,\rm model}(r, \phi)$, and the true circular velocity from the potential $v_{c}(r, \phi)$, shown as class-stacked medians with $50^\mathrm{th}$-percentile envelopes (both weighted) as a function of radius. Lines correspond to the centrifugal model (dark orange/dashed), the pressure-supported model (dark red/dash dotted), and the equilibrium model (dark green/dotted) -- i.e., the same colour and line-style schemes as in Fig.~\ref{fig:RC Reconstruction}.
    Despite similar $\langle v_c \rangle_\phi$ shapes across classes, the residuals with respect to $\langle v_\phi \rangle_\phi$ reveal distinct dynamical states, which are then reflected in the structure of the fractional error distributions across different models, including the hydrodynamical equilibrium-based, reinforcing the link between centrifugal balance and hydrodynamical equilibrium.
    }
    \label{fig:classification}
\end{figure*}

The condition of centrifugal balance (Equation~(\ref{eq:NaiveModel})) directly follows from the hydrodynamical equilibrium condition (Equation~(\ref{eq:EquilibriumModel})), under the assumptions of axisymmetry and negligible pressure gradients. The converse, however, is not strictly true: a galaxy may satisfy $v_\phi \approx v_c$ despite the presence of significant non-circular motions or pressure support. In practice, sustained departures from equilibrium are expected to produce systematic residuals between the azimuthally averaged rotational velocity, $\langle v_\phi \rangle_\phi$, and azimuthally averaged circular velocity, $\langle v_c \rangle_\phi$, which are then reflected in the radial behaviour of $\varepsilon$. The expectation that the morphology and amplitude of these residuals provide a useful first-order diagnostic of dynamical state is motivated by the quasi-spherical nature of CDM haloes dominating the potentials of low-mass systems; such geometry naturally biases the acceleration field towards centrifugal dominance through the approximate conservation of angular momentum.

We quantify this connection by characterising each galaxy with three complementary residual-based metrics computed from $\langle v_\phi \rangle_\phi$ and $\langle v_c \rangle_\phi$: a global integrated deviation ($L_1$ norm), a fractional pointwise deviation in amplitude (NRMSE-Amplitude), and a fractional pointwise deviation in slope (NRMSE-Slope). Applying $k$-means clustering in this three-dimensional metric space yields three statistically well-separated groups, interpreted as dynamical classes. For clarity, Appendix~\ref{app:classAlg} presents two case-study galaxies -- a stable and a disturbed disc -- showing the resulting metrics, together with a visualisation of the 3D parameter space used by our $k$-means clustering algorithm to identify the three dynamical classes and further details. 

Fig.~\ref{fig:classification} provides a consolidated view of the dynamical classification. The diversity of azimuthally averaged rotational velocities, $\langle v_\phi\rangle_\phi$, is striking (top row): Class 1 systems show clean, monotonic rises with minimal scatter; Class 2 systems exhibit coherent, wave-like modulations; and Class 3 systems rotate slowly, irregularly, or both. In contrast, the circular velocity profiles $\langle v_c\rangle_\phi$ are comparatively uniform across classes, demonstrating that differences in the gravitational potential -- whether mildly cored or cuspy -- are not the primary drivers of the observed rotational diversity. Instead, the variation originates almost entirely from gas dynamics: non-circular motions, pressure forces, and departures from steady-state flow.

The bottom row makes this explicit by directly comparing the reconstructed circular velocity to the true value via the mass-weighted fractional difference profiles for the three dynamical models introduced in Sec.~\ref{sec:ModelsResults:OneGal}:
(i) the centrifugal model (orange, dashed),
\begin{equation}
v_{c,{\rm model}}(r, \phi) = v_\phi(r, \phi);
\end{equation}
(ii) the pressure-supported model (red, dash-dotted),
\begin{equation}
v_{c,\rm model}(r,\phi) = \sqrt{r\bigg|\frac{1}{\rho}\frac{\partial P_{\mathrm{eff}}}{\partial r}-v_\phi^2/r\bigg|}(r,\phi);
\end{equation}
(iii) the full hydrodynamical-equilibrium model (green, dotted),
\begin{equation}
v_{c,\rm model}(r,\phi) = \sqrt{r\bigg|\frac{1}{\rho}\frac{\partial P_{\mathrm{eff}}}{\partial r} + \left[(\boldsymbol{v}\cdot \boldsymbol{\nabla}) \boldsymbol{v}\right]_r\bigg|}(r,\phi).
\end{equation}
For each class, the shaded regions denote the weighted $50^\mathrm{th}$-percentile interval, while the lines show the weighted median.

\subsubsection{Class 1 -- Coherent Rotators, $29$~per~cent of the sample (35/122)}
These galaxies exhibit near-perfect alignment between $\langle v_\phi\rangle_\phi$ and $\langle v_c\rangle_\phi$, with median residuals $\approx 0$~per~cent across most radii and only mild ($\approx -5$~per~cent) under-rotation beyond $r \gtrsim 12$kpc. This bias largely vanishes when pressure or advective terms are included. Inside $\approx 4~\mathrm{kpc}$ -- and especially within $\approx 2~\mathrm{kpc}$ -- all models show significantly broadened error distributions, with $50^\mathrm{th}$-percentile ranges extending beyond the $\pm10$~per~cent threshold, which might be critical to discuss local features (or bumps) in the rotation curve and determine whether a halo is cuspy or cored \citep[e.g.][]{Sands2024}. Even for dynamically coherent discs, therefore, the innermost kiloparsec remains a somewhat challenging regime for reliable mass modelling.

\subsubsection{Class 2 -- Perturbed Rotators, $37$~per~cent of the sample (46/122)}
These systems display a baseline under-rotation of $\approx -5$~per~cent at $r \gtrsim 4~\mathrm{kpc}$, steepening to $\approx -10$~per~cent at $r \approx 2~\mathrm{kpc}$ and $\approx -20$~per~cent at $r \approx 1~\mathrm{kpc}$, with considerable scatter skewed further negative. For this class, including pressure support or the full equilibrium terms substantially improves the median recovery, yielding values near $0$~per~cent across most radii. However, the scatter remains large -- with $50^\mathrm{th}$-percentile widths exceeding $\pm10$~per~cent for $r \lesssim 6~\mathrm{kpc}$ for both corrected models, and for $r \lesssim 10~\mathrm{kpc}$ in the case of the equilibrium model. Thus, even though the median can be corrected, the intrinsic variation remains quite large for robust inference in the inner disc.

\subsubsection{Class 3 -- Slow/Erratic Rotators, $34$~per~cent of the sample (41/122)}
This class yields the most biased and dispersed estimates. The centrifugal model never enters the $\pm10$~per~cent accuracy band at any radius, with median errors reaching $\sim -60$~per~cent. Pressure support reduces the bias, especially within the inner $5~\mathrm{kpc}$, but still leaves systematic offsets of $\approx -20$ to $-30$~per~cent at larger radii. The equilibrium model achieves the largest improvement, but remains significantly biased at all radii and exhibits the widest dispersion of any model (typically $\pm 20$–$30$~per~cent). For these galaxies, no estimator provides a reliable reconstruction of the true circular velocity.

Taken together, these results clarify the dynamical meaning of the three classes. Coherent rotators lie close to both centrifugal and hydrodynamical equilibrium; perturbed rotators contain substantial but structured departures that can be partially corrected; and slow/erratic rotators are dominated by strong disequilibrium that no steady-state model can adequately capture. The close correspondence between the morphology of the $\langle v_\phi\rangle_\phi$–$\langle v_c\rangle_\phi$ residuals and the radial structure of the model-dependent fractional errors demonstrates that centrifugal balance provides a practical, physically motivated proxy for hydrodynamical equilibrium. Crucially, only a minority of galaxies -- less than one third centrals in the halo mass range $10.75 \leq \log_{10}(M_{200\mathrm{c}}/\mathrm{M_\odot}) \leq 11.00$ -- reside in a regime where circular velocity recovery is unbiased and tightly constrained. This underscores the importance of dynamical classification when interpreting rotation curves of low-mass systems and motivates the use of coherent rotators as the subset best suited for robust dark-matter mass distribution inference.

\subsection{Galaxy-wide correlations with dynamical class}
\label{sec:Classification:Class_Features}

\begin{figure*}
    \centering
    \includegraphics[width=\textwidth]{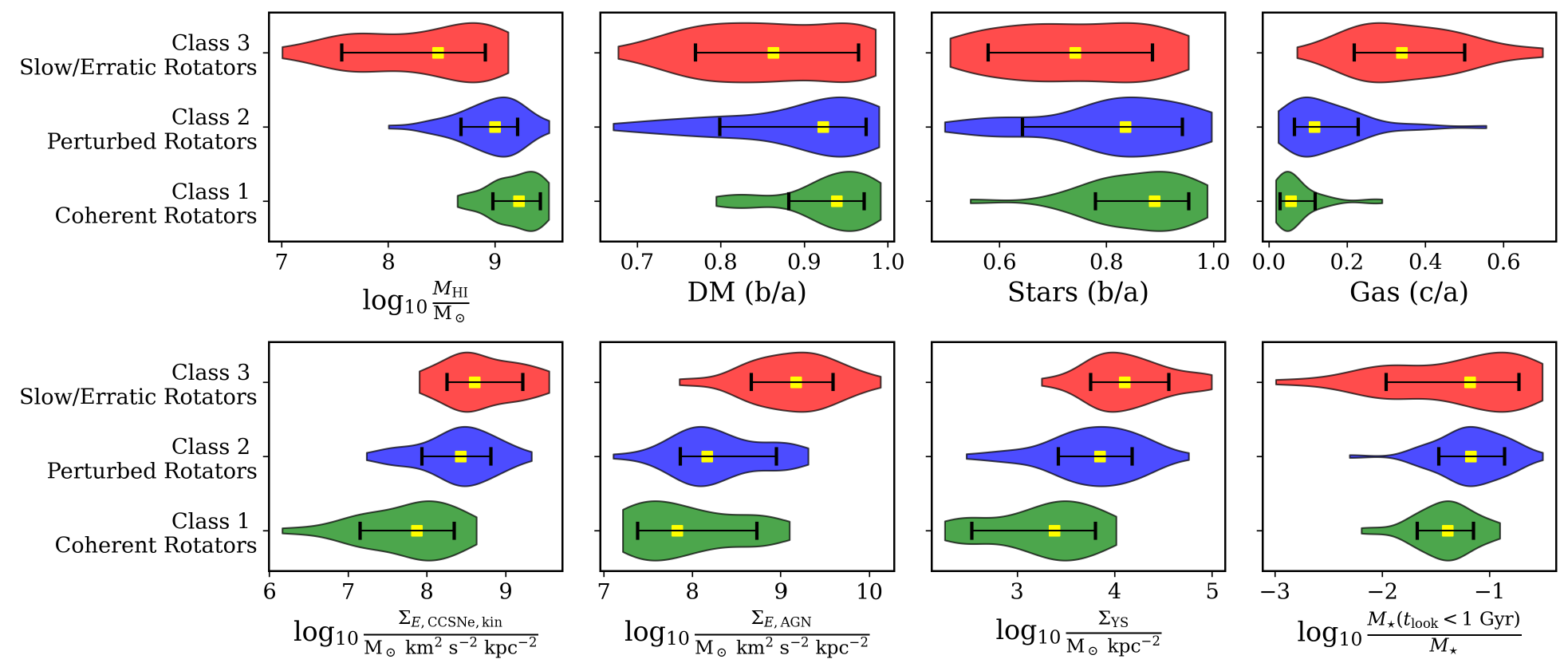}
    \caption{Distributions of global galaxy and disc properties across the three classes: coherent rotators, perturbed rotators, and slow rotators. Each panel shows a violin plot with class on the y-axis and a given physical property on the x-axis. Yellow squares indicate the class median; black horizontal lines span the 68th percentile range. 
    \textit{Top row, left to right:} Total neutral hydrogen mass ($M_{\rm H\textsc{i}}$), DM halo intermediate-to-major axis ratio ($b/a$), stellar distribution intermediate-to-major axis ratio ($b/a$) and gas distribution minor-to-major axis ratio ($c/a$). \textit{Bottom row, left to right:} Relative strength of kinetic feedback from CCSNe, AGN feedback, and generic stellar feedback from young stars (YS), as quantified in Sec.~\ref{sec:Methods:FeedbackTracing} and discussed in the text. The rightmost panel instead shows a proxy for recent star formation: the fraction of stellar mass formed within the last gigayear relative to the total stellar mass at $z=0$. 
    Systematic trends across classes offer insight into how global physical conditions and feedback processes relate to gas disc perturbation states.}
    \label{fig:class_global_features}
\end{figure*}

Having established the existence of three distinct dynamical classes, we now examine whether they correlate with broader galactic properties. While many of the quantities considered here are not directly measurable observationally, they serve two purposes: (i) to identify the dominant physical processes driving disc-wide perturbations, and (ii) to guide the development of observational pre-selection criteria for identifying systems dynamically suitable for rotation curve modelling and dark matter inference.

Fig.~\ref{fig:class_global_features} shows the distributions of selected global and disc-scale properties for coherent rotators (Class 1, green), perturbed rotators (Class 2, blue), and slow rotators (Class 3, red). Each panel presents a violin plot for one property, allowing visual comparison of trends and scatter between classes. Our aim is to link a galaxy’s perturbation state -- quantified via residual kinematics -- to structural and feedback-related properties that may drive or sustain dynamical disequilibrium. 

A clear trend is seen in total H\textsc{i} mass, i.e. $M_{\mathrm{H\textsc{i}}} = M_{\mathrm{H\textsc{i}}}(r \leq R_{\rm ext, H\textsc{i}})$. Coherent rotators have the highest median $M_{\rm H\textsc{i}}$; this is consistent with the physical picture of more inertia yielding enhanced resistance to perturbative forces, and the maintenance of well-ordered discs. Perturbed rotators occupy an intermediate regime with broader scatter, indicating that moderately gas-rich systems can retain some rotational support despite disturbances. Slow rotators show systematically lower $M_{\rm H\textsc{i}}$, suggesting that reduced H\textsc{i} content is strongly associated with disrupted kinematics. This correlation is evident in our narrow halo-mass-selected sample ($M_{200\mathrm{c}}$) but may be obscured in $V_{\rm max, ~rot}$–selected samples, where lower rotation speeds in disturbed systems bias them out of the selection -- i.e., the observationally accessible $V_{\rm max, ~rot} \equiv \max(\langle v_\phi \rangle_\phi)$ will be a biased estimate of $V_{\rm max} \equiv \max(\langle v_c \rangle_\phi)$ for this specific class. Similar trends are seen in the NIHAO simulations \citep{Dutton2019} and in the FIRE/FIRE-3 simulations \citep{ElBadry2018, Sands2024}, which find that lower-mass, H\textsc{i}-poor galaxies are less rotationally supported and more strongly affected by non-circular motions. As noted in \citet{Dutton2019}, this is confirmed observationally from the skewness of linewidths in bins of H\textsc{i} mass in both ALFALFA \citep{Haynes2018} and HIPASS \citep{Koribalski2004} observations.

We now examine the thickness and triaxiality of the gas, stellar, and dark matter components. To ensure that the shape parameters of the distributions are both comparable across galaxies and tied to the spatial region where the gas kinematics are classified, we measure their principal axes within a spherical aperture defined by the disc extent radius, $R_{\rm ext, H\textsc{i}}$.

We quantify the gas disc thickness with the minor-to-major axis ratio ($c/a$). Thin discs are notoriously difficult to capture accurately in cosmological hydrodynamical simulations \citep[e.g.][]{Agertz2011,Tacchella2019,Ludlow2023,Benavides2025}. COLIBRE seems to fare better in this regard than past efforts: our sample includes $39/122$ galaxies with stellar discs thinner than $c/a=0.3$, and $7$ galaxies with $c/a\leq0.2$. This still falls short of the frequency of (very) thin stellar discs in dwarf galaxies inferred from observed shape distributions, but is a substantial improvement over other models including TNG50, FIREbox and ROMULUS25 (Benavides et al. in preparation). This is encouraging: thin gas discs are a prerequisite to the formation of thin stellar discs, and the gas $c/a$ distributions for our three dynamical classes show a monotonic progression that parallels the trend with total H\textsc{i} mass, indicating a tight connection between gas disc morphology and dynamical state. In vertical hydrostatic equilibrium, greater turbulence implies a thicker disc, and observationally $V_{\rm rot}/\sigma$ correlates tightly with disc aspect ratio \citep[see sec.~5.3.1 of][]{Pina2025}. Here, coherent rotators are thin (median $c/a \sim 0.05$) with minimal scatter; perturbed rotators are thicker (median $\sim 0.15$) with a broader range; slow rotators are systematically thick (median $\sim 0.3$, none below $\sim 0.15$). These results suggest that $c/a$ is a strong morphological indicator of disequilibrium in this halo-mass regime, and that its diagnostic potential in rotation curve reliability assessments should be considered of high fidelity.

For stellar and dark matter shapes, the distributions overlap between classes but show a bias toward greater prolateness (lower intermediate-to-major axis ratio $b/a$) in perturbed and slow rotators. This trend is consistent with findings from \citet{DowningOman2023}, who identified similar correlations between halo shape and dynamical perturbations in high-resolution APOSTLE dwarf galaxies. Theoretically, this connection is well motivated: a galaxy embedded in a prolate halo often resides in a non-axisymmetric potential, which naturally excites strong non-circular motions in the gas kinematics \citep{Hayashi2006, Hayashi2007, Marasco2018, Oman2019} -- a behaviour our classification is explicitly sensitive to. Moreover, when the gas disc is significantly misaligned with the plane defined by the halo's intermediate and major axes, no stable configuration can be sustained \citep[see, e.g., the supplementary videos in][]{DowningOman2023}, further reinforcing the link between halo shape and kinematic disequilibrium. 

The bottom row of panels in Fig.~\ref{fig:class_global_features} quantifies the recent baryonic feedback impact from three channels: CCSNe kinetic feedback, AGN thermal feedback, and a proxy for generic stellar feedback traced by the mass of young stars. We measure these by selecting gas (or young stars for the latter) within a spherical aperture of $R_{\rm ext, H\textsc{i}}$ and with the most recent event within 100 Myr (or age for young stars), integrating the relevant energy proxy and normalising by disc area. This choice ensures fair comparisons across discs of different sizes; full details are given in Sec.~\ref{sec:Methods:FeedbackTracing}. All three feedback modes show systematic increases from coherent to perturbed to slow rotators, with median values rising by more than an order of magnitude in several cases, supporting a picture in which recent, concentrated feedback stirs turbulence, thickens the disc, and drives sustained departures from hydrodynamical equilibrium. Our SNe proxy tracks only the kinetic channel of core-collapse explosions; although this constitutes roughly $10$~per~cent of the total SNe energy budget, the remaining thermal component scales proportionally (when averaged over the relevant timescales) owing to the numerical implementation of the feedback model. As a result, including the thermal contribution would shift the absolute values but would not alter the relative trends across the dynamical classes. 

In particular, coherent rotators exhibit uniformly low levels of feedback impact, consistent with their dynamically stable, thin discs and quiescent gas kinematics. Perturbed rotators show moderate enhancement in all feedback measures, reflecting a plausible scenario where recent or ongoing feedback contributes to the development of wave-like residuals and asymmetric distortions in the gas velocity field. Finally, slow rotators display the highest overall feedback levels, especially for AGN and generic stellar feedback channels, aligning with their strongly disrupted, thickened, and erratic gas discs. 

These trends are both intuitive and well-aligned with existing literature. For instance, \citet{Pontzen2012} demonstrated, albeit under different subgrid prescriptions, that repeated momentum injection from supernova explosions can cause localised gas parcels to migrate substantially within the gravitational potential. This process induces rapid orbital fluctuations and delays the re-establishment of equilibrium in the disc; moreover, it aligns nicely with what we have shown and discussed in Sec.~\ref{sec:Origins_of_diseq:Feedback} and Sec.~\ref{sec:Origins_of_diseq:Budget}.

Furthermore, while AGN feedback has often been regarded as energetically unimportant in low-mass galaxies, recent work suggests that even modest AGN episodes can induce significant dynamical disturbances -- see, e.g., the analysis in \citet{Koudmani2025} and the detailed discussion in their sec.~4.2. High-resolution simulations show that intermittent AGN outbursts can drive cycles of expansion and re-accretion in the central gas, generating potential fluctuations and delaying the re-establishment of equilibrium \citep[e.g.][]{Martizzi2013, Peirani2017} -- i.e. a mechanism similar to what described in \citet{Pontzen2012} regarding supernova explosions\footnote{As noted by \citet{Martizzi2013}, this mechanism is effective only when the AGN-driven cycle of gas expansion and subsequent re-accretion occurs on timescales comparable to the local dynamical time. In Sec.~\ref{sec:Origins_of_diseq:AGN} we indeed find $|\partial_t v_r| \sim |g_r|$ (i.e. $|\varepsilon| \sim 1$) in regions adjacent to an active AGN, indicating that the relevant gas motions occur on precisely the timescales required for this mechanism to operate efficiently.}.
Observational evidence reinforces this picture: \citet{ManzanoKing2020} reported a strong association between AGN activity and disturbed ionized-gas kinematics in dwarf galaxies, suggesting that AGN may play a more substantial role in shaping disc structure and rotation-curve morphology than previously assumed. This interpretation accords with our results, where slow rotators exhibit the highest AGN feedback signatures -- with median energy proxies more than $1~\mathrm{dex}$ above those of perturbed and coherent rotators -- as well as with the broader trends discussed in Sec.~\ref{sec:Origins_of_diseq:AGN} and Sec.~\ref{sec:Origins_of_diseq:Budget}.

In summary, these correlations point to a coherent physical picture in which halo shape, gas disc structure, and recent feedback collectively determine a galaxy’s dynamical state. This not only clarifies the mechanisms that drive and sustain disequilibrium, but also begins to suggest concrete, physically motivated diagnostics for identifying systems whose rotation curves are likely unreliable for dynamical mass modeling in observations.

\section{Conclusions} \label{sec:Conclusions}

We have used the currently available highest resolution, largest boxsize COLIBRE cosmological hydrodynamical simulations (i.e. a $25 ~\mathrm{cMpc}$ box at $\sim 10^{5}~\mathrm{M_\odot}$ particle resolution) to quantify departures from hydrodynamical and centrifugal equilibrium in the gas discs of dwarf galaxies, defined as systems with virial mass in the range $10^{10.75}<M_\mathrm{200c}/\mathrm{M}_\odot<10^{11}$. By evaluating the full Eulerian acceleration balance locally in warped midplane frames, we have assessed -- for the first time on a large simulated sample -- the spatial and radial prevalence of disequilibrium physics, and its implications for rotation-curve-based mass inference. We carried out our entire analysis for both the thermal and hybrid AGN feedback models and found no substantial differences affecting our conclusions; for clarity, we have presented only results from the fiducial thermal runs throughout. Our main conclusions are as follows:
\begin{enumerate}
    \item Disequilibrium is ubiquitous. Across our sample, both the full equilibrium model (Equation~(\ref{eq:EquilibriumModel})) and the naïve centrifugal balance assumption (Equation~(\ref{eq:NaiveModel})) frequently fail to recover the true circular velocity, with typical fractional deviations of $\geq 10$~per~cent. The inner few kiloparsecs, which provide the greatest leverage for dark matter profile constraints, are also the most strongly perturbed with fractional deviations typically much greater than $10$~per~cent. In particular, we find a population-wide tendency for the rotational velocity to systematically underestimate the true circular velocity in the inner disc, while a minority of regions exhibit the opposite bias; this bidirectional scatter can naturally result in a wide diversity of inferred inner rotation-curve shapes, including both cored and centrally concentrated profiles.
    \item Related to the first point, more complete fluid models do not guarantee improved mass recovery. Even when including pressure-gradient and convective terms, equilibrium-based reconstructions often perform worse than the simple assumption of $v_\phi = v_c$. This reflects the fact that unobservable time-dependent terms limit the validity of analytic corrections. Understanding the dynamical state of the system should have higher priority than attempting corrections to the rotation curve.
    \item Three physical drivers account for most strong disequilibria. Stellar feedback, self-gravitating gas clumps, and AGN energy injection collectively explain $\approx 85$~per~cent of the gas mass in extreme disequilibrium ($|\varepsilon| > 1$), often acting in concert. Large-scale gravitational asymmetries provide a persistent structural background that modulates and sustains these perturbations, and might also explain some residual mass unaccounted for by these three processes, especially in the $0.2 <|\varepsilon| < 1$ regime (mild disequilibrium).
    \item A physically grounded classification of rotation-curve reliability is possible. Using residual-based metrics linking centrifugal and hydrodynamical equilibrium, we identify three dynamical classes -- coherent rotators, perturbed rotators, and slow/erratic rotators -- which differ primarily in gas dynamics, not underlying potential shapes. Population trends at a class level in structural- and feedback-related properties reveal that recent feedback activity, gas vertical thickness and total H\textsc{i} mass are powerful indicators of the galaxy's dynamical state, albeit for a sample `unrealistically' selected in a narrow halo mass range. 
\end{enumerate}

The quantitative fractions that we report -- such as the $\approx 75$~per~cent of gas mass in non-equilibrium ($|\varepsilon| > 0.2$) and the $\approx 70$~per~cent of systems at $\log_{10}(M_{200\mathrm{c}}/\mathrm{M_\odot}) = 10.75$–$11$ unsuitable for rotation-curve analysis -- should be interpreted as indicators of prevalence rather than universal numbers. Our $25~\mathrm{cMpc}$ box is the largest volume that simultaneously reaches the gas-mass resolution required to resolve dwarf discs with $\gtrsim 10^3$ gas particles, enabling spatially resolved midplane kinematics. Larger-volume, lower-resolution simulations would inevitably wash out the disequilibrium features that we measure and inject numerical noise, while higher-resolution runs in even larger boxes remain computationally prohibitive.

A second limitation is the numerical implementation of feedback. While COLIBRE incorporates a self-consistent treatment of the cold and dense phases of the ISM, and an enhanced sampling of the gravitational potential, mitigating spurious baryon–DM energy transfer, hence making it an ideal choice for analyses of gas dynamics, our results will inevitably depend on the strength and coupling of feedback, hence remain model-dependent. Our cross-checks with the hybrid AGN feedback dual versions confirm that our main results are robust to these numerical choices: despite different calibration values and injection schemes, the nature and spatial pattern of disequilibrium are qualitatively unchanged. Ultimately, verifying the generality of our findings will require similarly resolved analyses in other simulation suites. Comparisons with recent FIRE-3 work using similar diagnostics \citep{Sands2024} further suggest that our qualitative conclusions are not simulation specific. Related evidence for systematic departures from equilibrium and biased circular-velocity recovery has also been reported by \citet{Jahn2023} within idealised simulations using the SMUGGLE model, indicating that these effects arise across a range of numerical implementations and modelling strategies, even outside fully cosmological contexts. Within the EDGE simulation suite, \citet{Rey2024} found that inner circular velocities are recovered only in a rare subset of quiescent dwarf galaxies that temporarily host well-ordered H\textsc{i} discs. In particular, their results reinforce our conclusions that rotation-curve–based mass estimates critically depend on the dynamical state of the gas -- which they also found tightly linked to morphology and star formation history, and that corrections for pressure gradients do not systematically improve recovery of the underlying gravitational potential. 

To summarise, once hydrodynamical equilibrium is broken, the foundational assumption required to interpret the rotation curve as a circular velocity -- or, equivalently, enclosed mass -- tracer is itself invalid, irrespective of any detailed analytic correction.  Given the prevalence of out-of-equilibrium gas ($\approx 75 $~per~cent of the midplane gas mass out of equilibrium), the disproportionate impact in the innermost regions of galaxies, and the $\approx 70$~per~cent of systems at $\log_{10}(M_{200\mathrm{c}}/\mathrm{M_\odot}) = 10.75$–$11$ that we find to be ill-suited for rotation-curve analysis (with this fraction expected to rise at lower halo mass), it is unavoidable that at least part of the observed diversity in dwarf galaxy rotation curves reflects mistaking non-equilibrium gas dynamics for variation in dark matter structure. This reframes long-standing issues such as the core–cusp and rotation-curve diversity problems primarily as matters of tracer reliability, implying that steady-state mass modelling requires either stringent dynamical pre-selection or methods that explicitly incorporate time-dependent gas physics.

Crucially, the near-symmetry of the $\varepsilon$ distribution (Fig.~\ref{fig:Dis_Summary}) does not imply that disequilibrium effects cancel in the mean. The observationally inferred rotation curve will inevitably depend sensitively on the detailed local structure of each individual galaxy and on how its disequilibrium features project onto the line-of-sight, so ensemble averages cannot be assumed to recover the true underlying mass -- similar arguments apply for the pressure supported and purely centrifugal models shown in the bottom row of Fig~\ref{fig:classification}. Also note that, by construction of the warped midplane geometry, we are evaluating the degree of disequilibrium in the local plane of maximal rotation; upper and lower vertical layers likely exhibit even stronger effects, enhancing these issues in thicker discs. This highlights the need to treat disequilibrium not as a nuisance that cancels in the mean, but as a first-order systematic shaping the kinematic signal itself.

Quantifying precisely the contribution of disequilibrium to rotation-curve diversity remains an open question.
The next step is to confront our theoretical diagnostics and predictions directly with mock and real observations. By generating synthetic radio (21~cm) and IFU emission-line data from our simulations \citep[e.g., respectively \textsc{MARTINI} and \textsc{SimSpin},][]{Oman_Martini2024, Harborne_SimSpin2019}, we will assess the extent to which the disequilibrium drivers and classification criteria identified here can be recognised in practice, and how they propagate into dark matter inferences. Such work will ultimately determine whether the disequilibrium physics identified in simulations is a major contributor to small-scale tensions in $\Lambda$CDM, or a challenge peculiar to certain galaxy formation models.

\section*{Software}

The following software packages were used in this work:
\textsc{numpy} \citep{Numpy}, 
\textsc{scipy}\footnote{docs.scipy.org/doc/},
\textsc{scikit-learn} \citep{Scikit-learn},
\textsc{numba} \citep{Numba},
\textsc{astropy} \citep{Astropy}, 
\textsc{h5py} \citep{h5py},
\textsc{hbt-herons} \citep{ForouharMoreno2025},
\textsc{soap} \citep{McGibbon2025},
\textsc{SwiftSimIO} \citep{swiftsimio},
\textsc{SWIFTGalaxy} \citep{SWIFTGalaxy}.

\section*{Acknowledgements} 
The authors thank the researchers working within the broader COLIBRE collaboration for helpful discussions, and providing insights that helped refining the final version of the manuscript.

DD, KAO \& KEH acknowledge support by the Royal Society through a Dorothy Hodgkin Fellowship (DHF/R1/231105) held by KAO.
FF is supported by a UKRI Future Leaders Fellowship (grant no. MR/X033740/1).
ABL acknowledges support by the Italian Ministry for Universities (MUR) program ``Dipartimenti di Eccellenza 2023-2027'' within the Centro Bicocca di Cosmologia Quantitativa (BiCoQ), and support by UNIMIB’s Fondo Di Ateneo Quota Competitiva (project 2024-ATEQC-0050).
EC and FH acknowledge funding from the Netherlands Organization for Scientific Research (NWO) through research programme Athena 184.034.002.
CSF was supported by the European Research Council (ERC) Advanced Investigator grant DMIDAS (GA 786910) and the STFC Consolidated Grant ST/T000244/1. SP acknowledges support by the Austrian Science Fund (FWF) through grant-DOI: 10.55776/V982.

This work used the DiRAC@Durham facility managed by the Institute for Computational Cosmology on behalf of the STFC DiRAC HPC Facility (www.dirac.ac.uk). The equipment was funded by BEIS capital funding via STFC capital grants ST/K00042X/1, ST/P002293/1, ST/R002371/1 and ST/S002502/1, Durham University and STFC operations grant ST/R000832/1. DiRAC is part of the National e-Infrastructure. This work has made use of NASA’s Astrophysics Data System Bibliographic Services. 

For the purpose of open access, the author has applied a Creative Commons Attribution (CC BY) licence to any Author Accepted Manuscript version arising from this submission.

Author contributions: DD led the analysis, interpretation and presentation of all results under the supervision of KAO, KEH and FF. All other authors contributed to creating the COLIBRE simulations, and provided input on methodology, interpretation of results, and draft versions of the manuscript.

\section*{Data availability}

The data supporting the plots within this article are available from the corresponding author upon request. The COLIBRE simulation suite will eventually be released publicly as part of the collaboration’s planned data-release programme, although its multi-petabyte volume will likely prevent the raw data from being hosted directly online. Until then, researchers wishing to use the simulations may contact the COLIBRE team. Further information on the project is available at \url{https://colibre.strw.leidenuniv.nl/}.

\bibliographystyle{mnras}
\bibliography{main_refs}

@ARTICLE{Navarro1997,
       author = {{Navarro}, Julio F. and {Frenk}, Carlos S. and {White}, Simon D.~M.},
        title = "{A Universal Density Profile from Hierarchical Clustering}",
      journal = {\apj},
         year = 1997,
        month = dec,
       volume = {490},
       number = {2},
        pages = {493-508},
          doi = {10.1086/304888},
archivePrefix = {arXiv},
       eprint = {astro-ph/9611107},
 primaryClass = {astro-ph},
       adsurl = {https://ui.adsabs.harvard.edu/abs/1997ApJ...490..493N}
}

@ARTICLE{Flores1994,
       author = {{Flores}, Ricardo A. and {Primack}, Joel R.},
        title = "{Observational and Theoretical Constraints on Singular Dark Matter Halos}",
      journal = {\apjl},
         year = 1994,
        month = may,
       volume = {427},
        pages = {L1},
          doi = {10.1086/187350},
archivePrefix = {arXiv},
       eprint = {astro-ph/9402004},
 primaryClass = {astro-ph},
       adsurl = {https://ui.adsabs.harvard.edu/abs/1994ApJ...427L...1F}
}

@ARTICLE{Moore1994,
       author = {{Moore}, Ben},
        title = "{Evidence against dissipation-less dark matter from observations of galaxy haloes}",
      journal = {\nat},
         year = 1994,
        month = aug,
       volume = {370},
       number = {6491},
        pages = {629-631},
          doi = {10.1038/370629a0},
       adsurl = {https://ui.adsabs.harvard.edu/abs/1994Natur.370..629M}
}

@ARTICLE{DeBlok2010,
       author = {{\VAN{De Blok}{de}{de}} Blok, W.~J.~G.},
        title = "{The Core-Cusp Problem}",
      journal = {Advances in Astronomy},
         year = 2010,
        month = jan,
       volume = {2010},
          eid = {789293},
        pages = {789293},
          doi = {10.1155/2010/789293},
archivePrefix = {arXiv},
       eprint = {0910.3538},
 primaryClass = {astro-ph.CO},
       adsurl = {https://ui.adsabs.harvard.edu/abs/2010AdAst2010E...5D}
}

@ARTICLE{Planck2016,
       author = {{Planck Collaboration} and {Ade}, P.~A.~R. and {Aghanim}, N. and {Arnaud}, M. and {Ashdown}, M. and {Aumont}, J. and {Baccigalupi}, C. and {Banday}, A.~J. and {Barreiro}, R.~B. and {Bartlett}, J.~G. and {Bartolo}, N. and {Battaner}, E. and {Battye}, R. and {Benabed}, K. and {Beno{\^\i}t}, A. and {Benoit-L{\'e}vy}, A. and {Bernard}, J.-P. and {Bersanelli}, M. and {Bielewicz}, P. and {Bock}, J.~J. and {Bonaldi}, A. and {Bonavera}, L. and {Bond}, J.~R. and {Borrill}, J. and {Bouchet}, F.~R. and {Boulanger}, F. and {Bucher}, M. and {Burigana}, C. and {Butler}, R.~C. and {Calabrese}, E. and {Cardoso}, J.-F. and {Catalano}, A. and {Challinor}, A. and {Chamballu}, A. and {Chary}, R.-R. and {Chiang}, H.~C. and {Chluba}, J. and {Christensen}, P.~R. and {Church}, S. and {Clements}, D.~L. and {Colombi}, S. and {Colombo}, L.~P.~L. and {Combet}, C. and {Coulais}, A. and {Crill}, B.~P. and {Curto}, A. and {Cuttaia}, F. and {Danese}, L. and {Davies}, R.~D. and {Davis}, R.~J. and {de Bernardis}, P. and {de Rosa}, A. and {de Zotti}, G. and {Delabrouille}, J. and {D{\'e}sert}, F.-X. and {Di Valentino}, E. and {Dickinson}, C. and {Diego}, J.~M. and {Dolag}, K. and {Dole}, H. and {Donzelli}, S. and {Dor{\'e}}, O. and {Douspis}, M. and {Ducout}, A. and {Dunkley}, J. and {Dupac}, X. and {Efstathiou}, G. and {Elsner}, F. and {En{\ss}lin}, T.~A. and {Eriksen}, H.~K. and {Farhang}, M. and {Fergusson}, J. and {Finelli}, F. and {Forni}, O. and {Frailis}, M. and {Fraisse}, A.~A. and {Franceschi}, E. and {Frejsel}, A. and {Galeotta}, S. and {Galli}, S. and {Ganga}, K. and {Gauthier}, C. and {Gerbino}, M. and {Ghosh}, T. and {Giard}, M. and {Giraud-H{\'e}raud}, Y. and {Giusarma}, E. and {Gjerl{\o}w}, E. and {Gonz{\'a}lez-Nuevo}, J. and {G{\'o}rski}, K.~M. and {Gratton}, S. and {Gregorio}, A. and {Gruppuso}, A. and {Gudmundsson}, J.~E. and {Hamann}, J. and {Hansen}, F.~K. and {Hanson}, D. and {Harrison}, D.~L. and {Helou}, G. and {Henrot-Versill{\'e}}, S. and {Hern{\'a}ndez-Monteagudo}, C. and {Herranz}, D. and {Hildebrandt}, S.~R. and {Hivon}, E. and {Hobson}, M. and {Holmes}, W.~A. and {Hornstrup}, A. and {Hovest}, W. and {Huang}, Z. and {Huffenberger}, K.~M. and {Hurier}, G. and {Jaffe}, A.~H. and {Jaffe}, T.~R. and {Jones}, W.~C. and {Juvela}, M. and {Keih{\"a}nen}, E. and {Keskitalo}, R. and {Kisner}, T.~S. and {Kneissl}, R. and {Knoche}, J. and {Knox}, L. and {Kunz}, M. and {Kurki-Suonio}, H. and {Lagache}, G. and {L{\"a}hteenm{\"a}ki}, A. and {Lamarre}, J.-M. and {Lasenby}, A. and {Lattanzi}, M. and {Lawrence}, C.~R. and {Leahy}, J.~P. and {Leonardi}, R. and {Lesgourgues}, J. and {Levrier}, F. and {Lewis}, A. and {Liguori}, M. and {Lilje}, P.~B. and {Linden-V{\o}rnle}, M. and {L{\'o}pez-Caniego}, M. and {Lubin}, P.~M. and {Mac{\'\i}as-P{\'e}rez}, J.~F. and {Maggio}, G. and {Maino}, D. and {Mandolesi}, N. and {Mangilli}, A. and {Marchini}, A. and {Maris}, M. and {Martin}, P.~G. and {Martinelli}, M. and {Mart{\'\i}nez-Gonz{\'a}lez}, E. and {Masi}, S. and {Matarrese}, S. and {McGehee}, P. and {Meinhold}, P.~R. and {Melchiorri}, A. and {Melin}, J.-B. and {Mendes}, L. and {Mennella}, A. and {Migliaccio}, M. and {Millea}, M. and {Mitra}, S. and {Miville-Desch{\^e}nes}, M.-A. and {Moneti}, A. and {Montier}, L. and {Morgante}, G. and {Mortlock}, D. and {Moss}, A. and {Munshi}, D. and {Murphy}, J.~A. and {Naselsky}, P. and {Nati}, F. and {Natoli}, P. and {Netterfield}, C.~B. and {N{\o}rgaard-Nielsen}, H.~U. and {Noviello}, F. and {Novikov}, D. and {Novikov}, I. and {Oxborrow}, C.~A. and {Paci}, F. and {Pagano}, L. and {Pajot}, F. and {Paladini}, R. and {Paoletti}, D. and {Partridge}, B. and {Pasian}, F. and {Patanchon}, G. and {Pearson}, T.~J. and {Perdereau}, O. and {Perotto}, L. and {Perrotta}, F. and {Pettorino}, V. and {Piacentini}, F. and {Piat}, M. and {Pierpaoli}, E. and {Pietrobon}, D. and {Plaszczynski}, S. and {Pointecouteau}, E. and {Polenta}, G. and {Popa}, L. and {Pratt}, G.~W. and {Pr{\'e}zeau}, G.},
        title = "{Planck 2015 results. XIII. Cosmological parameters}",
      journal = {\aap},
         year = 2016,
        month = sep,
       volume = {594},
          eid = {A13},
        pages = {A13},
          doi = {10.1051/0004-6361/201525830},
archivePrefix = {arXiv},
       eprint = {1502.01589},
 primaryClass = {astro-ph.CO},
       adsurl = {https://ui.adsabs.harvard.edu/abs/2016A&A...594A..13P}
}

@ARTICLE{Ludlow2016,
       author = {{Ludlow}, Aaron D. and {Bose}, Sownak and {Angulo}, Ra{\'u}l E. and {Wang}, Lan and {Hellwing}, Wojciech A. and {Navarro}, Julio F. and {Cole}, Shaun and {Frenk}, Carlos S.},
        title = "{The mass-concentration-redshift relation of cold and warm dark matter haloes}",
      journal = {\mnras},
         year = 2016,
        month = aug,
       volume = {460},
       number = {2},
        pages = {1214-1232},
          doi = {10.1093/mnras/stw1046},
archivePrefix = {arXiv},
       eprint = {1601.02624},
 primaryClass = {astro-ph.CO},
       adsurl = {https://ui.adsabs.harvard.edu/abs/2016MNRAS.460.1214L}
}

@ARTICLE{Oman2015,
       author = {{Oman}, Kyle A. and {Navarro}, Julio F. and {Fattahi}, Azadeh and {Frenk}, Carlos S. and {Sawala}, Till and {White}, Simon D.~M. and {Bower}, Richard and {Crain}, Robert A. and {Furlong}, Michelle and {Schaller}, Matthieu and {Schaye}, Joop and {Theuns}, Tom},
        title = "{The unexpected diversity of dwarf galaxy rotation curves}",
      journal = {\mnras},
         year = 2015,
        month = oct,
       volume = {452},
       number = {4},
        pages = {3650-3665},
          doi = {10.1093/mnras/stv1504},
archivePrefix = {arXiv},
       eprint = {1504.01437},
 primaryClass = {astro-ph.GA},
       adsurl = {https://ui.adsabs.harvard.edu/abs/2015MNRAS.452.3650O}
}

@ARTICLE{Mashchenko2008,
       author = {{Mashchenko}, Sergey and {Wadsley}, James and {Couchman}, H.~M.~P.},
        title = "{Stellar Feedback in Dwarf Galaxy Formation}",
      journal = {Science},
         year = 2008,
        month = jan,
       volume = {319},
       number = {5860},
        pages = {174},
          doi = {10.1126/science.1148666},
archivePrefix = {arXiv},
       eprint = {0711.4803},
 primaryClass = {astro-ph},
       adsurl = {https://ui.adsabs.harvard.edu/abs/2008Sci...319..174M}
}

@ARTICLE{Governato2012,
       author = {{Governato}, F. and {Zolotov}, A. and {Pontzen}, A. and {Christensen}, C. and {Oh}, S.~H. and {Brooks}, A.~M. and {Quinn}, T. and {Shen}, S. and {Wadsley}, J.},
        title = "{Cuspy no more: how outflows affect the central dark matter and baryon distribution in {\ensuremath{\Lambda}} cold dark matter galaxies}",
      journal = {\mnras},
         year = 2012,
        month = may,
       volume = {422},
       number = {2},
        pages = {1231-1240},
          doi = {10.1111/j.1365-2966.2012.20696.x},
archivePrefix = {arXiv},
       eprint = {1202.0554},
 primaryClass = {astro-ph.CO},
       adsurl = {https://ui.adsabs.harvard.edu/abs/2012MNRAS.422.1231G}
}

@ARTICLE{Pontzen2012,
       author = {{Pontzen}, Andrew and {Governato}, Fabio},
        title = "{How supernova feedback turns dark matter cusps into cores}",
      journal = {\mnras},
         year = 2012,
        month = apr,
       volume = {421},
       number = {4},
        pages = {3464-3471},
          doi = {10.1111/j.1365-2966.2012.20571.x},
archivePrefix = {arXiv},
       eprint = {1106.0499},
 primaryClass = {astro-ph.CO},
       adsurl = {https://ui.adsabs.harvard.edu/abs/2012MNRAS.421.3464P}
}

@ARTICLE{DiCintio2014,
       author = {{Di Cintio}, Arianna and {Brook}, Chris B. and {Macci{\`o}}, Andrea V. and {Stinson}, Greg S. and {Knebe}, Alexander and {Dutton}, Aaron A. and {Wadsley}, James},
        title = "{The dependence of dark matter profiles on the stellar-to-halo mass ratio: a prediction for cusps versus cores}",
      journal = {\mnras},
         year = 2014,
        month = jan,
       volume = {437},
       number = {1},
        pages = {415-423},
          doi = {10.1093/mnras/stt1891},
archivePrefix = {arXiv},
       eprint = {1306.0898},
 primaryClass = {astro-ph.CO},
       adsurl = {https://ui.adsabs.harvard.edu/abs/2014MNRAS.437..415D}
}

@ARTICLE{Chan2015,
       author = {{Chan}, T.~K. and {Kere{\v{s}}}, D. and {O{\~n}orbe}, J. and {Hopkins}, P.~F. and {Muratov}, A.~L. and {Faucher-Gigu{\`e}re}, C.-A. and {Quataert}, E.},
        title = "{The impact of baryonic physics on the structure of dark matter haloes: the view from the FIRE cosmological simulations}",
      journal = {\mnras},
         year = 2015,
        month = dec,
       volume = {454},
       number = {3},
        pages = {2981-3001},
          doi = {10.1093/mnras/stv2165},
archivePrefix = {arXiv},
       eprint = {1507.02282},
 primaryClass = {astro-ph.GA},
       adsurl = {https://ui.adsabs.harvard.edu/abs/2015MNRAS.454.2981C}
}

@ARTICLE{Schaye2014,
       author = {{Schaye}, Joop and {Crain}, Robert A. and {Bower}, Richard G. and {Furlong}, Michelle and {Schaller}, Matthieu and {Theuns}, Tom and {Dalla Vecchia}, Claudio and {Frenk}, Carlos S. and {McCarthy}, I.~G. and {Helly}, John C. and {Jenkins}, Adrian and {Rosas-Guevara}, Y.~M. and {White}, Simon D.~M. and {Baes}, Maarten and {Booth}, C.~M. and {Camps}, Peter and {Navarro}, Julio F. and {Qu}, Yan and {Rahmati}, Alireza and {Sawala}, Till and {Thomas}, Peter A. and {Trayford}, James},
        title = "{The EAGLE project: simulating the evolution and assembly of galaxies and their environments}",
      journal = {\mnras},
         year = 2015,
        month = jan,
       volume = {446},
       number = {1},
        pages = {521-554},
          doi = {10.1093/mnras/stu2058},
archivePrefix = {arXiv},
       eprint = {1407.7040},
 primaryClass = {astro-ph.GA},
       adsurl = {https://ui.adsabs.harvard.edu/abs/2015MNRAS.446..521S}
}

@ARTICLE{Vogelsberger2014,
       author = {{Vogelsberger}, Mark and {Genel}, Shy and {Springel}, Volker and {Torrey}, Paul and {Sijacki}, Debora and {Xu}, Dandan and {Snyder}, Greg and {Nelson}, Dylan and {Hernquist}, Lars},
        title = "{Introducing the Illustris Project: simulating the coevolution of dark and visible matter in the Universe}",
      journal = {\mnras},
         year = 2014,
        month = oct,
       volume = {444},
       number = {2},
        pages = {1518-1547},
          doi = {10.1093/mnras/stu1536},
archivePrefix = {arXiv},
       eprint = {1405.2921},
 primaryClass = {astro-ph.CO},
       adsurl = {https://ui.adsabs.harvard.edu/abs/2014MNRAS.444.1518V}
}

@ARTICLE{Spergel2000,
       author = {{Spergel}, David N. and {Steinhardt}, Paul J.},
        title = "{Observational Evidence for Self-Interacting Cold Dark Matter}",
      journal = {\prl},
         year = 2000,
        month = apr,
       volume = {84},
       number = {17},
        pages = {3760-3763},
          doi = {10.1103/PhysRevLett.84.3760},
archivePrefix = {arXiv},
       eprint = {astro-ph/9909386},
 primaryClass = {astro-ph},
       adsurl = {https://ui.adsabs.harvard.edu/abs/2000PhRvL..84.3760S}
}

@ARTICLE{Rocha2013,
       author = {{Rocha}, Miguel and {Peter}, Annika H.~G. and {Bullock}, James S. and {Kaplinghat}, Manoj and {Garrison-Kimmel}, Shea and {O{\~n}orbe}, Jose and {Moustakas}, Leonidas A.},
        title = "{Cosmological simulations with self-interacting dark matter - I. Constant-density cores and substructure}",
      journal = {\mnras},
         year = 2013,
        month = mar,
       volume = {430},
       number = {1},
        pages = {81-104},
          doi = {10.1093/mnras/sts514},
archivePrefix = {arXiv},
       eprint = {1208.3025},
 primaryClass = {astro-ph.CO},
       adsurl = {https://ui.adsabs.harvard.edu/abs/2013MNRAS.430...81R}
}

@ARTICLE{Elbert2015,
       author = {{Elbert}, Oliver D. and {Bullock}, James S. and {Garrison-Kimmel}, Shea and {Rocha}, Miguel and {O{\~n}orbe}, Jose and {Peter}, Annika H.~G.},
        title = "{Core formation in dwarf haloes with self-interacting dark matter: no fine-tuning necessary}",
      journal = {\mnras},
         year = 2015,
        month = oct,
       volume = {453},
       number = {1},
        pages = {29-37},
          doi = {10.1093/mnras/stv1470},
archivePrefix = {arXiv},
       eprint = {1412.1477},
 primaryClass = {astro-ph.GA},
       adsurl = {https://ui.adsabs.harvard.edu/abs/2015MNRAS.453...29E}
}

@ARTICLE{Ren2019,
       author = {{Ren}, Tao and {Kwa}, Anna and {Kaplinghat}, Manoj and {Yu}, Hai-Bo},
        title = "{Reconciling the Diversity and Uniformity of Galactic Rotation Curves with Self-Interacting Dark Matter}",
      journal = {Physical Review X},
         year = 2019,
        month = jul,
       volume = {9},
       number = {3},
          eid = {031020},
        pages = {031020},
          doi = {10.1103/PhysRevX.9.031020},
archivePrefix = {arXiv},
       eprint = {1808.05695},
 primaryClass = {astro-ph.GA},
       adsurl = {https://ui.adsabs.harvard.edu/abs/2019PhRvX...9c1020R}
}

@ARTICLE{Creasey2017,
       author = {{Creasey}, Peter and {Sameie}, Omid and {Sales}, Laura V. and {Yu}, Hai-Bo and {Vogelsberger}, Mark and {Zavala}, Jes{\'u}s},
        title = "{Spreading out and staying sharp - creating diverse rotation curves via baryonic and self-interaction effects}",
      journal = {\mnras},
         year = 2017,
        month = jun,
       volume = {468},
       number = {2},
        pages = {2283-2295},
          doi = {10.1093/mnras/stx522},
archivePrefix = {arXiv},
       eprint = {1612.03903},
 primaryClass = {astro-ph.GA},
       adsurl = {https://ui.adsabs.harvard.edu/abs/2017MNRAS.468.2283C}
}

@ARTICLE{Kaplinghat2020,
       author = {{Kaplinghat}, Manoj and {Ren}, Tao and {Yu}, Hai-Bo},
        title = "{Dark matter cores and cusps in spiral galaxies and their explanations}",
      journal = {\jcap},
         year = 2020,
        month = jun,
       volume = {2020},
       number = {6},
          eid = {027},
        pages = {027},
          doi = {10.1088/1475-7516/2020/06/027},
archivePrefix = {arXiv},
       eprint = {1911.00544},
 primaryClass = {astro-ph.GA},
       adsurl = {https://ui.adsabs.harvard.edu/abs/2020JCAP...06..027K}
}

@ARTICLE{Swaters2009,
       author = {{Swaters}, R.~A. and {Sancisi}, R. and {van Albada}, T.~S. and {van der Hulst}, J.~M.},
        title = "{The rotation curves shapes of late-type dwarf galaxies}",
      journal = {\aap},
         year = 2009,
        month = jan,
       volume = {493},
       number = {3},
        pages = {871-892},
          doi = {10.1051/0004-6361:200810516},
archivePrefix = {arXiv},
       eprint = {0901.4222},
 primaryClass = {astro-ph.CO},
       adsurl = {https://ui.adsabs.harvard.edu/abs/2009A&A...493..871S}
}

@ARTICLE{Swaters2003,
       author = {{Swaters}, R.~A. and {Madore}, B.~F. and {van den Bosch}, Frank C. and {Balcells}, M.},
        title = "{The Central Mass Distribution in Dwarf and Low Surface Brightness Galaxies}",
      journal = {\apj},
         year = 2003,
        month = feb,
       volume = {583},
       number = {2},
        pages = {732-751},
          doi = {10.1086/345426},
archivePrefix = {arXiv},
       eprint = {astro-ph/0210152},
 primaryClass = {astro-ph},
       adsurl = {https://ui.adsabs.harvard.edu/abs/2003ApJ...583..732S}
}

@ARTICLE{Spekkens2005,
       author = {{Spekkens}, Kristine and {Giovanelli}, Riccardo and {Haynes}, Martha P.},
        title = "{The Cusp/Core Problem in Galactic Halos: Long-Slit Spectra for a Large Dwarf Galaxy Sample}",
      journal = {\aj},
         year = 2005,
        month = may,
       volume = {129},
       number = {5},
        pages = {2119-2137},
          doi = {10.1086/429592},
archivePrefix = {arXiv},
       eprint = {astro-ph/0502166},
 primaryClass = {astro-ph},
       adsurl = {https://ui.adsabs.harvard.edu/abs/2005AJ....129.2119S}
}

@ARTICLE{Walter2008,
       author = {{Walter}, Fabian and {Brinks}, Elias and {de Blok}, W.~J.~G. and {Bigiel}, Frank and {Kennicutt}, Jr., Robert C. and {Thornley}, Michele D. and {Leroy}, Adam},
        title = "{THINGS: The H I Nearby Galaxy Survey}",
      journal = {\aj},
         year = 2008,
        month = dec,
       volume = {136},
       number = {6},
        pages = {2563-2647},
          doi = {10.1088/0004-6256/136/6/2563},
archivePrefix = {arXiv},
       eprint = {0810.2125},
 primaryClass = {astro-ph},
       adsurl = {https://ui.adsabs.harvard.edu/abs/2008AJ....136.2563W}
}

@ARTICLE{Hunter2012,
       author = {{Hunter}, Deidre A. and {Ficut-Vicas}, Dana and {Ashley}, Trisha and {Brinks}, Elias and {Cigan}, Phil and {Elmegreen}, Bruce G. and {Heesen}, Volker and {Herrmann}, Kimberly A. and {Johnson}, Megan and {Oh}, Se-Heon and {Rupen}, Michael P. and {Schruba}, Andreas and {Simpson}, Caroline E. and {Walter}, Fabian and {Westpfahl}, David J. and {Young}, Lisa M. and {Zhang}, Hong-Xin},
        title = "{Little Things}",
      journal = {\aj},
         year = 2012,
        month = nov,
       volume = {144},
       number = {5},
          eid = {134},
        pages = {134},
          doi = {10.1088/0004-6256/144/5/134},
archivePrefix = {arXiv},
       eprint = {1208.5834},
 primaryClass = {astro-ph.GA},
       adsurl = {https://ui.adsabs.harvard.edu/abs/2012AJ....144..134H}
}

@ARTICLE{Lelli2016,
       author = {{Lelli}, Federico and {McGaugh}, Stacy S. and {Schombert}, James M.},
        title = "{SPARC: Mass Models for 175 Disk Galaxies with Spitzer Photometry and Accurate Rotation Curves}",
      journal = {\aj},
         year = 2016,
        month = dec,
       volume = {152},
       number = {6},
          eid = {157},
        pages = {157},
          doi = {10.3847/0004-6256/152/6/157},
archivePrefix = {arXiv},
       eprint = {1606.09251},
 primaryClass = {astro-ph.GA},
       adsurl = {https://ui.adsabs.harvard.edu/abs/2016AJ....152..157L}
}

@ARTICLE{Bell2001,
       author = {{Bell}, Eric F. and {de Jong}, Roelof S.},
        title = "{Stellar Mass-to-Light Ratios and the Tully-Fisher Relation}",
      journal = {\apj},
         year = 2001,
        month = mar,
       volume = {550},
       number = {1},
        pages = {212-229},
          doi = {10.1086/319728},
archivePrefix = {arXiv},
       eprint = {astro-ph/0011493},
 primaryClass = {astro-ph},
       adsurl = {https://ui.adsabs.harvard.edu/abs/2001ApJ...550..212B}
}

@ARTICLE{Bruzual2003,
       author = {{Bruzual}, G. and {Charlot}, S.},
        title = "{Stellar population synthesis at the resolution of 2003}",
      journal = {\mnras},
         year = 2003,
        month = oct,
       volume = {344},
       number = {4},
        pages = {1000-1028},
          doi = {10.1046/j.1365-8711.2003.06897.x},
archivePrefix = {arXiv},
       eprint = {astro-ph/0309134},
 primaryClass = {astro-ph},
       adsurl = {https://ui.adsabs.harvard.edu/abs/2003MNRAS.344.1000B}
}

@ARTICLE{Oh2011,
       author = {{Oh}, Se-Heon and {Brook}, Chris and {Governato}, Fabio and {Brinks}, Elias and {Mayer}, Lucio and {de Blok}, W.~J.~G. and {Brooks}, Alyson and {Walter}, Fabian},
        title = "{The Central Slope of Dark Matter Cores in Dwarf Galaxies: Simulations versus THINGS}",
      journal = {\aj},
         year = 2011,
        month = jul,
       volume = {142},
       number = {1},
          eid = {24},
        pages = {24},
          doi = {10.1088/0004-6256/142/1/24},
archivePrefix = {arXiv},
       eprint = {1011.2777},
 primaryClass = {astro-ph.CO},
       adsurl = {https://ui.adsabs.harvard.edu/abs/2011AJ....142...24O}
}

@ARTICLE{Davis1985,
       author = {{Davis}, M. and {Efstathiou}, G. and {Frenk}, C.~S. and {White}, S.~D.~M.},
        title = "{The evolution of large-scale structure in a universe dominated by cold dark matter}",
      journal = {\apj},
         year = 1985,
        month = may,
       volume = {292},
        pages = {371-394},
          doi = {10.1086/163168},
       adsurl = {https://ui.adsabs.harvard.edu/abs/1985ApJ...292..371D}
}

@ARTICLE{Frenk88,
       author = {{Frenk}, Carlos S. and {White}, Simon D.~M. and {Davis}, Marc and {Efstathiou}, George},
        title = "{The Formation of Dark Halos in a Universe Dominated by Cold Dark Matter}",
      journal = {\apj},
         year = 1988,
        month = apr,
       volume = {327},
        pages = {507},
          doi = {10.1086/166213},
       adsurl = {https://ui.adsabs.harvard.edu/abs/1988ApJ...327..507F}
}

@ARTICLE{Hayashi2007,
       author = {{Hayashi}, Eric and {Navarro}, Julio F. and {Springel}, Volker},
        title = "{The shape of the gravitational potential in cold dark matter haloes}",
      journal = {\mnras},
         year = 2007,
        month = may,
       volume = {377},
       number = {1},
        pages = {50-62},
          doi = {10.1111/j.1365-2966.2007.11599.x},
archivePrefix = {arXiv},
       eprint = {astro-ph/0612327},
 primaryClass = {astro-ph},
       adsurl = {https://ui.adsabs.harvard.edu/abs/2007MNRAS.377...50H}
}

@ARTICLE{Hayashi2006,
       author = {{Hayashi}, Eric and {Navarro}, Julio F.},
        title = "{Hiding cusps in cores: kinematics of disc galaxies in triaxial dark matter haloes}",
      journal = {\mnras},
         year = 2006,
        month = dec,
       volume = {373},
       number = {3},
        pages = {1117-1124},
          doi = {10.1111/j.1365-2966.2006.10927.x},
archivePrefix = {arXiv},
       eprint = {astro-ph/0608376},
 primaryClass = {astro-ph},
       adsurl = {https://ui.adsabs.harvard.edu/abs/2006MNRAS.373.1117H}
}

@ARTICLE{Oman2019,
       author = {{Oman}, Kyle A. and {Marasco}, Antonino and {Navarro}, Julio F. and {Frenk}, Carlos S. and {Schaye}, Joop and {Ben{\'\i}tez-Llambay}, Alejandro},
        title = "{Non-circular motions and the diversity of dwarf galaxy rotation curves}",
      journal = {\mnras},
         year = 2019,
        month = jan,
       volume = {482},
       number = {1},
        pages = {821-847},
          doi = {10.1093/mnras/sty2687},
archivePrefix = {arXiv},
       eprint = {1706.07478},
 primaryClass = {astro-ph.GA},
       adsurl = {https://ui.adsabs.harvard.edu/abs/2019MNRAS.482..821O}
}

@ARTICLE{Marasco2018,
       author = {{Marasco}, A. and {Oman}, K.~A. and {Navarro}, J.~F. and {Frenk}, C.~S. and {Oosterloo}, T.},
        title = "{Bars in dark-matter-dominated dwarf galaxy discs}",
      journal = {\mnras},
         year = 2018,
        month = may,
       volume = {476},
       number = {2},
        pages = {2168-2176},
          doi = {10.1093/mnras/sty354},
archivePrefix = {arXiv},
       eprint = {1711.09914},
 primaryClass = {astro-ph.GA},
       adsurl = {https://ui.adsabs.harvard.edu/abs/2018MNRAS.476.2168M}
}

@ARTICLE{Trachternach2008,
       author = {{Trachternach}, C. and {de Blok}, W.~J.~G. and {Walter}, F. and {Brinks}, E. and {Kennicutt}, Jr., R.~C.},
        title = "{Dynamical Centers and Noncircular Motions in THINGS Galaxies: Implications for Dark Matter Halos}",
      journal = {\aj},
         year = 2008,
        month = dec,
       volume = {136},
       number = {6},
        pages = {2720-2760},
          doi = {10.1088/0004-6256/136/6/2720},
archivePrefix = {arXiv},
       eprint = {0810.2116},
 primaryClass = {astro-ph},
       adsurl = {https://ui.adsabs.harvard.edu/abs/2008AJ....136.2720T}
}

@ARTICLE{Roper2023,
       author = {{Roper}, Finn A. and {Oman}, Kyle A. and {Frenk}, Carlos S. and {Ben{\'\i}tez-Llambay}, Alejandro and {Navarro}, Julio F. and {Santos-Santos}, Isabel M.~E.},
        title = "{The diversity of rotation curves of simulated galaxies with cusps and cores}",
      journal = {\mnras},
         year = 2023,
        month = may,
       volume = {521},
       number = {1},
        pages = {1316-1336},
          doi = {10.1093/mnras/stad549},
archivePrefix = {arXiv},
       eprint = {2203.16652},
 primaryClass = {astro-ph.GA},
       adsurl = {https://ui.adsabs.harvard.edu/abs/2023MNRAS.521.1316R}
}

@ARTICLE{Valenzuela2007,
       author = {{Valenzuela}, Octavio and {Rhee}, George and {Klypin}, Anatoly and {Governato}, Fabio and {Stinson}, Gregory and {Quinn}, Thomas and {Wadsley}, James},
        title = "{Is There Evidence for Flat Cores in the Halos of Dwarf Galaxies? The Case of NGC 3109 and NGC 6822}",
      journal = {\apj},
         year = 2007,
        month = mar,
       volume = {657},
       number = {2},
        pages = {773-789},
          doi = {10.1086/508674},
archivePrefix = {arXiv},
       eprint = {astro-ph/0509644},
 primaryClass = {astro-ph},
       adsurl = {https://ui.adsabs.harvard.edu/abs/2007ApJ...657..773V}
}

@ARTICLE{Read2016,
       author = {{Read}, J.~I. and {Iorio}, G. and {Agertz}, O. and {Fraternali}, F.},
        title = "{Understanding the shape and diversity of dwarf galaxy rotation curves in {\ensuremath{\Lambda}}CDM}",
      journal = {\mnras},
         year = 2016,
        month = nov,
       volume = {462},
       number = {4},
        pages = {3628-3645},
          doi = {10.1093/mnras/stw1876},
archivePrefix = {arXiv},
       eprint = {1601.05821},
 primaryClass = {astro-ph.GA},
       adsurl = {https://ui.adsabs.harvard.edu/abs/2016MNRAS.462.3628R}
}

@ARTICLE{Pina2025,
       author = {{Mancera Pi{\~n}a}, Pavel E. and {Read}, Justin I. and {Kim}, Stacy and {Marasco}, Antonino and {Benavides}, Jos{\'e} A. and {Glowacki}, Marcin and {Pezzulli}, Gabriele and {Lagos}, Claudia del P.},
        title = "{The galaxy-halo connection of disc galaxies over six orders of magnitude in stellar mass}",
      journal = {\aap},
         year = 2025,
        month = jul,
       volume = {699},
          eid = {A311},
        pages = {A311},
          doi = {10.1051/0004-6361/202554381},
archivePrefix = {arXiv},
       eprint = {2505.22727},
 primaryClass = {astro-ph.GA},
       adsurl = {https://ui.adsabs.harvard.edu/abs/2025A&A...699A.311M}
}

@ARTICLE{Santos2020,
       author = {{Santos-Santos}, Isabel M.~E. and {Navarro}, Julio F. and {Robertson}, Andrew and {Ben{\'\i}tez-Llambay}, Alejandro and {Oman}, Kyle A. and {Lovell}, Mark R. and {Frenk}, Carlos S. and {Ludlow}, Aaron D. and {Fattahi}, Azadeh and {Ritz}, Adam},
        title = "{Baryonic clues to the puzzling diversity of dwarf galaxy rotation curves}",
      journal = {\mnras},
         year = 2020,
        month = jun,
       volume = {495},
       number = {1},
        pages = {58-77},
          doi = {10.1093/mnras/staa1072},
archivePrefix = {arXiv},
       eprint = {1911.09116},
 primaryClass = {astro-ph.GA},
       adsurl = {https://ui.adsabs.harvard.edu/abs/2020MNRAS.495...58S}
}

@ARTICLE{Sales2023,
       author = {{Sales}, Laura V. and {Wetzel}, Andrew and {Fattahi}, Azadeh},
        title = "{Baryonic solutions and challenges for cosmological models of dwarf galaxies}",
      journal = {Nature Astronomy},
         year = 2022,
        month = jun,
       volume = {6},
        pages = {897-910},
          doi = {10.1038/s41550-022-01689-w},
archivePrefix = {arXiv},
       eprint = {2206.05295},
 primaryClass = {astro-ph.GA},
       adsurl = {https://ui.adsabs.harvard.edu/abs/2022NatAs...6..897S}
}

@ARTICLE{ElBadry2018,
       author = {{El-Badry}, Kareem and {Quataert}, Eliot and {Wetzel}, Andrew and {Hopkins}, Philip F. and {Weisz}, Daniel R. and {Chan}, T.~K. and {Fitts}, Alex and {Boylan-Kolchin}, Michael and {Kere{\v{s}}}, Du{\v{s}}an and {Faucher-Gigu{\`e}re}, Claude-Andr{\'e} and {Garrison-Kimmel}, Shea},
        title = "{Gas kinematics, morphology and angular momentum in the FIRE simulations}",
      journal = {\mnras},
         year = 2018,
        month = jan,
       volume = {473},
       number = {2},
        pages = {1930-1955},
          doi = {10.1093/mnras/stx2482},
archivePrefix = {arXiv},
       eprint = {1705.10321},
 primaryClass = {astro-ph.GA},
       adsurl = {https://ui.adsabs.harvard.edu/abs/2018MNRAS.473.1930E}
}

@ARTICLE{DowningOman2023,
       author = {{Downing}, Eleanor R. and {Oman}, Kyle A.},
        title = "{The many reasons that the rotation curves of low-mass galaxies can fail as tracers of their matter distributions}",
      journal = {\mnras},
         year = 2023,
        month = jul,
       volume = {522},
       number = {3},
        pages = {3318-3336},
          doi = {10.1093/mnras/stad868},
archivePrefix = {arXiv},
       eprint = {2301.05242},
 primaryClass = {astro-ph.GA},
       adsurl = {https://ui.adsabs.harvard.edu/abs/2023MNRAS.522.3318D}
}

@ARTICLE{Sands2024,
       author = {{Sands}, Isabel S. and {Hopkins}, Philip F. and {Shen}, Xuejian and {Boylan-Kolchin}, Michael and {Bullock}, James and {Faucher-Giguere}, Claude-Andre and {Mercado}, Francisco J. and {Moreno}, Jorge and {Necib}, Lina and {Ou}, Xiaowei and {Wellons}, Sarah and {Wetzel}, Andrew},
        title = "{Confronting the Diversity Problem: The Limits of Galaxy Rotation Curves as a tool to Understand Dark Matter Profiles}",
      journal = {arXiv e-prints},
         year = 2024,
        month = apr,
          eid = {arXiv:2404.16247},
        pages = {arXiv:2404.16247},
          doi = {10.48550/arXiv.2404.16247},
archivePrefix = {arXiv},
       eprint = {2404.16247},
 primaryClass = {astro-ph.GA},
       adsurl = {https://ui.adsabs.harvard.edu/abs/2024arXiv240416247S}
}

@ARTICLE{Schaye2025,
       author = {{Schaye}, Joop and {Chaikin}, Evgenii and {Schaller}, Matthieu and {Ploeckinger}, Sylvia and {Hu{\v{s}}ko}, Filip and {McGibbon}, Rob and {Trayford}, James W. and {Ben{\'\i}tez-Llambay}, Alejandro and {Correa}, Camila and {Frenk}, Carlos S. and {Richings}, Alexander J. and {Forouhar Moreno}, Victor J. and {Bah{\'e}}, Yannick M. and {Borrow}, Josh and {Durrant}, Anna and {Gebek}, Andrea and {Helly}, John C. and {Jenkins}, Adrian and {Lacey}, Cedric G. and {Ludlow}, Aaron and {Nobels}, Folkert S.~J.},
        title = "{The COLIBRE project: cosmological hydrodynamical simulations of galaxy formation and evolution}",
      journal = {arXiv e-prints},
         year = 2025,
        month = aug,
          eid = {arXiv:2508.21126},
        pages = {arXiv:2508.21126},
          doi = {10.48550/arXiv.2508.21126},
archivePrefix = {arXiv},
       eprint = {2508.21126},
 primaryClass = {astro-ph.GA},
       adsurl = {https://ui.adsabs.harvard.edu/abs/2025arXiv250821126S}
}

@ARTICLE{Chaikin2025,
       author = {{Chaikin}, Evgenii and {Schaye}, Joop and {Schaller}, Matthieu and {Ploeckinger}, Sylvia and {Bah{\'e}}, Yannick M. and {Ben{\'\i}tez-Llambay}, Alejandro and {Correa}, Camila and {Forouhar Moreno}, Victor J. and {Frenk}, Carlos S. and {Hu{\v{s}}ko}, Filip and {Kugel}, Roi and {McGibbon}, Robert and {Richings}, Alexander J. and {Trayford}, James W. and {Borrow}, Josh and {Crain}, Robert A. and {Helly}, John C. and {Lacey}, Cedric G. and {Ludlow}, Aaron and {Nobels}, Folkert S.~J.},
        title = "{COLIBRE: calibrating subgrid feedback in cosmological simulations that include a cold gas phase}",
      journal = {arXiv e-prints},
         year = 2025,
        month = sep,
          eid = {arXiv:2509.04067},
        pages = {arXiv:2509.04067},
          doi = {10.48550/arXiv.2509.04067},
archivePrefix = {arXiv},
       eprint = {2509.04067},
 primaryClass = {astro-ph.GA},
       adsurl = {https://ui.adsabs.harvard.edu/abs/2025arXiv250904067C}
}

@ARTICLE{Schaller2024,
       author = {{Schaller}, Matthieu and {Borrow}, Josh and {Draper}, Peter W. and {Ivkovic}, Mladen and {McAlpine}, Stuart and {Vandenbroucke}, Bert and {Bah{\'e}}, Yannick and {Chaikin}, Evgenii and {Chalk}, Aidan B.~G. and {Chan}, Tsang Keung and {Correa}, Camila and {van Daalen}, Marcel and {Elbers}, Willem and {Gonnet}, Pedro and {Hausammann}, Lo{\"\i}c and {Helly}, John and {Hu{\v{s}}ko}, Filip and {Kegerreis}, Jacob A. and {Nobels}, Folkert S.~J. and {Ploeckinger}, Sylvia and {Revaz}, Yves and {Roper}, William J. and {Ruiz-Bonilla}, Sergio and {Sandnes}, Thomas D. and {Uyttenhove}, Yolan and {Willis}, James S. and {Xiang}, Zhen},
        title = "{SWIFT: A modern highly-parallel gravity and smoothed particle hydrodynamics solver for astrophysical and cosmological applications}",
      journal = {\mnras},
         year = 2024,
        month = may,
       volume = {530},
       number = {2},
        pages = {2378-2419},
          doi = {10.1093/mnras/stae922},
archivePrefix = {arXiv},
       eprint = {2305.13380},
 primaryClass = {astro-ph.IM},
       adsurl = {https://ui.adsabs.harvard.edu/abs/2024MNRAS.530.2378S}
}

@ARTICLE{Abbott2022,
       author = {{Abbott}, T.~M.~C. and {Aguena}, M. and {Alarcon}, A. and {Allam}, S. and {Alves}, O. and {Amon}, A. and {Andrade-Oliveira}, F. and {Annis}, J. and {Avila}, S. and {Bacon}, D. and {Baxter}, E. and {Bechtol}, K. and {Becker}, M.~R. and {Bernstein}, G.~M. and {Bhargava}, S. and {Birrer}, S. and {Blazek}, J. and {Brandao-Souza}, A. and {Bridle}, S.~L. and {Brooks}, D. and {Buckley-Geer}, E. and {Burke}, D.~L. and {Camacho}, H. and {Campos}, A. and {Carnero Rosell}, A. and {Carrasco Kind}, M. and {Carretero}, J. and {Castander}, F.~J. and {Cawthon}, R. and {Chang}, C. and {Chen}, A. and {Chen}, R. and {Choi}, A. and {Conselice}, C. and {Cordero}, J. and {Costanzi}, M. and {Crocce}, M. and {da Costa}, L.~N. and {da Silva Pereira}, M.~E. and {Davis}, C. and {Davis}, T.~M. and {De Vicente}, J. and {DeRose}, J. and {Desai}, S. and {Di Valentino}, E. and {Diehl}, H.~T. and {Dietrich}, J.~P. and {Dodelson}, S. and {Doel}, P. and {Doux}, C. and {Drlica-Wagner}, A. and {Eckert}, K. and {Eifler}, T.~F. and {Elsner}, F. and {Elvin-Poole}, J. and {Everett}, S. and {Evrard}, A.~E. and {Fang}, X. and {Farahi}, A. and {Fernandez}, E. and {Ferrero}, I. and {Fert{\'e}}, A. and {Fosalba}, P. and {Friedrich}, O. and {Frieman}, J. and {Garc{\'\i}a-Bellido}, J. and {Gatti}, M. and {Gaztanaga}, E. and {Gerdes}, D.~W. and {Giannantonio}, T. and {Giannini}, G. and {Gruen}, D. and {Gruendl}, R.~A. and {Gschwend}, J. and {Gutierrez}, G. and {Harrison}, I. and {Hartley}, W.~G. and {Herner}, K. and {Hinton}, S.~R. and {Hollowood}, D.~L. and {Honscheid}, K. and {Hoyle}, B. and {Huff}, E.~M. and {Huterer}, D. and {Jain}, B. and {James}, D.~J. and {Jarvis}, M. and {Jeffrey}, N. and {Jeltema}, T. and {Kovacs}, A. and {Krause}, E. and {Kron}, R. and {Kuehn}, K. and {Kuropatkin}, N. and {Lahav}, O. and {Leget}, P. -F. and {Lemos}, P. and {Liddle}, A.~R. and {Lidman}, C. and {Lima}, M. and {Lin}, H. and {MacCrann}, N. and {Maia}, M.~A.~G. and {Marshall}, J.~L. and {Martini}, P. and {McCullough}, J. and {Melchior}, P. and {Mena-Fern{\'a}ndez}, J. and {Menanteau}, F. and {Miquel}, R. and {Mohr}, J.~J. and {Morgan}, R. and {Muir}, J. and {Myles}, J. and {Nadathur}, S. and {Navarro-Alsina}, A. and {Nichol}, R.~C. and {Ogando}, R.~L.~C. and {Omori}, Y. and {Palmese}, A. and {Pandey}, S. and {Park}, Y. and {Paz-Chinch{\'o}n}, F. and {Petravick}, D. and {Pieres}, A. and {Plazas Malag{\'o}n}, A.~A. and {Porredon}, A. and {Prat}, J. and {Raveri}, M. and {Rodriguez-Monroy}, M. and {Rollins}, R.~P. and {Romer}, A.~K. and {Roodman}, A. and {Rosenfeld}, R. and {Ross}, A.~J. and {Rykoff}, E.~S. and {Samuroff}, S. and {S{\'a}nchez}, C. and {Sanchez}, E. and {Sanchez}, J. and {Sanchez Cid}, D. and {Scarpine}, V. and {Schubnell}, M. and {Scolnic}, D. and {Secco}, L.~F. and {Serrano}, S. and {Sevilla-Noarbe}, I. and {Sheldon}, E. and {Shin}, T. and {Smith}, M. and {Soares-Santos}, M. and {Suchyta}, E. and {Swanson}, M.~E.~C. and {Tabbutt}, M. and {Tarle}, G. and {Thomas}, D. and {To}, C. and {Troja}, A. and {Troxel}, M.~A. and {Tucker}, D.~L. and {Tutusaus}, I. and {Varga}, T.~N. and {Walker}, A.~R. and {Weaverdyck}, N. and {Wechsler}, R. and {Weller}, J. and {Yanny}, B. and {Yin}, B. and {Zhang}, Y. and {Zuntz}, J. and {DES Collaboration}},
        title = "{Dark Energy Survey Year 3 results: Cosmological constraints from galaxy clustering and weak lensing}",
      journal = {\prd},
         year = 2022,
        month = jan,
       volume = {105},
       number = {2},
          eid = {023520},
        pages = {023520},
          doi = {10.1103/PhysRevD.105.023520},
archivePrefix = {arXiv},
       eprint = {2105.13549},
 primaryClass = {astro-ph.CO},
       adsurl = {https://ui.adsabs.harvard.edu/abs/2022PhRvD.105b3520A}
}

@ARTICLE{Power2003,
       author = {{Power}, C. and {Navarro}, J.~F. and {Jenkins}, A. and {Frenk}, C.~S. and {White}, S.~D.~M. and {Springel}, V. and {Stadel}, J. and {Quinn}, T.},
        title = "{The inner structure of {\ensuremath{\Lambda}}CDM haloes - I. A numerical convergence study}",
      journal = {\mnras},
         year = 2003,
        month = jan,
       volume = {338},
       number = {1},
        pages = {14-34},
          doi = {10.1046/j.1365-8711.2003.05925.x},
archivePrefix = {arXiv},
       eprint = {astro-ph/0201544},
 primaryClass = {astro-ph},
       adsurl = {https://ui.adsabs.harvard.edu/abs/2003MNRAS.338...14P}
}

@ARTICLE{Schaye2023,
       author = {{Schaye}, Joop and {Kugel}, Roi and {Schaller}, Matthieu and {Helly}, John C. and {Braspenning}, Joey and {Elbers}, Willem and {McCarthy}, Ian G. and {van Daalen}, Marcel P. and {Vandenbroucke}, Bert and {Frenk}, Carlos S. and {Kwan}, Juliana and {Salcido}, Jaime and {Bah{\'e}}, Yannick M. and {Borrow}, Josh and {Chaikin}, Evgenii and {Hahn}, Oliver and {Hu{\v{s}}ko}, Filip and {Jenkins}, Adrian and {Lacey}, Cedric G. and {Nobels}, Folkert S.~J.},
        title = "{The FLAMINGO project: cosmological hydrodynamical simulations for large-scale structure and galaxy cluster surveys}",
      journal = {\mnras},
         year = 2023,
        month = dec,
       volume = {526},
       number = {4},
        pages = {4978-5020},
          doi = {10.1093/mnras/stad2419},
archivePrefix = {arXiv},
       eprint = {2306.04024},
 primaryClass = {astro-ph.CO},
       adsurl = {https://ui.adsabs.harvard.edu/abs/2023MNRAS.526.4978S}
}

@ARTICLE{Richings2014a,
       author = {{Richings}, A.~J. and {Schaye}, J. and {Oppenheimer}, B.~D.},
        title = "{Non-equilibrium chemistry and cooling in the diffuse interstellar medium - I. Optically thin regime}",
      journal = {\mnras},
         year = 2014,
        month = jun,
       volume = {440},
       number = {4},
        pages = {3349-3369},
          doi = {10.1093/mnras/stu525},
archivePrefix = {arXiv},
       eprint = {1401.4719},
 primaryClass = {astro-ph.GA},
       adsurl = {https://ui.adsabs.harvard.edu/abs/2014MNRAS.440.3349R}
}

@ARTICLE{Richings2014b,
       author = {{Richings}, A.~J. and {Schaye}, J. and {Oppenheimer}, B.~D.},
        title = "{Non-equilibrium chemistry and cooling in the diffuse interstellar medium - II. Shielded gas}",
      journal = {\mnras},
         year = 2014,
        month = aug,
       volume = {442},
       number = {3},
        pages = {2780-2796},
          doi = {10.1093/mnras/stu1046},
archivePrefix = {arXiv},
       eprint = {1403.6155},
 primaryClass = {astro-ph.GA},
       adsurl = {https://ui.adsabs.harvard.edu/abs/2014MNRAS.442.2780R}
}

@ARTICLE{Ploeckinger2025,
       author = {{Ploeckinger}, Sylvia and {Richings}, Alexander J. and {Schaye}, Joop and {Trayford}, James W. and {Schaller}, Matthieu and {Chaikin}, Evgenii},
        title = "{HYBRID-CHIMES: a model for radiative cooling and the abundances of ions and molecules in simulations of galaxy formation}",
      journal = {\mnras},
         year = 2025,
        month = oct,
       volume = {543},
       number = {2},
        pages = {891-916},
          doi = {10.1093/mnras/staf1402},
archivePrefix = {arXiv},
       eprint = {2506.15773},
 primaryClass = {astro-ph.GA},
       adsurl = {https://ui.adsabs.harvard.edu/abs/2025MNRAS.543..891P}
}

@ARTICLE{Trayford2025,
       author = {{Trayford}, James W. and {Schaye}, Joop and {Correa}, Camila and {Ploeckinger}, Sylvia and {Richings}, Alexander J. and {Chaikin}, Evgenii and {Schaller}, Matthieu and {Ben{\'\i}tez-Llambay}, Alejandro and {Frenk}, Carlos and {Hu{\v{s}}ko}, Filip},
        title = "{Modelling the evolution and influence of dust in cosmological simulations that include the cold phase of the interstellar medium}",
      journal = {\mnras},
         year = 2025,
        month = nov,
          doi = {10.1093/mnras/staf2040},
archivePrefix = {arXiv},
       eprint = {2505.13056},
 primaryClass = {astro-ph.GA},
       adsurl = {https://ui.adsabs.harvard.edu/abs/2025MNRAS.tmp.1935T}
}

@ARTICLE{Nobels2024,
       author = {{Nobels}, Folkert S.~J. and {Schaye}, Joop and {Schaller}, Matthieu and {Ploeckinger}, Sylvia and {Chaikin}, Evgenii and {Richings}, Alexander J.},
        title = "{Tests of subgrid models for star formation using simulations of isolated disc galaxies}",
      journal = {\mnras},
         year = 2024,
        month = aug,
       volume = {532},
       number = {3},
        pages = {3299-3321},
          doi = {10.1093/mnras/stae1390},
archivePrefix = {arXiv},
       eprint = {2309.13750},
 primaryClass = {astro-ph.GA},
       adsurl = {https://ui.adsabs.harvard.edu/abs/2024MNRAS.532.3299N}
}

@ARTICLE{ForouharMoreno2025,
       author = {{Forouhar Moreno}, Victor J. and {Helly}, John and {McGibbon}, Robert and {Schaye}, Joop and {Schaller}, Matthieu and {Han}, Jiaxin and {Kugel}, Roi and {Bah{\'e}}, Yannick M.},
        title = "{Assessing subhalo finders in cosmological hydrodynamical simulations}",
      journal = {\mnras},
         year = 2025,
        month = oct,
       volume = {543},
       number = {2},
        pages = {1339-1372},
          doi = {10.1093/mnras/staf1478},
archivePrefix = {arXiv},
       eprint = {2502.06932},
 primaryClass = {astro-ph.CO},
       adsurl = {https://ui.adsabs.harvard.edu/abs/2025MNRAS.543.1339F}
}

@ARTICLE{Han2018,
       author = {{Han}, Jiaxin and {Cole}, Shaun and {Frenk}, Carlos S. and {Benitez-Llambay}, Alejandro and {Helly}, John},
        title = "{HBT+: an improved code for finding subhaloes and building merger trees in cosmological simulations}",
      journal = {\mnras},
         year = 2018,
        month = feb,
       volume = {474},
       number = {1},
        pages = {604-617},
          doi = {10.1093/mnras/stx2792},
archivePrefix = {arXiv},
       eprint = {1708.03646},
 primaryClass = {astro-ph.CO},
       adsurl = {https://ui.adsabs.harvard.edu/abs/2018MNRAS.474..604H}
}

@ARTICLE{McGibbon2025,
       author = {{McGibbon}, Robert and {Helly}, John and {Schaye}, Joop and {Schaller}, Matthieu and {Vandenbroucke}, Bert},
        title = "{SOAP: A Python Package for Calculating the Properties of Galaxies and Halos Formed in Cosmological Simulations}",
      journal = {The Journal of Open Source Software},
         year = 2025,
        month = jul,
       volume = {10},
       number = {111},
          eid = {8252},
        pages = {8252},
          doi = {10.21105/joss.08252},
archivePrefix = {arXiv},
       eprint = {2507.22669},
 primaryClass = {astro-ph.IM},
       adsurl = {https://ui.adsabs.harvard.edu/abs/2025JOSS...10.8252M}
}

@ARTICLE{BoothSchaye2009,
       author = {{Booth}, C.~M. and {Schaye}, Joop},
        title = "{Cosmological simulations of the growth of supermassive black holes and feedback from active galactic nuclei: method and tests}",
      journal = {\mnras},
         year = 2009,
        month = sep,
       volume = {398},
       number = {1},
        pages = {53-74},
          doi = {10.1111/j.1365-2966.2009.15043.x},
archivePrefix = {arXiv},
       eprint = {0904.2572},
 primaryClass = {astro-ph.CO},
       adsurl = {https://ui.adsabs.harvard.edu/abs/2009MNRAS.398...53B}
}

@ARTICLE{Schoenmakers1997,
       author = {{Schoenmakers}, R.~H.~M. and {Franx}, M. and {de Zeeuw}, P.~T.},
        title = "{Measuring non-axisymmetry in spiral galaxies}",
      journal = {\mnras},
         year = 1997,
        month = dec,
       volume = {292},
       number = {2},
        pages = {349-364},
          doi = {10.1093/mnras/292.2.349},
archivePrefix = {arXiv},
       eprint = {astro-ph/9707332},
 primaryClass = {astro-ph},
       adsurl = {https://ui.adsabs.harvard.edu/abs/1997MNRAS.292..349S}
}

@ARTICLE{Sellwood2010,
       author = {{Sellwood}, J.~A. and {S{\'a}nchez}, Ricardo Z{\'a}nmar},
        title = "{Quantifying non-circular streaming motions in disc galaxies}",
      journal = {\mnras},
         year = 2010,
        month = jun,
       volume = {404},
       number = {4},
        pages = {1733-1744},
          doi = {10.1111/j.1365-2966.2010.16430.x},
archivePrefix = {arXiv},
       eprint = {0912.5493},
 primaryClass = {astro-ph.CO},
       adsurl = {https://ui.adsabs.harvard.edu/abs/2010MNRAS.404.1733S}
}

@ARTICLE{Hopkins2023,
       author = {{Hopkins}, Philip F. and {Wetzel}, Andrew and {Wheeler}, Coral and {Sanderson}, Robyn and {Grudi{\'c}}, Michael Y. and {Sameie}, Omid and {Boylan-Kolchin}, Michael and {Orr}, Matthew and {Ma}, Xiangcheng and {Faucher-Gigu{\`e}re}, Claude-Andr{\'e} and {Kere{\v{s}}}, Du{\v{s}}an and {Quataert}, Eliot and {Su}, Kung-Yi and {Moreno}, Jorge and {Feldmann}, Robert and {Bullock}, James S. and {Loebman}, Sarah R. and {Angl{\'e}s-Alc{\'a}zar}, Daniel and {Stern}, Jonathan and {Necib}, Lina and {Choban}, Caleb R. and {Hayward}, Christopher C.},
        title = "{FIRE-3: updated stellar evolution models, yields, and microphysics and fitting functions for applications in galaxy simulations}",
      journal = {\mnras},
         year = 2023,
        month = feb,
       volume = {519},
       number = {2},
        pages = {3154-3181},
          doi = {10.1093/mnras/stac3489},
archivePrefix = {arXiv},
       eprint = {2203.00040},
 primaryClass = {astro-ph.GA},
       adsurl = {https://ui.adsabs.harvard.edu/abs/2023MNRAS.519.3154H}
}

@ARTICLE{Borrow2022,
       author = {{Borrow}, Josh and {Schaller}, Matthieu and {Bower}, Richard G. and {Schaye}, Joop},
        title = "{SPHENIX: smoothed particle hydrodynamics for the next generation of galaxy formation simulations}",
      journal = {\mnras},
         year = 2022,
        month = apr,
       volume = {511},
       number = {2},
        pages = {2367-2389},
          doi = {10.1093/mnras/stab3166},
archivePrefix = {arXiv},
       eprint = {2012.03974},
 primaryClass = {astro-ph.GA},
       adsurl = {https://ui.adsabs.harvard.edu/abs/2022MNRAS.511.2367B}
}

@ARTICLE{McGaugh2016,
       author = {{McGaugh}, Stacy S. and {Lelli}, Federico and {Schombert}, James M.},
        title = "{Radial Acceleration Relation in Rotationally Supported Galaxies}",
      journal = {\prl},
         year = 2016,
        month = nov,
       volume = {117},
       number = {20},
          eid = {201101},
        pages = {201101},
          doi = {10.1103/PhysRevLett.117.201101},
archivePrefix = {arXiv},
       eprint = {1609.05917},
 primaryClass = {astro-ph.GA},
       adsurl = {https://ui.adsabs.harvard.edu/abs/2016PhRvL.117t1101M}
}

@ARTICLE{Oman_Martini2024,
       author = {{Oman}, Kyle},
        title = "{MARTINI: Mock Array Radio Telescope Interferometry of the Neutral ISM}",
      journal = {The Journal of Open Source Software},
         year = 2024,
        month = jun,
       volume = {9},
       number = {98},
          eid = {6860},
        pages = {6860},
          doi = {10.21105/joss.06860},
archivePrefix = {arXiv},
       eprint = {2406.05574},
 primaryClass = {astro-ph.GA},
       adsurl = {https://ui.adsabs.harvard.edu/abs/2024JOSS....9.6860O}
}

@ARTICLE{Koudmani2025,
       author = {{Koudmani}, Sophie and {Rennehan}, Douglas and {Somerville}, Rachel S. and {Hayward}, Christopher C. and {Angl{\'e}s-Alc{\'a}zar}, Daniel and {Orr}, Matthew E. and {Sands}, Isabel S. and {Wellons}, Sarah},
        title = "{Diverse dark matter profiles in FIRE dwarfs: black holes, cosmic rays and the cusp{\textendash}core enigma}",
      journal = {\mnras},
         year = 2025,
        month = jun,
       volume = {540},
       number = {2},
        pages = {1928-1950},
          doi = {10.1093/mnras/staf778},
archivePrefix = {arXiv},
       eprint = {2409.02172},
 primaryClass = {astro-ph.GA},
       adsurl = {https://ui.adsabs.harvard.edu/abs/2025MNRAS.540.1928K}
}

@ARTICLE{ManzanoKing2020,
       author = {{Manzano-King}, Christina M. and {Canalizo}, Gabriela},
        title = "{Active galactic nucleus and dwarf galaxy gas kinematics}",
      journal = {\mnras},
         year = 2020,
        month = nov,
       volume = {498},
       number = {3},
        pages = {4562-4576},
          doi = {10.1093/mnras/staa2654},
archivePrefix = {arXiv},
       eprint = {2009.01389},
 primaryClass = {astro-ph.GA},
       adsurl = {https://ui.adsabs.harvard.edu/abs/2020MNRAS.498.4562M}
}

@software{Harborne_SimSpin2019,
       author = {{Harborne}, Katherine},
        title = "{SimSpin: Kinematic analysis of galaxy simulations}",
 howpublished = {Astrophysics Source Code Library, record ascl:1903.006},
         year = 2019,
        month = mar,
          eid = {ascl:1903.006},
       adsurl = {https://ui.adsabs.harvard.edu/abs/2019ascl.soft03006H}
}

@ARTICLE{Dekel1986,
       author = {{Dekel}, A. and {Silk}, J.},
        title = "{The Origin of Dwarf Galaxies, Cold Dark Matter, and Biased Galaxy Formation}",
      journal = {\apj},
         year = 1986,
        month = apr,
       volume = {303},
        pages = {39},
          doi = {10.1086/164050},
       adsurl = {https://ui.adsabs.harvard.edu/abs/1986ApJ...303...39D}
}

@ARTICLE{Stinson2006,
       author = {{Stinson}, Greg and {Seth}, Anil and {Katz}, Neal and {Wadsley}, James and {Governato}, Fabio and {Quinn}, Tom},
        title = "{Star formation and feedback in smoothed particle hydrodynamic simulations - I. Isolated galaxies}",
      journal = {\mnras},
         year = 2006,
        month = dec,
       volume = {373},
       number = {3},
        pages = {1074-1090},
          doi = {10.1111/j.1365-2966.2006.11097.x},
archivePrefix = {arXiv},
       eprint = {astro-ph/0602350},
 primaryClass = {astro-ph},
       adsurl = {https://ui.adsabs.harvard.edu/abs/2006MNRAS.373.1074S}
}

@ARTICLE{Hopkins2012,
       author = {{Hopkins}, Philip F. and {Quataert}, Eliot and {Murray}, Norman},
        title = "{Stellar feedback in galaxies and the origin of galaxy-scale winds}",
      journal = {\mnras},
         year = 2012,
        month = apr,
       volume = {421},
       number = {4},
        pages = {3522-3537},
          doi = {10.1111/j.1365-2966.2012.20593.x},
archivePrefix = {arXiv},
       eprint = {1110.4638},
 primaryClass = {astro-ph.CO},
       adsurl = {https://ui.adsabs.harvard.edu/abs/2012MNRAS.421.3522H}
}

@ARTICLE{KimOstriker2017,
       author = {{Kim}, Chang-Goo and {Ostriker}, Eve C.},
        title = "{Three-phase Interstellar Medium in Galaxies Resolving Evolution with Star Formation and Supernova Feedback (TIGRESS): Algorithms, Fiducial Model, and Convergence}",
      journal = {\apj},
         year = 2017,
        month = sep,
       volume = {846},
       number = {2},
          eid = {133},
        pages = {133},
          doi = {10.3847/1538-4357/aa8599},
archivePrefix = {arXiv},
       eprint = {1612.03918},
 primaryClass = {astro-ph.GA},
       adsurl = {https://ui.adsabs.harvard.edu/abs/2017ApJ...846..133K}
}

@ARTICLE{Martizzi2015,
       author = {{Martizzi}, Davide and {Faucher-Gigu{\`e}re}, Claude-Andr{\'e} and {Quataert}, Eliot},
        title = "{Supernova feedback in an inhomogeneous interstellar medium}",
      journal = {\mnras},
         year = 2015,
        month = jun,
       volume = {450},
       number = {1},
        pages = {504-522},
          doi = {10.1093/mnras/stv562},
archivePrefix = {arXiv},
       eprint = {1409.4425},
 primaryClass = {astro-ph.GA},
       adsurl = {https://ui.adsabs.harvard.edu/abs/2015MNRAS.450..504M}
}

@ARTICLE{Orr2020,
       author = {{Orr}, Matthew E. and {Hayward}, Christopher C. and {Medling}, Anne M. and {Gurvich}, Alexander B. and {Hopkins}, Philip F. and {Murray}, Norman and {Pineda}, Jorge L. and {Faucher-Gigu{\`e}re}, Claude-Andr{\'e} and {Kere{\v{s}}}, Du{\v{s}}an and {Wetzel}, Andrew and {Su}, Kung-Yi},
        title = "{Swirls of FIRE: spatially resolved gas velocity dispersions and star formation rates in FIRE-2 disc environments}",
      journal = {\mnras},
         year = 2020,
        month = aug,
       volume = {496},
       number = {2},
        pages = {1620-1637},
          doi = {10.1093/mnras/staa1619},
archivePrefix = {arXiv},
       eprint = {1911.00020},
 primaryClass = {astro-ph.GA},
       adsurl = {https://ui.adsabs.harvard.edu/abs/2020MNRAS.496.1620O}
}

@ARTICLE{Stilp2013,
       author = {{Stilp}, Adrienne M. and {Dalcanton}, Julianne J. and {Skillman}, Evan and {Warren}, Steven R. and {Ott}, J{\"u}rgen and {Koribalski}, B{\"a}rbel},
        title = "{Drivers of H I Turbulence in Dwarf Galaxies}",
      journal = {\apj},
         year = 2013,
        month = aug,
       volume = {773},
       number = {2},
          eid = {88},
        pages = {88},
          doi = {10.1088/0004-637X/773/2/88},
archivePrefix = {arXiv},
       eprint = {1306.2321},
 primaryClass = {astro-ph.GA},
       adsurl = {https://ui.adsabs.harvard.edu/abs/2013ApJ...773...88S}
}

@ARTICLE{Boomsma2008,
       author = {{Boomsma}, R. and {Oosterloo}, T.~A. and {Fraternali}, F. and {van der Hulst}, J.~M. and {Sancisi}, R.},
        title = "{HI holes and high-velocity clouds in the spiral galaxy NGC 6946}",
      journal = {\aap},
         year = 2008,
        month = nov,
       volume = {490},
       number = {2},
        pages = {555-570},
          doi = {10.1051/0004-6361:200810120},
archivePrefix = {arXiv},
       eprint = {0807.3339},
 primaryClass = {astro-ph},
       adsurl = {https://ui.adsabs.harvard.edu/abs/2008A&A...490..555B}
}

@ARTICLE{Bacchini2020,
       author = {{Bacchini}, Cecilia and {Fraternali}, Filippo and {Iorio}, Giuliano and {Pezzulli}, Gabriele and {Marasco}, Antonino and {Nipoti}, Carlo},
        title = "{Evidence for supernova feedback sustaining gas turbulence in nearby star-forming galaxies}",
      journal = {\aap},
         year = 2020,
        month = sep,
       volume = {641},
          eid = {A70},
        pages = {A70},
          doi = {10.1051/0004-6361/202038223},
archivePrefix = {arXiv},
       eprint = {2006.10764},
 primaryClass = {astro-ph.GA},
       adsurl = {https://ui.adsabs.harvard.edu/abs/2020A&A...641A..70B}
}

@ARTICLE{Noguchi1999,
       author = {{Noguchi}, Masafumi},
        title = "{Early Evolution of Disk Galaxies: Formation of Bulges in Clumpy Young Galactic Disks}",
      journal = {\apj},
         year = 1999,
        month = mar,
       volume = {514},
       number = {1},
        pages = {77-95},
          doi = {10.1086/306932},
archivePrefix = {arXiv},
       eprint = {astro-ph/9806355},
 primaryClass = {astro-ph},
       adsurl = {https://ui.adsabs.harvard.edu/abs/1999ApJ...514...77N}
}

@ARTICLE{Dekel2009,
       author = {{Dekel}, Avishai and {Sari}, Re'em and {Ceverino}, Daniel},
        title = "{Formation of Massive Galaxies at High Redshift: Cold Streams, Clumpy Disks, and Compact Spheroids}",
      journal = {\apj},
         year = 2009,
        month = sep,
       volume = {703},
       number = {1},
        pages = {785-801},
          doi = {10.1088/0004-637X/703/1/785},
archivePrefix = {arXiv},
       eprint = {0901.2458},
 primaryClass = {astro-ph.GA},
       adsurl = {https://ui.adsabs.harvard.edu/abs/2009ApJ...703..785D}
}

@ARTICLE{Genzel2011,
       author = {{Genzel}, R. and {Newman}, S. and {Jones}, T. and {F{\"o}rster Schreiber}, N.~M. and {Shapiro}, K. and {Genel}, S. and {Lilly}, S.~J. and {Renzini}, A. and {Tacconi}, L.~J. and {Bouch{\'e}}, N. and {Burkert}, A. and {Cresci}, G. and {Buschkamp}, P. and {Carollo}, C.~M. and {Ceverino}, D. and {Davies}, R. and {Dekel}, A. and {Eisenhauer}, F. and {Hicks}, E. and {Kurk}, J. and {Lutz}, D. and {Mancini}, C. and {Naab}, T. and {Peng}, Y. and {Sternberg}, A. and {Vergani}, D. and {Zamorani}, G.},
        title = "{The Sins Survey of z \raisebox{-0.5ex}\textasciitilde 2 Galaxy Kinematics: Properties of the Giant Star-forming Clumps}",
      journal = {\apj},
         year = 2011,
        month = jun,
       volume = {733},
       number = {2},
          eid = {101},
        pages = {101},
          doi = {10.1088/0004-637X/733/2/101},
archivePrefix = {arXiv},
       eprint = {1011.5360},
 primaryClass = {astro-ph.CO},
       adsurl = {https://ui.adsabs.harvard.edu/abs/2011ApJ...733..101G}
}

@ARTICLE{Goldbaum2015,
       author = {{Goldbaum}, Nathan J. and {Krumholz}, Mark R. and {Forbes}, John C.},
        title = "{Mass Transport and Turbulence in Gravitationally Unstable Disk Galaxies. I. The Case of Pure Self-gravity}",
      journal = {\apj},
         year = 2015,
        month = dec,
       volume = {814},
       number = {2},
          eid = {131},
        pages = {131},
          doi = {10.1088/0004-637X/814/2/131},
archivePrefix = {arXiv},
       eprint = {1510.08458},
 primaryClass = {astro-ph.GA},
       adsurl = {https://ui.adsabs.harvard.edu/abs/2015ApJ...814..131G}
}

@ARTICLE{Elmegreen2007,
       author = {{Elmegreen}, Debra Meloy and {Elmegreen}, Bruce G. and {Ravindranath}, Swara and {Coe}, Daniel A.},
        title = "{Resolved Galaxies in the Hubble Ultra Deep Field: Star Formation in Disks at High Redshift}",
      journal = {\apj},
         year = 2007,
        month = apr,
       volume = {658},
       number = {2},
        pages = {763-777},
          doi = {10.1086/511667},
archivePrefix = {arXiv},
       eprint = {astro-ph/0701121},
 primaryClass = {astro-ph},
       adsurl = {https://ui.adsabs.harvard.edu/abs/2007ApJ...658..763E}
}

@ARTICLE{Rix1995,
       author = {{Rix}, Hans-Walter and {Zaritsky}, Dennis},
        title = "{Nonaxisymmetric Structures in the Stellar Disks of Galaxies}",
      journal = {\apj},
         year = 1995,
        month = jul,
       volume = {447},
        pages = {82},
          doi = {10.1086/175858},
archivePrefix = {arXiv},
       eprint = {astro-ph/9505111},
 primaryClass = {astro-ph},
       adsurl = {https://ui.adsabs.harvard.edu/abs/1995ApJ...447...82R}
}

@ARTICLE{Zaritsky2013,
       author = {{Zaritsky}, Dennis and {Salo}, Heikki and {Laurikainen}, Eija and {Elmegreen}, Debra and {Athanassoula}, E. and {Bosma}, Albert and {Comer{\'o}n}, S{\'e}bastien and {Erroz-Ferrer}, Santiago and {Elmegreen}, Bruce and {Gadotti}, Dimitri A. and {Gil de Paz}, Armando and {Hinz}, Joannah L. and {Ho}, Luis C. and {Holwerda}, Benne W. and {Kim}, Taehyun and {Knapen}, Johan H. and {Laine}, Jarkko and {Laine}, Seppo and {Madore}, Barry F. and {Meidt}, Sharon and {Menendez-Delmestre}, Karin and {Mizusawa}, Trisha and {Mu{\~n}oz-Mateos}, Juan Carlos and {Regan}, Michael W. and {Seibert}, Mark and {Sheth}, Kartik},
        title = "{On the Origin of Lopsidedness in Galaxies as Determined from the Spitzer Survey of Stellar Structure in Galaxies (S$^{4}$G)}",
      journal = {\apj},
         year = 2013,
        month = aug,
       volume = {772},
       number = {2},
          eid = {135},
        pages = {135},
          doi = {10.1088/0004-637X/772/2/135},
archivePrefix = {arXiv},
       eprint = {1305.2940},
 primaryClass = {astro-ph.CO},
       adsurl = {https://ui.adsabs.harvard.edu/abs/2013ApJ...772..135Z}
}

@ARTICLE{Kazantzidis2008,
       author = {{Kazantzidis}, Stelios and {Bullock}, James S. and {Zentner}, Andrew R. and {Kravtsov}, Andrey V. and {Moustakas}, Leonidas A.},
        title = "{Cold Dark Matter Substructure and Galactic Disks. I. Morphological Signatures of Hierarchical Satellite Accretion}",
      journal = {\apj},
         year = 2008,
        month = nov,
       volume = {688},
       number = {1},
        pages = {254-276},
          doi = {10.1086/591958},
archivePrefix = {arXiv},
       eprint = {0708.1949},
 primaryClass = {astro-ph},
       adsurl = {https://ui.adsabs.harvard.edu/abs/2008ApJ...688..254K}
}

@ARTICLE{Peschken2020,
       author = {{Peschken}, Nicolas and {{\L}okas}, Ewa L. and {Athanassoula}, E.},
        title = "{Disc galaxies formed from major mergers in Illustris}",
      journal = {\mnras},
         year = 2020,
        month = mar,
       volume = {493},
       number = {1},
        pages = {1375-1387},
          doi = {10.1093/mnras/staa299},
archivePrefix = {arXiv},
       eprint = {1909.01033},
 primaryClass = {astro-ph.GA},
       adsurl = {https://ui.adsabs.harvard.edu/abs/2020MNRAS.493.1375P}
}

@ARTICLE{Oh2015,
       author = {{Oh}, Se-Heon and {Hunter}, Deidre A. and {Brinks}, Elias and {Elmegreen}, Bruce G. and {Schruba}, Andreas and {Walter}, Fabian and {Rupen}, Michael P. and {Young}, Lisa M. and {Simpson}, Caroline E. and {Johnson}, Megan C. and {Herrmann}, Kimberly A. and {Ficut-Vicas}, Dana and {Cigan}, Phil and {Heesen}, Volker and {Ashley}, Trisha and {Zhang}, Hong-Xin},
        title = "{High-resolution Mass Models of Dwarf Galaxies from LITTLE THINGS}",
      journal = {\aj},
         year = 2015,
        month = jun,
       volume = {149},
       number = {6},
          eid = {180},
        pages = {180},
          doi = {10.1088/0004-6256/149/6/180},
archivePrefix = {arXiv},
       eprint = {1502.01281},
 primaryClass = {astro-ph.GA},
       adsurl = {https://ui.adsabs.harvard.edu/abs/2015AJ....149..180O}
}

@ARTICLE{Schaller2015,
       author = {{Schaller}, Matthieu and {Frenk}, Carlos S. and {Bower}, Richard G. and {Theuns}, Tom and {Jenkins}, Adrian and {Schaye}, Joop and {Crain}, Robert A. and {Furlong}, Michelle and {Dalla Vecchia}, Claudio and {McCarthy}, I.~G.},
        title = "{Baryon effects on the internal structure of {\ensuremath{\Lambda}}CDM haloes in the EAGLE simulations}",
      journal = {\mnras},
         year = 2015,
        month = aug,
       volume = {451},
       number = {2},
        pages = {1247-1267},
          doi = {10.1093/mnras/stv1067},
archivePrefix = {arXiv},
       eprint = {1409.8617},
 primaryClass = {astro-ph.CO},
       adsurl = {https://ui.adsabs.harvard.edu/abs/2015MNRAS.451.1247S}
}

@ARTICLE{Llambay2019,
       author = {{Ben{\'\i}tez-Llambay}, Alejandro and {Frenk}, Carlos S. and {Ludlow}, Aaron D. and {Navarro}, Julio F.},
        title = "{Baryon-induced dark matter cores in the EAGLE simulations}",
      journal = {\mnras},
         year = 2019,
        month = sep,
       volume = {488},
       number = {2},
        pages = {2387-2404},
          doi = {10.1093/mnras/stz1890},
archivePrefix = {arXiv},
       eprint = {1810.04186},
 primaryClass = {astro-ph.GA},
       adsurl = {https://ui.adsabs.harvard.edu/abs/2019MNRAS.488.2387B}
}

@ARTICLE{Chua2019,
       author = {{Chua}, Kun Ting Eddie and {Pillepich}, Annalisa and {Vogelsberger}, Mark and {Hernquist}, Lars},
        title = "{Shape of dark matter haloes in the Illustris simulation: effects of baryons}",
      journal = {\mnras},
         year = 2019,
        month = mar,
       volume = {484},
       number = {1},
        pages = {476-493},
          doi = {10.1093/mnras/sty3531},
archivePrefix = {arXiv},
       eprint = {1809.07255},
 primaryClass = {astro-ph.GA},
       adsurl = {https://ui.adsabs.harvard.edu/abs/2019MNRAS.484..476C}
}

@ARTICLE{Chemin2020,
       author = {{Chemin}, L. and {Amram}, P. and {Carignan}, C. and {Balkowski}, C. and {{\'E}pinat}, B.},
        title = "{Kinematic asymmetries in dark matter dominated galaxies}",
      journal = {Boletin de la Asociacion Argentina de Astronomia La Plata Argentina},
         year = 2020,
        month = aug,
       volume = {61C},
        pages = {45-45},
       adsurl = {https://ui.adsabs.harvard.edu/abs/2020BAAA...61C..45C}
}

@ARTICLE{Chemin2016,
       author = {{Chemin}, Laurent and {Hur{\'e}}, Jean-Marc and {Soubiran}, Caroline and {Zibetti}, Stefano and {Charlot}, St{\'e}phane and {Kawata}, Daisuke},
        title = "{Asymmetric mass models of disk galaxies. I. Messier 99}",
      journal = {\aap},
         year = 2016,
        month = apr,
       volume = {588},
          eid = {A48},
        pages = {A48},
          doi = {10.1051/0004-6361/201527730},
archivePrefix = {arXiv},
       eprint = {1601.01601},
 primaryClass = {astro-ph.GA},
       adsurl = {https://ui.adsabs.harvard.edu/abs/2016A&A...588A..48C}
}

@ARTICLE{Brinks2003,
       author = {{Brinks}, Elias and {Walter}, Fabian and {Kerp}, J{\"u}rgen},
        title = "{X-ray emission from dwarf galaxies: IC 2574 revisited}",
      journal = {\apss},
         year = 2003,
        month = apr,
       volume = {284},
       number = {2},
        pages = {627-630},
          doi = {10.1023/A:1024026610546},
       adsurl = {https://ui.adsabs.harvard.edu/abs/2003Ap&SS.284..627B}
}

@ARTICLE{Banerjee2011,
       author = {{Banerjee}, Arunima and {Jog}, Chanda J. and {Brinks}, Elias and {Bagetakos}, Ioannis},
        title = "{Theoretical determination of H I vertical scale heights in the dwarf galaxies DDO 154, Ho II, IC 2574 and NGC 2366}",
      journal = {\mnras},
         year = 2011,
        month = jul,
       volume = {415},
       number = {1},
        pages = {687-694},
          doi = {10.1111/j.1365-2966.2011.18745.x},
archivePrefix = {arXiv},
       eprint = {1103.4494},
 primaryClass = {astro-ph.CO},
       adsurl = {https://ui.adsabs.harvard.edu/abs/2011MNRAS.415..687B}
}

@ARTICLE{Albers2019,
       author = {{Albers}, Saundra M. and {Weisz}, Daniel R. and {Cole}, Andrew A. and {Dolphin}, Andrew E. and {Skillman}, Evan D. and {Williams}, Benjamin F. and {Boylan-Kolchin}, Michael and {Bullock}, James S. and {Dalcanton}, Julianne J. and {Hopkins}, Philip F. and {Leaman}, Ryan and {McConnachie}, Alan W. and {Vogelsberger}, Mark and {Wetzel}, Andrew},
        title = "{Star formation at the edge of the Local Group: a rising star formation history in the isolated galaxy WLM}",
      journal = {\mnras},
         year = 2019,
        month = dec,
       volume = {490},
       number = {4},
        pages = {5538-5550},
          doi = {10.1093/mnras/stz2903},
archivePrefix = {arXiv},
       eprint = {1909.04040},
 primaryClass = {astro-ph.GA},
       adsurl = {https://ui.adsabs.harvard.edu/abs/2019MNRAS.490.5538A}
}

@ARTICLE{Mondal2021,
       author = {{Mondal}, Chayan and {Subramaniam}, Annapurni and {George}, Koshy},
        title = "{A tale of two nearby dwarf irregular galaxies WLM and IC 2574: As revealed by UVIT}",
      journal = {Journal of Astrophysics and Astronomy},
         year = 2021,
        month = oct,
       volume = {42},
       number = {2},
          eid = {50},
        pages = {50},
          doi = {10.1007/s12036-021-09761-z},
archivePrefix = {arXiv},
       eprint = {2105.13048},
 primaryClass = {astro-ph.GA},
       adsurl = {https://ui.adsabs.harvard.edu/abs/2021JApA...42...50M}
}

@ARTICLE{Dutton2019,
       author = {{Dutton}, Aaron A. and {Obreja}, Aura and {Macci{\`o}}, Andrea V.},
        title = "{NIHAO - XVII. The diversity of dwarf galaxy kinematics and implications for the H I velocity function}",
      journal = {\mnras},
         year = 2019,
        month = feb,
       volume = {482},
       number = {4},
        pages = {5606-5624},
          doi = {10.1093/mnras/sty3064},
archivePrefix = {arXiv},
       eprint = {1807.10518},
 primaryClass = {astro-ph.GA},
       adsurl = {https://ui.adsabs.harvard.edu/abs/2019MNRAS.482.5606D}
}

@ARTICLE{Koribalski2004,
       author = {{Koribalski}, B.~S. and {Staveley-Smith}, L. and {Kilborn}, V.~A. and {Ryder}, S.~D. and {Kraan-Korteweg}, R.~C. and {Ryan-Weber}, E.~V. and {Ekers}, R.~D. and {Jerjen}, H. and {Henning}, P.~A. and {Putman}, M.~E. and {Zwaan}, M.~A. and {de Blok}, W.~J.~G. and {Calabretta}, M.~R. and {Disney}, M.~J. and {Minchin}, R.~F. and {Bhathal}, R. and {Boyce}, P.~J. and {Drinkwater}, M.~J. and {Freeman}, K.~C. and {Gibson}, B.~K. and {Green}, A.~J. and {Haynes}, R.~F. and {Juraszek}, S. and {Kesteven}, M.~J. and {Knezek}, P.~M. and {Mader}, S. and {Marquarding}, M. and {Meyer}, M. and {Mould}, J.~R. and {Oosterloo}, T. and {O'Brien}, J. and {Price}, R.~M. and {Sadler}, E.~M. and {Schr{\"o}der}, A. and {Stewart}, I.~M. and {Stootman}, F. and {Waugh}, M. and {Warren}, B.~E. and {Webster}, R.~L. and {Wright}, A.~E.},
        title = "{The 1000 Brightest HIPASS Galaxies: H I Properties}",
      journal = {\aj},
         year = 2004,
        month = jul,
       volume = {128},
       number = {1},
        pages = {16-46},
          doi = {10.1086/421744},
archivePrefix = {arXiv},
       eprint = {astro-ph/0404436},
 primaryClass = {astro-ph},
       adsurl = {https://ui.adsabs.harvard.edu/abs/2004AJ....128...16K}
}

@ARTICLE{Haynes2018,
       author = {{Haynes}, Martha P. and {Giovanelli}, Riccardo and {Kent}, Brian R. and {Adams}, Elizabeth A.~K. and {Balonek}, Thomas J. and {Craig}, David W. and {Fertig}, Derek and {Finn}, Rose and {Giovanardi}, Carlo and {Hallenbeck}, Gregory and {Hess}, Kelley M. and {Hoffman}, G. Lyle and {Huang}, Shan and {Jones}, Michael G. and {Koopmann}, Rebecca A. and {Kornreich}, David A. and {Leisman}, Lukas and {Miller}, Jeffrey and {Moorman}, Crystal and {O'Connor}, Jessica and {O'Donoghue}, Aileen and {Papastergis}, Emmanouil and {Troischt}, Parker and {Stark}, David and {Xiao}, Li},
        title = "{The Arecibo Legacy Fast ALFA Survey: The ALFALFA Extragalactic H I Source Catalog}",
      journal = {\apj},
         year = 2018,
        month = jul,
       volume = {861},
       number = {1},
          eid = {49},
        pages = {49},
          doi = {10.3847/1538-4357/aac956},
archivePrefix = {arXiv},
       eprint = {1805.11499},
 primaryClass = {astro-ph.GA},
       adsurl = {https://ui.adsabs.harvard.edu/abs/2018ApJ...861...49H}
}

@ARTICLE{Correa2015c,
       author = {{Correa}, Camila A. and {Wyithe}, J. Stuart B. and {Schaye}, Joop and {Duffy}, Alan R.},
        title = "{The accretion history of dark matter haloes - III. A physical model for the concentration-mass relation}",
      journal = {\mnras},
         year = 2015,
        month = sep,
       volume = {452},
       number = {2},
        pages = {1217-1232},
          doi = {10.1093/mnras/stv1363},
archivePrefix = {arXiv},
       eprint = {1502.00391},
 primaryClass = {astro-ph.CO},
       adsurl = {https://ui.adsabs.harvard.edu/abs/2015MNRAS.452.1217C}
}

@ARTICLE{DiemerJoyce2019,
       author = {{Diemer}, Benedikt and {Joyce}, Michael},
        title = "{An Accurate Physical Model for Halo Concentrations}",
      journal = {\apj},
         year = 2019,
        month = feb,
       volume = {871},
       number = {2},
          eid = {168},
        pages = {168},
          doi = {10.3847/1538-4357/aafad6},
archivePrefix = {arXiv},
       eprint = {1809.07326},
 primaryClass = {astro-ph.CO},
       adsurl = {https://ui.adsabs.harvard.edu/abs/2019ApJ...871..168D}
}

@ARTICLE{Correa2025,
       author = {{Correa}, Camila A. and {Schaller}, Matthieu and {Schaye}, Joop and {Ploeckinger}, Sylvia and {Borrow}, Josh and {Bah{\'e}}, Yannick},
        title = "{TangoSIDM Project: is the stellar mass Tully-Fisher relation consistent with SIDM?}",
      journal = {\mnras},
         year = 2025,
        month = feb,
       volume = {536},
       number = {4},
        pages = {3338-3356},
          doi = {10.1093/mnras/stae2811},
archivePrefix = {arXiv},
       eprint = {2403.09186},
 primaryClass = {astro-ph.CO},
       adsurl = {https://ui.adsabs.harvard.edu/abs/2025MNRAS.536.3338C}
}

@ARTICLE{Burger2022,
       author = {{Burger}, Jan D. and {Zavala}, Jes{\'u}s and {Sales}, Laura V. and {Vogelsberger}, Mark and {Marinacci}, Federico and {Torrey}, Paul},
        title = "{Degeneracies between self-interacting dark matter and supernova feedback as cusp-core transformation mechanisms}",
      journal = {\mnras},
         year = 2022,
        month = jul,
       volume = {513},
       number = {3},
        pages = {3458-3481},
          doi = {10.1093/mnras/stac994},
archivePrefix = {arXiv},
       eprint = {2108.07358},
 primaryClass = {astro-ph.GA},
       adsurl = {https://ui.adsabs.harvard.edu/abs/2022MNRAS.513.3458B}
}

@ARTICLE{Chaikin2023,
       author = {{Chaikin}, Evgenii and {Schaye}, Joop and {Schaller}, Matthieu and {Ben{\'\i}tez-Llambay}, Alejandro and {Nobels}, Folkert S.~J. and {Ploeckinger}, Sylvia},
        title = "{A thermal-kinetic subgrid model for supernova feedback in simulations of galaxy formation}",
      journal = {\mnras},
         year = 2023,
        month = aug,
       volume = {523},
       number = {3},
        pages = {3709-3731},
          doi = {10.1093/mnras/stad1626},
archivePrefix = {arXiv},
       eprint = {2211.04619},
 primaryClass = {astro-ph.GA},
       adsurl = {https://ui.adsabs.harvard.edu/abs/2023MNRAS.523.3709C}
}

@ARTICLE{Llambay2025,
       author = {{Ben{\'\i}tez-Llambay}, Alejandro and {Ploeckinger}, Sylvia and {Schaye}, Joop and {Richings}, Alexander J. and {Chaikin}, Evgenii and {Schaller}, Matthieu and {Trayford}, James W. and {Frenk}, Carlos S. and {Hu{\v{s}}ko}, Filip and {Correa}, Camila},
        title = "{Non-explosive pre-supernova feedback in the COLIBRE model of galaxy formation}",
      journal = {arXiv e-prints},
         year = 2025,
        month = sep,
          eid = {arXiv:2509.25309},
        pages = {arXiv:2509.25309},
          doi = {10.48550/arXiv.2509.25309},
archivePrefix = {arXiv},
       eprint = {2509.25309},
 primaryClass = {astro-ph.GA},
       adsurl = {https://ui.adsabs.harvard.edu/abs/2025arXiv250925309B}
}

@ARTICLE{DallaVecchia2012,
       author = {{Dalla Vecchia}, Claudio and {Schaye}, Joop},
        title = "{Simulating galactic outflows with thermal supernova feedback}",
      journal = {\mnras},
         year = 2012,
        month = oct,
       volume = {426},
       number = {1},
        pages = {140-158},
          doi = {10.1111/j.1365-2966.2012.21704.x},
archivePrefix = {arXiv},
       eprint = {1203.5667},
 primaryClass = {astro-ph.GA},
       adsurl = {https://ui.adsabs.harvard.edu/abs/2012MNRAS.426..140D}
}

@ARTICLE{Ludlow2019a,
       author = {{Ludlow}, Aaron D. and {Schaye}, Joop and {Schaller}, Matthieu and {Richings}, Jack},
        title = "{Energy equipartition between stellar and dark matter particles in cosmological simulations results in spurious growth of galaxy sizes}",
      journal = {\mnras},
         year = 2019,
        month = sep,
       volume = {488},
       number = {1},
        pages = {L123-L128},
          doi = {10.1093/mnrasl/slz110},
archivePrefix = {arXiv},
       eprint = {1903.10110},
 primaryClass = {astro-ph.GA},
       adsurl = {https://ui.adsabs.harvard.edu/abs/2019MNRAS.488L.123L}
}

@ARTICLE{Ludlow2021,
       author = {{Ludlow}, Aaron D. and {Fall}, S. Michael and {Schaye}, Joop and {Obreschkow}, Danail},
        title = "{Spurious heating of stellar motions in simulated galactic discs by dark matter halo particles}",
      journal = {\mnras},
         year = 2021,
        month = dec,
       volume = {508},
       number = {4},
        pages = {5114-5137},
          doi = {10.1093/mnras/stab2770},
archivePrefix = {arXiv},
       eprint = {2105.03561},
 primaryClass = {astro-ph.GA},
       adsurl = {https://ui.adsabs.harvard.edu/abs/2021MNRAS.508.5114L}
}

@ARTICLE{Ludlow2023,
       author = {{Ludlow}, Aaron D. and {Fall}, S. Michael and {Wilkinson}, Matthew J. and {Schaye}, Joop and {Obreschkow}, Danail},
        title = "{Spurious heating of stellar motions by dark matter particles in cosmological simulations of galaxy formation}",
      journal = {\mnras},
         year = 2023,
        month = nov,
       volume = {525},
       number = {4},
        pages = {5614-5630},
          doi = {10.1093/mnras/stad2615},
archivePrefix = {arXiv},
       eprint = {2306.05753},
 primaryClass = {astro-ph.GA},
       adsurl = {https://ui.adsabs.harvard.edu/abs/2023MNRAS.525.5614L}
}

@ARTICLE{Husko2025b,
       author = {{Hu{\v{s}}ko}, Filip and {Lacey}, Cedric G. and {Schaye}, Joop and {Schaller}, Matthieu and {Chaikin}, Evgenii and {Ploeckinger}, Sylvia and {Ben{\'\i}tez Llambay}, Alejandro and {Richings}, Alexander J. and {Trayford}, James W.},
        title = "{A hybrid active galactic nucleus feedback model with spinning black holes, winds and jets}",
      journal = {arXiv e-prints},
         year = 2025,
        month = sep,
          eid = {arXiv:2509.05179},
        pages = {arXiv:2509.05179},
          doi = {10.48550/arXiv.2509.05179},
archivePrefix = {arXiv},
       eprint = {2509.05179},
 primaryClass = {astro-ph.GA},
       adsurl = {https://ui.adsabs.harvard.edu/abs/2025arXiv250905179H}
}

@ARTICLE{Taylor1950,
       author = {{Taylor}, Geoffrey},
        title = "{The Formation of a Blast Wave by a Very Intense Explosion. I. Theoretical Discussion}",
      journal = {Proceedings of the Royal Society of London Series A},
         year = 1950,
        month = mar,
       volume = {201},
       number = {1065},
        pages = {159-174},
          doi = {10.1098/rspa.1950.0049},
       adsurl = {https://ui.adsabs.harvard.edu/abs/1950RSPSA.201..159T}
}

@ARTICLE{Sedov1946,
       author = {{Sedov}, Leonid Ivanovich},
        title = "{Propagation of strong shock waves}",
      journal = {Journal of Applied Mathematics and Mechanics},
         year = 1946,
        month = jan,
       volume = {10},
        pages = {241-250},
       adsurl = {https://ui.adsabs.harvard.edu/abs/1946JApMM..10..241S}
}

@ARTICLE{VonNeumann1964,
  title={John von Neumann Collected Works:@@@Volume V: Design of Computers, Theory of Automata and Numerical Analysis@@@Volume VI: Theory of Games, Astrophysics, Hydrodynamics and Meteorology.},
  author={{\VAN{Von Neumann}{von}{von}} Neumann, John and A. H. Taub},
  journal={Journal of the American Statistical Association},
  year={1964},
  volume={59},
  pages={981},
  url={https://api.semanticscholar.org/CorpusID:124649622}
}

@ARTICLE{Penston1969,
       author = {{Penston}, M.~V.},
        title = "{Dynamics of self-gravitating gaseous spheres-II. Collapses of gas spheres with cooling and the behaviour of polytropic gas spheres}",
      journal = {\mnras},
         year = 1969,
        month = jan,
       volume = {145},
        pages = {457},
          doi = {10.1093/mnras/145.4.457},
       adsurl = {https://ui.adsabs.harvard.edu/abs/1969MNRAS.145..457P}
}

@ARTICLE{Reines2013,
       author = {{Reines}, Amy E. and {Greene}, Jenny E. and {Geha}, Marla},
        title = "{Dwarf Galaxies with Optical Signatures of Active Massive Black Holes}",
      journal = {\apj},
         year = 2013,
        month = oct,
       volume = {775},
       number = {2},
          eid = {116},
        pages = {116},
          doi = {10.1088/0004-637X/775/2/116},
archivePrefix = {arXiv},
       eprint = {1308.0328},
 primaryClass = {astro-ph.CO},
       adsurl = {https://ui.adsabs.harvard.edu/abs/2013ApJ...775..116R}
}

@ARTICLE{Baldassare2017,
       author = {{Baldassare}, Vivienne F. and {Reines}, Amy E. and {Gallo}, Elena and {Greene}, Jenny E.},
        title = "{X-ray and Ultraviolet Properties of AGNs in Nearby Dwarf Galaxies}",
      journal = {\apj},
         year = 2017,
        month = feb,
       volume = {836},
       number = {1},
          eid = {20},
        pages = {20},
          doi = {10.3847/1538-4357/836/1/20},
archivePrefix = {arXiv},
       eprint = {1609.07148},
 primaryClass = {astro-ph.HE},
       adsurl = {https://ui.adsabs.harvard.edu/abs/2017ApJ...836...20B}
}

@ARTICLE{Koudmani2022,
       author = {{Koudmani}, Sophie and {Sijacki}, Debora and {Smith}, Matthew C.},
        title = "{Two can play at that game: constraining the role of supernova and AGN feedback in dwarf galaxies with cosmological zoom-in simulations}",
      journal = {\mnras},
         year = 2022,
        month = oct,
       volume = {516},
       number = {2},
        pages = {2112-2141},
          doi = {10.1093/mnras/stac2252},
archivePrefix = {arXiv},
       eprint = {2206.11274},
 primaryClass = {astro-ph.GA},
       adsurl = {https://ui.adsabs.harvard.edu/abs/2022MNRAS.516.2112K}
}

@ARTICLE{ArjonaGalvez2024,
       author = {{Arjona-G{\'a}lvez}, Elena and {Di Cintio}, Arianna and {Grand}, Robert J.~J.},
        title = "{The role of active galactic nucleus feedback on the evolution of dwarf galaxies from cosmological simulations: Supermassive black holes suppress star formation in low-mass galaxies}",
      journal = {\aap},
         year = 2024,
        month = oct,
       volume = {690},
          eid = {A286},
        pages = {A286},
          doi = {10.1051/0004-6361/202449439},
archivePrefix = {arXiv},
       eprint = {2402.00929},
 primaryClass = {astro-ph.GA},
       adsurl = {https://ui.adsabs.harvard.edu/abs/2024A&A...690A.286A}
}

@ARTICLE{Martizzi2013,
       author = {{Martizzi}, Davide and {Teyssier}, Romain and {Moore}, Ben},
        title = "{Cusp-core transformations induced by AGN feedback in the progenitors of cluster galaxies}",
      journal = {\mnras},
         year = 2013,
        month = jul,
       volume = {432},
       number = {3},
        pages = {1947-1954},
          doi = {10.1093/mnras/stt297},
archivePrefix = {arXiv},
       eprint = {1211.2648},
 primaryClass = {astro-ph.CO},
       adsurl = {https://ui.adsabs.harvard.edu/abs/2013MNRAS.432.1947M}
}

@ARTICLE{Peirani2017,
       author = {{Peirani}, S{\'e}bastien and {Dubois}, Yohan and {Volonteri}, Marta and {Devriendt}, Julien and {Bundy}, Kevin and {Silk}, Joe and {Pichon}, Christophe and {Kaviraj}, Sugata and {Gavazzi}, Rapha{\"e}l and {Habouzit}, M{\'e}lanie},
        title = "{Density profile of dark matter haloes and galaxies in the HORIZON-AGN simulation: the impact of AGN feedback}",
      journal = {\mnras},
         year = 2017,
        month = dec,
       volume = {472},
       number = {2},
        pages = {2153-2169},
          doi = {10.1093/mnras/stx2099},
archivePrefix = {arXiv},
       eprint = {1611.09922},
 primaryClass = {astro-ph.GA},
       adsurl = {https://ui.adsabs.harvard.edu/abs/2017MNRAS.472.2153P}
}

@ARTICLE{Eisenstein2005,
       author = {{Eisenstein}, Daniel J. and {Zehavi}, Idit and {Hogg}, David W. and {Scoccimarro}, Roman and {Blanton}, Michael R. and {Nichol}, Robert C. and {Scranton}, Ryan and {Seo}, Hee-Jong and {Tegmark}, Max and {Zheng}, Zheng and {Anderson}, Scott F. and {Annis}, Jim and {Bahcall}, Neta and {Brinkmann}, Jon and {Burles}, Scott and {Castander}, Francisco J. and {Connolly}, Andrew and {Csabai}, Istvan and {Doi}, Mamoru and {Fukugita}, Masataka and {Frieman}, Joshua A. and {Glazebrook}, Karl and {Gunn}, James E. and {Hendry}, John S. and {Hennessy}, Gregory and {Ivezi{\'c}}, Zeljko and {Kent}, Stephen and {Knapp}, Gillian R. and {Lin}, Huan and {Loh}, Yeong-Shang and {Lupton}, Robert H. and {Margon}, Bruce and {McKay}, Timothy A. and {Meiksin}, Avery and {Munn}, Jeffery A. and {Pope}, Adrian and {Richmond}, Michael W. and {Schlegel}, David and {Schneider}, Donald P. and {Shimasaku}, Kazuhiro and {Stoughton}, Christopher and {Strauss}, Michael A. and {SubbaRao}, Mark and {Szalay}, Alexander S. and {Szapudi}, Istv{\'a}n and {Tucker}, Douglas L. and {Yanny}, Brian and {York}, Donald G.},
        title = "{Detection of the Baryon Acoustic Peak in the Large-Scale Correlation Function of SDSS Luminous Red Galaxies}",
      journal = {\apj},
         year = 2005,
        month = nov,
       volume = {633},
       number = {2},
        pages = {560-574},
          doi = {10.1086/466512},
archivePrefix = {arXiv},
       eprint = {astro-ph/0501171},
 primaryClass = {astro-ph},
       adsurl = {https://ui.adsabs.harvard.edu/abs/2005ApJ...633..560E}
}

@ARTICLE{Heymans2013,
       author = {{Heymans}, Catherine and {Grocutt}, Emma and {Heavens}, Alan and {Kilbinger}, Martin and {Kitching}, Thomas D. and {Simpson}, Fergus and {Benjamin}, Jonathan and {Erben}, Thomas and {Hildebrandt}, Hendrik and {Hoekstra}, Henk and {Mellier}, Yannick and {Miller}, Lance and {Van Waerbeke}, Ludovic and {Brown}, Michael L. and {Coupon}, Jean and {Fu}, Liping and {Harnois-D{\'e}raps}, Joachim and {Hudson}, Michael J. and {Kuijken}, Konrad and {Rowe}, Barnaby and {Schrabback}, Tim and {Semboloni}, Elisabetta and {Vafaei}, Sanaz and {Velander}, Malin},
        title = "{CFHTLenS tomographic weak lensing cosmological parameter constraints: Mitigating the impact of intrinsic galaxy alignments}",
      journal = {\mnras},
         year = 2013,
        month = jul,
       volume = {432},
       number = {3},
        pages = {2433-2453},
          doi = {10.1093/mnras/stt601},
archivePrefix = {arXiv},
       eprint = {1303.1808},
 primaryClass = {astro-ph.CO},
       adsurl = {https://ui.adsabs.harvard.edu/abs/2013MNRAS.432.2433H}
}

@ARTICLE{vandenBosch2001,
       author = {{\VAN{Van den Bosch}{van den}{van den}} Bosch, Frank C. and {Swaters}, Rob A.},
        title = "{Dwarf galaxy rotation curves and the core problem of dark matter haloes}",
      journal = {\mnras},
         year = 2001,
        month = aug,
       volume = {325},
       number = {3},
        pages = {1017-1038},
          doi = {10.1046/j.1365-8711.2001.04456.x},
archivePrefix = {arXiv},
       eprint = {astro-ph/0006048},
 primaryClass = {astro-ph},
       adsurl = {https://ui.adsabs.harvard.edu/abs/2001MNRAS.325.1017V}
}

@ARTICLE{Dutton2005,
       author = {{Dutton}, Aaron A. and {Courteau}, St{\'e}phane and {de Jong}, Roelof and {Carignan}, Claude},
        title = "{Mass Modeling of Disk Galaxies: Degeneracies, Constraints, and Adiabatic Contraction}",
      journal = {\apj},
         year = 2005,
        month = jan,
       volume = {619},
       number = {1},
        pages = {218-242},
          doi = {10.1086/426375},
archivePrefix = {arXiv},
       eprint = {astro-ph/0310001},
 primaryClass = {astro-ph},
       adsurl = {https://ui.adsabs.harvard.edu/abs/2005ApJ...619..218D}
}

@ARTICLE{Oh2011DM_L,
       author = {{Oh}, Se-Heon and {de Blok}, W.~J.~G. and {Brinks}, Elias and {Walter}, Fabian and {Kennicutt}, Jr., Robert C.},
        title = "{Dark and Luminous Matter in THINGS Dwarf Galaxies}",
      journal = {\aj},
         year = 2011,
        month = jun,
       volume = {141},
       number = {6},
          eid = {193},
        pages = {193},
          doi = {10.1088/0004-6256/141/6/193},
archivePrefix = {arXiv},
       eprint = {1011.0899},
 primaryClass = {astro-ph.CO},
       adsurl = {https://ui.adsabs.harvard.edu/abs/2011AJ....141..193O}
}

@ARTICLE{Robles2017,
       author = {{Robles}, Victor H. and {Bullock}, James S. and {Elbert}, Oliver D. and {Fitts}, Alex and {Gonz{\'a}lez-Samaniego}, Alejandro and {Boylan-Kolchin}, Michael and {Hopkins}, Philip F. and {Faucher-Gigu{\`e}re}, Claude-Andr{\'e} and {Kere{\v{s}}}, Du{\v{s}}an and {Hayward}, Christopher C.},
        title = "{SIDM on FIRE: hydrodynamical self-interacting dark matter simulations of low-mass dwarf galaxies}",
      journal = {\mnras},
         year = 2017,
        month = dec,
       volume = {472},
       number = {3},
        pages = {2945-2954},
          doi = {10.1093/mnras/stx2253},
archivePrefix = {arXiv},
       eprint = {1706.07514},
 primaryClass = {astro-ph.GA},
       adsurl = {https://ui.adsabs.harvard.edu/abs/2017MNRAS.472.2945R}
}

@ARTICLE{deBlok2001,
       author = {{\VAN{De Blok}{de}{de}} Blok, W.~J.~G. and {McGaugh}, Stacy S. and {Bosma}, Albert and {Rubin}, Vera C.},
        title = "{Mass Density Profiles of Low Surface Brightness Galaxies}",
      journal = {\apjl},
         year = 2001,
        month = may,
       volume = {552},
       number = {1},
        pages = {L23-L26},
          doi = {10.1086/320262},
archivePrefix = {arXiv},
       eprint = {astro-ph/0103102},
 primaryClass = {astro-ph},
       adsurl = {https://ui.adsabs.harvard.edu/abs/2001ApJ...552L..23D}
}

@ARTICLE{deBlok_Bosma2002,
       author = {{\VAN{De Blok}{de}{de}} Blok, W.~J.~G. and {Bosma}, A.},
        title = "{High-resolution rotation curves of low surface brightness galaxies}",
      journal = {\aap},
         year = 2002,
        month = apr,
       volume = {385},
        pages = {816-846},
          doi = {10.1051/0004-6361:20020080},
archivePrefix = {arXiv},
       eprint = {astro-ph/0201276},
 primaryClass = {astro-ph},
       adsurl = {https://ui.adsabs.harvard.edu/abs/2002A&A...385..816D}
}

@ARTICLE{Navarro_Eke_Frenk1996,
       author = {{Navarro}, Julio F. and {Eke}, Vincent R. and {Frenk}, Carlos S.},
        title = "{The cores of dwarf galaxy haloes}",
      journal = {\mnras},
         year = 1996,
        month = dec,
       volume = {283},
       number = {3},
        pages = {L72-L78},
          doi = {10.1093/mnras/283.3.L72},
archivePrefix = {arXiv},
       eprint = {astro-ph/9610187},
 primaryClass = {astro-ph},
       adsurl = {https://ui.adsabs.harvard.edu/abs/1996MNRAS.283L..72N}
}

@INPROCEEDINGS{Shepard1968,
  author    = {Shepard, D.},
  title     = {A two-dimensional interpolation function for irregularly-spaced data},
  booktitle = {Proceedings of the 23rd ACM National Conference},
  year      = {1968},
  pages     = {517--524},
  publisher = {ACM},
  doi       = {10.1145/800186.810616}
}

@ARTICLE{Reinhardt2019,
  author  = {Reinhardt, S. and Krake, T. and Eberhardt, B. and Weiskopf, D.},
  title   = {Consistent Shepard interpolation for SPH-based fluid animation},
  journal = {ACM Trans. Graph.},
  year    = {2019},
  month   = nov,
  volume  = {38},
  number  = {6},
  pages   = {189},
  doi     = {10.1145/3355089.3356503},
  url     = {https://doi.org/10.1145/3355089.3356503}
}

@ARTICLE{Numpy,
       author = {{Harris}, Charles R. and {Millman}, K. Jarrod and \VAN{Van der Walt}{van der}{van der} Walt, St{\'e}fan J. and {Gommers}, Ralf and {Virtanen}, Pauli and {Cournapeau}, David and {Wieser}, Eric and {Taylor}, Julian and {Berg}, Sebastian and {Smith}, Nathaniel J. and {Kern}, Robert and {Picus}, Matti and {Hoyer}, Stephan and \VAN{Van Kerkwijk}{van}{van} Kerkwijk, Marten H. and {Brett}, Matthew and {Haldane}, Allan and {del R{\'\i}o}, Jaime Fern{\'a}ndez and {Wiebe}, Mark and {Peterson}, Pearu and {G{\'e}rard-Marchant}, Pierre and {Sheppard}, Kevin and {Reddy}, Tyler and {Weckesser}, Warren and {Abbasi}, Hameer and {Gohlke}, Christoph and {Oliphant}, Travis E.},
        title = "{Array programming with NumPy}",
      journal = {\nat},
         year = 2020,
        month = sep,
       volume = {585},
       number = {7825},
        pages = {357-362},
          doi = {10.1038/s41586-020-2649-2},
archivePrefix = {arXiv},
       eprint = {2006.10256},
 primaryClass = {cs.MS},
       adsurl = {https://ui.adsabs.harvard.edu/abs/2020Natur.585..357H}
}

@ARTICLE{Astropy,
       author = {{Astropy Collaboration} and {Price-Whelan}, Adrian M. and {Lim}, Pey Lian and {Earl}, Nicholas and {Starkman}, Nathaniel and {Bradley}, Larry and {Shupe}, David L. and {Patil}, Aarya A. and {Corrales}, Lia and {Brasseur}, C.~E. and {N{\"o}the}, Maximilian and {Donath}, Axel and {Tollerud}, Erik and {Morris}, Brett M. and {Ginsburg}, Adam and {Vaher}, Eero and {Weaver}, Benjamin A. and {Tocknell}, James and {Jamieson}, William and \VAN{Van Kerkwijk}{van}{van} Kerkwijk, Marten H. and {Robitaille}, Thomas P. and {Merry}, Bruce and {Bachetti}, Matteo and {G{\"u}nther}, H. Moritz and {Aldcroft}, Thomas L. and {Alvarado-Montes}, Jaime A. and {Archibald}, Anne M. and {B{\'o}di}, Attila and {Bapat}, Shreyas and {Barentsen}, Geert and {Baz{\'a}n}, Juanjo and {Biswas}, Manish and {Boquien}, M{\'e}d{\'e}ric and {Burke}, D.~J. and {Cara}, Daria and {Cara}, Mihai and {Conroy}, Kyle E. and {Conseil}, Simon and {Craig}, Matthew W. and {Cross}, Robert M. and {Cruz}, Kelle L. and {D'Eugenio}, Francesco and {Dencheva}, Nadia and {Devillepoix}, Hadrien A.~R. and {Dietrich}, J{\"o}rg P. and {Eigenbrot}, Arthur Davis and {Erben}, Thomas and {Ferreira}, Leonardo and {Foreman-Mackey}, Daniel and {Fox}, Ryan and {Freij}, Nabil and {Garg}, Suyog and {Geda}, Robel and {Glattly}, Lauren and {Gondhalekar}, Yash and {Gordon}, Karl D. and {Grant}, David and {Greenfield}, Perry and {Groener}, Austen M. and {Guest}, Steve and {Gurovich}, Sebastian and {Handberg}, Rasmus and {Hart}, Akeem and {Hatfield-Dodds}, Zac and {Homeier}, Derek and {Hosseinzadeh}, Griffin and {Jenness}, Tim and {Jones}, Craig K. and {Joseph}, Prajwel and {Kalmbach}, J. Bryce and {Karamehmetoglu}, Emir and {Ka{\l}uszy{\'n}ski}, Miko{\l}aj and {Kelley}, Michael S.~P. and {Kern}, Nicholas and {Kerzendorf}, Wolfgang E. and {Koch}, Eric W. and {Kulumani}, Shankar and {Lee}, Antony and {Ly}, Chun and {Ma}, Zhiyuan and {MacBride}, Conor and {Maljaars}, Jakob M. and {Muna}, Demitri and {Murphy}, N.~A. and {Norman}, Henrik and {O'Steen}, Richard and {Oman}, Kyle A. and {Pacifici}, Camilla and {Pascual}, Sergio and {Pascual-Granado}, J. and {Patil}, Rohit R. and {Perren}, Gabriel I. and {Pickering}, Timothy E. and {Rastogi}, Tanuj and {Roulston}, Benjamin R. and {Ryan}, Daniel F. and {Rykoff}, Eli S. and {Sabater}, Jose and {Sakurikar}, Parikshit and {Salgado}, Jes{\'u}s and {Sanghi}, Aniket and {Saunders}, Nicholas and {Savchenko}, Volodymyr and {Schwardt}, Ludwig and {Seifert-Eckert}, Michael and {Shih}, Albert Y. and {Jain}, Anany Shrey and {Shukla}, Gyanendra and {Sick}, Jonathan and {Simpson}, Chris and {Singanamalla}, Sudheesh and {Singer}, Leo P. and {Singhal}, Jaladh and {Sinha}, Manodeep and {Sip{\H{o}}cz}, Brigitta M. and {Spitler}, Lee R. and {Stansby}, David and {Streicher}, Ole and {{\v{S}}umak}, Jani and {Swinbank}, John D. and {Taranu}, Dan S. and {Tewary}, Nikita and {Tremblay}, Grant R. and \VAN{De Val-Borro}{de}{de} Val-Borro, Miguel and \VAN{Van Kooten}{van}{van} Kooten, Samuel J. and {Vasovi{\'c}}, Zlatan and {Verma}, Shresth and \VAN{De Miranda Cardoso}{de}{de} Miranda Cardoso, Jos{\'e} Vin{\'\i}cius and {Williams}, Peter K.~G. and {Wilson}, Tom J. and {Winkel}, Benjamin and {Wood-Vasey}, W.~M. and {Xue}, Rui and {Yoachim}, Peter and {Zhang}, Chen and {Zonca}, Andrea and {Astropy Project Contributors}},
        title = "{The Astropy Project: Sustaining and Growing a Community-oriented Open-source Project and the Latest Major Release (v5.0) of the Core Package}",
      journal = {\apj},
         year = 2022,
        month = aug,
       volume = {935},
       number = {2},
          eid = {167},
        pages = {167},
          doi = {10.3847/1538-4357/ac7c74},
archivePrefix = {arXiv},
       eprint = {2206.14220},
 primaryClass = {astro-ph.IM},
       adsurl = {https://ui.adsabs.harvard.edu/abs/2022ApJ...935..167A}
}

@ARTICLE{Scikit-learn,
       author = {{Pedregosa}, Fabian and {Varoquaux}, Ga{\"e}l and {Gramfort}, Alexandre and {Michel}, Vincent and {Thirion}, Bertrand and {Grisel}, Olivier and {Blondel}, Mathieu and {M{\"u}ller}, Andreas and {Nothman}, Joel and {Louppe}, Gilles and {Prettenhofer}, Peter and {Weiss}, Ron and {Dubourg}, Vincent and {Vanderplas}, Jake and {Passos}, Alexandre and {Cournapeau}, David and {Brucher}, Matthieu and {Perrot}, Matthieu and {Duchesnay}, {\'E}douard},
        title = "{Scikit-learn: Machine Learning in Python}",
      journal = {Journal of Machine Learning Research},
         year = 2011,
        month = oct,
       volume = {12},
        pages = {2825-2830},
          doi = {10.48550/arXiv.1201.0490},
archivePrefix = {arXiv},
       eprint = {1201.0490},
 primaryClass = {cs.LG},
       adsurl = {https://ui.adsabs.harvard.edu/abs/2011JMLR...12.2825P}
}

@ARTICLE{SWIFTGalaxy,
       author = {{Oman}, Kyle},
        title = "{SWIFTGalaxy: a Python package to work with particle groups from SWIFT simulations}",
      journal = {The Journal of Open Source Software},
         year = 2025,
        month = oct,
       volume = {10},
       number = {114},
          eid = {9278},
        pages = {9278},
          doi = {10.21105/joss.09278},
archivePrefix = {arXiv},
       eprint = {2510.22328},
 primaryClass = {astro-ph.GA},
       adsurl = {https://ui.adsabs.harvard.edu/abs/2025JOSS...10.9278O}
}

@ARTICLE{swiftsimio,
       author = {{Borrow}, Josh and {Borrisov}, Alexei},
        title = "{swiftsimio: A Python library for reading SWIFT data}",
      journal = {The Journal of Open Source Software},
         year = 2020,
        month = aug,
       volume = {5},
       number = {52},
          eid = {2430},
        pages = {2430},
          doi = {10.21105/joss.02430},
       adsurl = {https://ui.adsabs.harvard.edu/abs/2020JOSS....5.2430B}
}

@INPROCEEDINGS{Numba,
       author = {{Lam}, Siu Kwan and {Pitrou}, Antoine and {Seibert}, Stanley},
        title = "{Numba: A LLVM-based Python JIT Compiler}",
    booktitle = {Proc. Second Workshop on the LLVM Compiler Infrastructure in HPC},
         year = 2015,
        month = nov,
        pages = {1-6},
          doi = {10.1145/2833157.2833162},
       adsurl = {https://ui.adsabs.harvard.edu/abs/2015llvm.confE...1L}
}

@software{h5py,
       author = {{Collette}, Andrew and {Kluyver}, Thomas and {Caswell}, Thomas A and {Tocknell}, James and {Kieffer}, Jerome and {Jelenak}, Aleksandar and {Scopatz}, Anthony and {Dale}, Darren and {Chen} and {VINCENT}, Thomas and {Einhorn}, Matt and {Payno} and {Juliagarriga} and {Sciarelli}, Pierlauro and {Valls}, Valentin and {Ghosh}, Satrajit and {Kofoed Pedersen}, Ulrik and {Jakirkham} and {Raspaud}, Martin and {Danilevski}, Cyril and {Abbasi}, Hameer and {Readey}, John and {M{\"u}hlbauer}, Kai and {Paramonov}, Andrey and {Chan}, Lawrence and {Sol{\'e}}, V. Armando and {Jialin} and {Hay Guest}, Daniel and {Feng}, Yu and {Kittisopikul}, Mark},
    title = "{h5py/h5py: 3.7.0}",
    year = 2022,
    month = may,
    eid = {10.5281/zenodo.6575970},
    doi = {10.5281/zenodo.6575970},
    version = {3.7.0},
    publisher = {Zenodo},
    adsurl = {https://ui.adsabs.harvard.edu/abs/2022zndo...6575970C}
}

@ARTICLE{Agertz2011,
       author = {{Agertz}, Oscar and {Teyssier}, Romain and {Moore}, Ben},
        title = "{The formation of disc galaxies in a {\ensuremath{\Lambda}}CDM universe}",
      journal = {\mnras},
         year = 2011,
        month = jan,
       volume = {410},
       number = {2},
        pages = {1391-1408},
          doi = {10.1111/j.1365-2966.2010.17530.x},
archivePrefix = {arXiv},
       eprint = {1004.0005},
 primaryClass = {astro-ph.CO},
       adsurl = {https://ui.adsabs.harvard.edu/abs/2011MNRAS.410.1391A}
}

@ARTICLE{Benavides2025,
       author = {{Benavides}, Jos{\'e} A. and {Sales}, Laura V. and {Wetzel}, Andrew and {Moreno}, Jorge and {Feldmann}, Robert and {Mercado}, Francisco J. and {Bullock}, James S. and {Hopkins}, Philip F. and {Faucher-Gigu{\'e}re}, Claude-Andr{\'e} and {Stern}, Jonathan and {Wheeler}, Coral and {Kere{\v{s}}}, Du{\v{s}}an},
        title = "{Disks no more: the morphology of low-mass simulated galaxies in FIREbox}",
      journal = {\mnras},
         year = 2025,
        month = oct,
          doi = {10.1093/mnras/staf1847},
archivePrefix = {arXiv},
       eprint = {2508.00991},
 primaryClass = {astro-ph.GA},
       adsurl = {https://ui.adsabs.harvard.edu/abs/2025MNRAS.tmp.1739B}
}

@ARTICLE{Tacchella2019,
       author = {{Tacchella}, Sandro and {Diemer}, Benedikt and {Hernquist}, Lars and {Genel}, Shy and {Marinacci}, Federico and {Nelson}, Dylan and {Pillepich}, Annalisa and {Rodriguez-Gomez}, Vicente and {Sales}, Laura V. and {Springel}, Volker and {Vogelsberger}, Mark},
        title = "{Morphology and star formation in IllustrisTNG: the build-up of spheroids and discs}",
      journal = {\mnras},
         year = 2019,
        month = aug,
       volume = {487},
       number = {4},
        pages = {5416-5440},
          doi = {10.1093/mnras/stz1657},
archivePrefix = {arXiv},
       eprint = {1904.12860},
 primaryClass = {astro-ph.GA},
       adsurl = {https://ui.adsabs.harvard.edu/abs/2019MNRAS.487.5416T}
}

@ARTICLE{Jahn2023,
       author = {{Jahn}, Ethan D. and {Sales}, Laura V. and {Marinacci}, Federico and {Vogelsberger}, Mark and {Torrey}, Paul and {Qi}, Jia and {Smith}, Aaron and {Li}, Hui and {Kannan}, Rahul and {Burger}, Jan D. and {Zavala}, Jes{\'u}s},
        title = "{Real and counterfeit cores: how feedback expands haloes and disrupts tracers of inner gravitational potential in dwarf galaxies}",
      journal = {\mnras},
         year = 2023,
        month = mar,
       volume = {520},
       number = {1},
        pages = {461-479},
          doi = {10.1093/mnras/stad109},
archivePrefix = {arXiv},
       eprint = {2110.00142},
 primaryClass = {astro-ph.GA},
       adsurl = {https://ui.adsabs.harvard.edu/abs/2023MNRAS.520..461J}
}

@ARTICLE{Rey2024,
       author = {{Rey}, Martin P. and {Orkney}, Matthew D.~A. and {Read}, Justin I. and {Das}, Payel and {Agertz}, Oscar and {Pontzen}, Andrew and {Ponomareva}, Anastasia A. and {Kim}, Stacy Y. and {McClymont}, William},
        title = "{EDGE - Dark matter or astrophysics? Breaking dark matter heating degeneracies with H I rotation in faint dwarf galaxies}",
      journal = {\mnras},
         year = 2024,
        month = apr,
       volume = {529},
       number = {3},
        pages = {2379-2398},
          doi = {10.1093/mnras/stae718},
archivePrefix = {arXiv},
       eprint = {2309.00041},
 primaryClass = {astro-ph.GA},
       adsurl = {https://ui.adsabs.harvard.edu/abs/2024MNRAS.529.2379R}
}

\appendix
\section{SPH Field Sampling in a Warped Gas Disc: A Representative Example} \label{app:MorphologySampling}

\begin{figure*}
    \centering
    \includegraphics[width=1.8\columnwidth, height = 0.46\textheight]{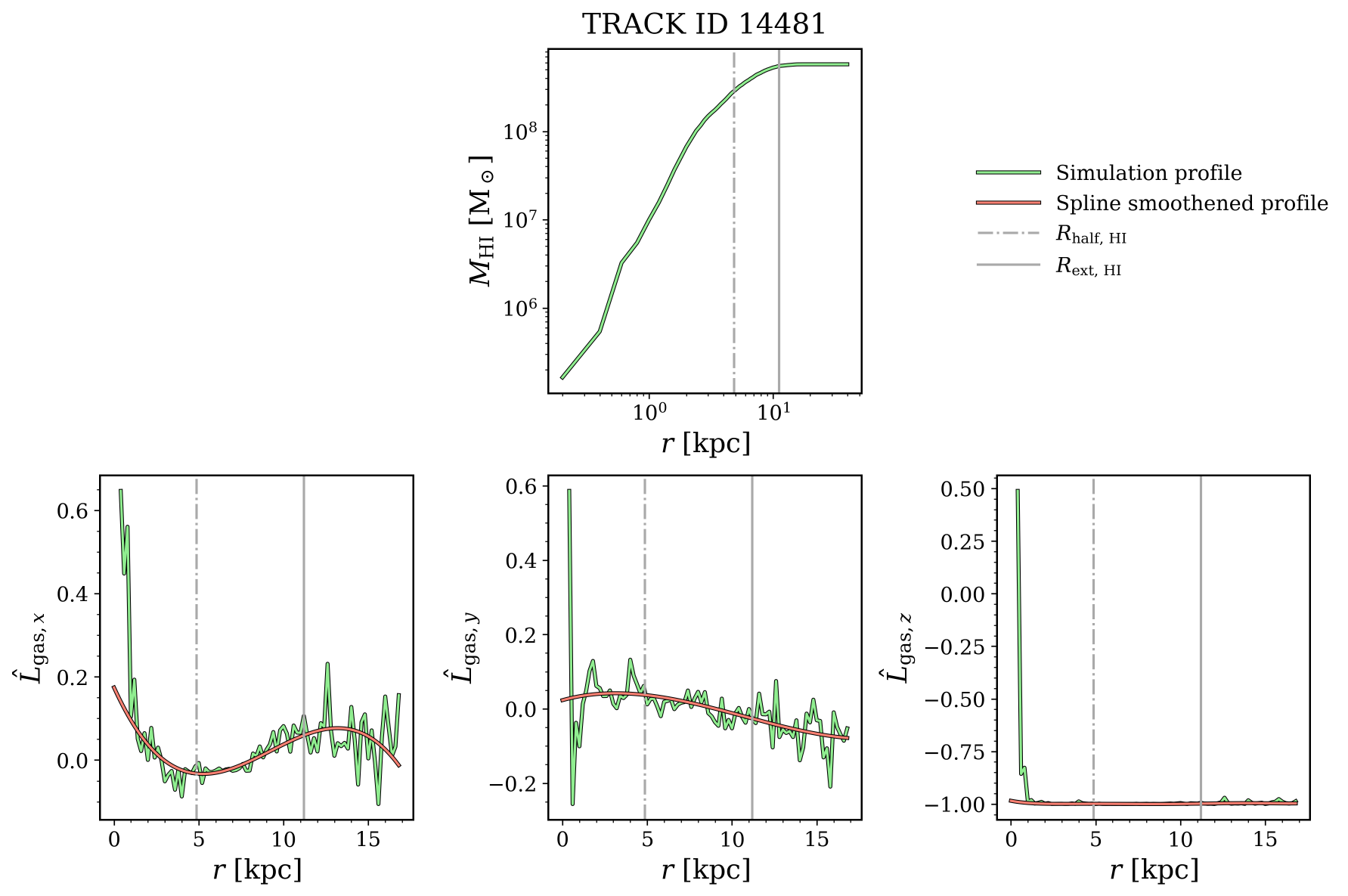}
    \caption{
Example of radial profiles for a representative halo in our sample. \textit{Top panel}: enclosed H\textsc{i} mass, $M_{\rm H\textsc{i}}(<r)$, as a function of radius (solid light green line). \textit{Bottom panels}: components of the normalised gas angular momentum vector in the global $x$, $y$, and $z$ directions (left to right), as functions of radius. Solid light green line show the profiles computed directly from the simulation particles (see Sec.~\ref{sec:Methods:GasDiscs}); solid light red lines show smoothed versions obtained via a cubic spline interpolation. The dash-dotted and solid vertical grey lines shown in both panels indicate the H\textsc{i} half-mass radius, $R_{\mathrm{half}, \rm H\textsc{i}}$, and the gas disc extent, $R_{\mathrm{ext}, \rm H\textsc{i}}$, respectively.
}
    \label{fig:AppendixMorphology1}
\end{figure*}

\begin{figure*}
    \centering
    \includegraphics[width=\textwidth, height=0.4\textheight]{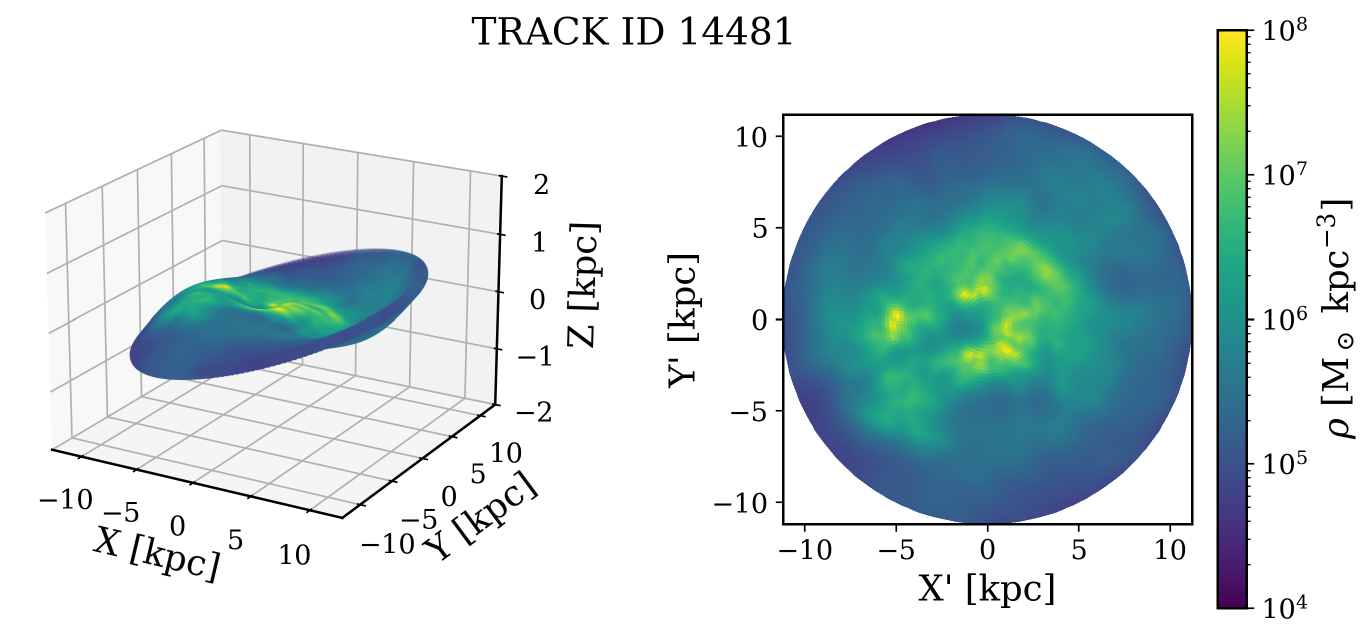}
    \caption{
Example of a warped disc midplane and the corresponding gas volume density field at the sampling points. \textit{Left panel}: 3D visualisation of the warped disc midplane in the global coordinate frame ($X$, $Y$, $Z$), with sampling points coloured by the gas density, $\rho$, computed via the SPH-based continuous field sampling method described in Sec.~\ref{sec:Methods:SPHSampling}. \textit{Right panel}: The same sampling points and gas density field, deprojected onto the local warped disc frame ($X'$, $Y'$), where $Z' = 0$ by construction. The same representative galaxy of Fig.~\ref{fig:AppendixMorphology1} is shown here, clarifying the angular momentum-based frame construction.
}
    \label{fig:AppendixMorphology2}
\end{figure*}

To complement the methodological description in Sections~\ref{sec:Methods:GasDiscs} and \ref{sec:Methods:SPHSampling}, here we show an example of how the warped disc midplane and associated SPH-sampled fields are constructed and interpreted, for a representative galaxy in our sample.

Fig.~\ref{fig:AppendixMorphology1} shows the cumulative H\textsc{i} mass profile, $M_{\rm H\textsc{i}}(<r)$, and the orientation of the gas angular momentum vector as a function of radius for a representative galaxy in our sample (\textsc{hbt-herons} track ID $= 14481$). The disc extent radius, $R_{\rm ext,H\textsc{i}}$, is identified from the flattening of the mass profile, using the slope criterion outlined in Sec.~\ref{sec:Methods:GasDiscs}, and is marked in both panels. The lower panel illustrates the radial evolution of the gas angular momentum direction in the global frame, revealing a modest but coherent warp between the half mass radius, $R_{\rm half, \rm H\textsc{i}}$, and the disc extent radius, $R_{\rm ext, H\textsc{i}}$, and a significantly stronger twist at radii $r < 5~\mathrm{kpc}$.

This radial variation -- in some cases significantly more pronounced -- in the orientation of the gaseous plane of rotation motivates the construction of a warped, radius-dependent local frame. Crucially, our definition of $R_{\rm ext,H\textsc{i}}$ -- based on the point where the cumulative mass profile flattens -- ensures that the number of gas particles per radial bin (or shell) remains sufficiently high within the disc. This mitigates noise in the angular momentum calculation, which could otherwise be dominated by a few outlier particles in low-density outer shells, thus improving the stability and accuracy of the local disc frame construction. Nevertheless, an interpolation of both the enclosed H\textsc{i} mass and gas angular momentum is preferable when eveluating slopes and constructing the geometry of the midplane.

Fig.~\ref{fig:AppendixMorphology2} provides a visual illustration of the warped disc frame construction motivated by the angular momentum profiles in Fig.~\ref{fig:AppendixMorphology1}; the same representative galaxy is shown in both figures to clarify the transition from a radially varying angular momentum direction to a warped midplane. The left panel shows the warped disc midplane and sampling points in the global coordinate frame ($X$, $Y$, $Z$), with points coloured by the local gas density, $\rho$. The \textit{right panel} shows the same points deprojected into the local warped disc frame ($X'$, $Y'$), where by construction the midplane lies at $Z' = 0$. This highlights how the local coordinate frame smoothly follows the radial variation in the gas angular momentum direction, effectively `unwarping' the disc morphology.

Importantly, here it is also possible to grasp the utility of our SPH-based continuous field sampling procedure: it enables evaluation of hydrodynamical quantities, such as the gas volume density, directly on the warped disc midplane. This would not be possible using only the raw gas particle data, which are irregularly distributed and not confined to the warped plane. Thus, our post-processing approach provides a physically meaningful, high-resolution sampling of continuous fields aligned with the true gas disc geometry. Note that in strongly disturbed systems the ‘midplane’ may still be ill-defined, as large warps or oscillations in the angular momentum profile prevent the existence of a singular stable configuration -- indeed the entire `tilted ring' approximation may break down. In such cases, our method still recovers the locally defined surface of maximum rotation, which represents the most meaningful reference for kinematic analysis.

\section{Accounting for Disequilibrium: Sensitivity to Threshold Variations}
\label{app:varThresh}

As discussed in Sec.~\ref{sec:Origins_of_diseq:Budget}, the quantitative contribution of individual physical drivers to the disequilibrium budget depends to some extent on the adopted criteria used to associate gas elements as affected by stellar feedback, AGN activity, and self-gravitating clumps. While our fiducial selections are physically motivated and empirically calibrated, it is important to assess the robustness of the resulting trends to reasonable variations in these somewhat arbitrary choices.

To this end, we repeat the full analysis using two alternative sets of tagging criteria. In a \emph{permissive} configuration, we loosen the spatial apertures and selection thresholds such that the total disc area associated with each driver is increased by approximately a factor of two relative to the fiducial case. Conversely, in a \emph{restrictive} configuration, we tighten the criteria so that the tagged area is reduced by a comparable factor. In all cases, the same physical definitions are retained, and only the extent of the tagged regions is modified. 

\begin{table*}
\centering
\caption{
Summary of the tagging criteria used to associate gas with the different disequilibrium drivers in the fiducial analysis and in the permissive and restrictive variants explored in Appendix~\ref{app:varThresh}. The permissive and restrictive configurations are constructed to roughly double or halve, respectively, the total disc area tagged by each mechanism relative to the fiducial case, while preserving the same physical definitions.
}
\label{tab:threshold_variations}
\begin{tabular}{l l c c c c}
\hline\hline
Mechanism & Parameter description & Symbol [unit] & Restrictive & Fiducial & Permissive \\
\hline
Stellar feedback 
& Maximum distance from YS or from CCSNe kicked gas
& $R_{\rm SF} [\rm kpc]$ 
& $\approx 0.423$ 
& $= 1.000$ 
& $\approx 2.450$ \\

\hline
Self-gravitating clumps 
& Free-fall to orbital time ratio
& $t_{\rm ff}/t_{\rm orb} [\rm dimensionless]$ 
& $\approx 0.094$ 
& $\approx 0.111$ 
& $\approx 0.131$ \\

\hline
AGN feedback 
& Maximum distance from active SMBH 
& $R_{\rm AGN} [\rm kpc]$ 
& $\approx 1.414$ 
& $= 2.000$ 
& $\approx 2.828$ \\ 

\hline\hline
\end{tabular}
\end{table*}

\begin{figure*}
    \centering
    \includegraphics[width=\textwidth, height=0.25\textheight]{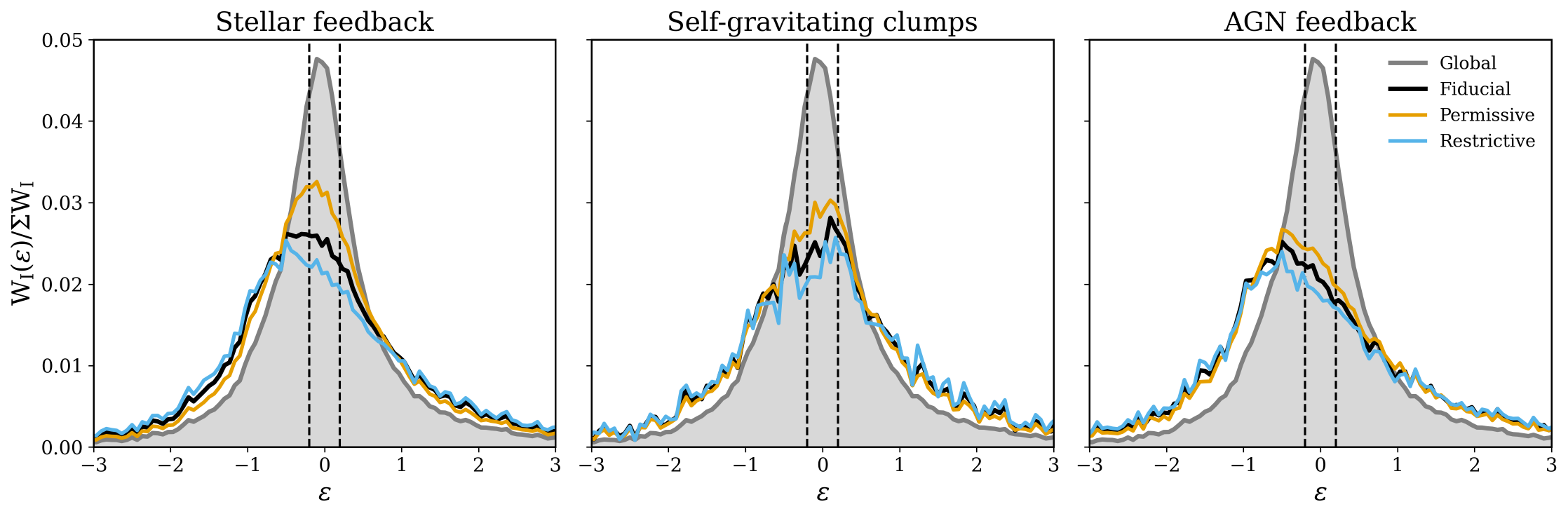}
    \caption{
    Self-normalised mass-weighted distributions of the disequilibrium parameter $\varepsilon$, $\mathrm{W}_{\rm I}(\varepsilon)/\Sigma\mathrm{W}_{\rm I}$, for gas associated with each physical driver under different tagging criteria (summarised in Table~\ref{tab:threshold_variations}). Each panel shows the distribution for one driver (stellar feedback, self-gravitating clumps, AGN feedback), comparing the fiducial selection (black) to permissive (orange) and restrictive (blue) variants. The grey shaded curve indicates the global gas distribution. Vertical dashed lines mark the quasi-equilibrium range, $|\varepsilon|\leq0.2$. While the absolute normalisation and degree of overlap with the low-$|\varepsilon|$ regime vary with the tagging criteria, the qualitative association between each driver and characteristic disequilibrium regimes is robust.
    }
    \label{fig:self_dist_variations}
\end{figure*}

Figure~\ref{fig:self_dist_variations} shows the resulting mass-weighted distributions of the disequilibrium parameter $\varepsilon$ for each driver under the three configurations, compared to the global distribution. As expected, loosening the cuts increases the overlap with the near-equilibrium regime, while tightening them preferentially selects gas at larger $|\varepsilon|$. Importantly, however, the overall shapes of the distributions are preserved to a good degree. Self-gravitating clumps preferentially populate the positive side of the distribution, while AGN feedback remains strongly associated with large negative $\varepsilon$. 

Stellar feedback exhibits the largest shape variation, particularly in the transition from the fiducial to the permissive configuration: in the latter case, the distribution becomes increasingly weighted toward the near-equilibrium regime, with a reduced relative contribution at negative $\varepsilon$. This behaviour indicates that overly permissive spatial selections (e.g. circular masks of order $\approx 2.5 ~\rm kpc$ around young stellar particles or feedback-affected gas) begin to include substantial amounts of ambient or recovering gas. This effect naturally drives the mild differences in recovery fractions observed across configurations and directly motivates the calibration procedure adopted for the fiducial cuts. 

\begin{figure*}
    \includegraphics[width=1.2\columnwidth, height=0.30\textheight]{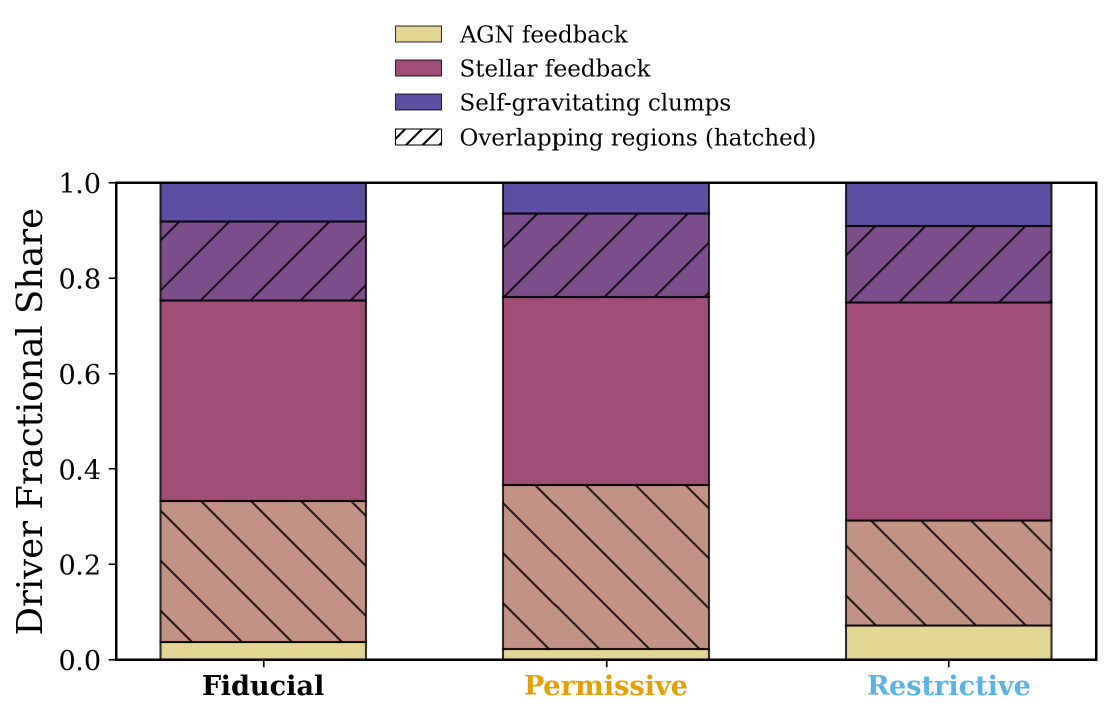}
    \caption{
    Fractional contribution of each physical driver to the total disequilibrium budget under fiducial, permissive, and restrictive tagging criteria (see Table~\ref{tab:threshold_variations}). Coloured regions indicate the contributions from stellar feedback, AGN feedback, and self-gravitating clumps, while hatched regions mark overlaps between drivers as indicated in the legend. Although the absolute fractions vary with the adopted cuts in different $\varepsilon$ regimes, the hierarchy of drivers and of overlap regions remain stable across all configurations.
    }
    \label{fig:driver_share_variations}
\end{figure*}

The impact of these variations on the integrated disequilibrium budget is summarised in Fig.~\ref{fig:driver_share_variations}, which shows the fractional contribution of each driver -- including overlaps -- to the total gas mass tagged by the union of the three mechanisms. As anticipated, the relative ranking of the drivers is robust to changes in the tagging criteria: stellar feedback consistently dominates the budget, while AGN feedback and self-gravitating clumps contribute at a secondary but non-negligible level. The substantial overlap fractions are likewise largely preserved across configurations, reinforcing the conclusion that disequilibrium is frequently driven by multiple mechanisms acting simultaneously.

The only clear systematic difference induced by varying the thresholds is a redistribution of mass between the individual stellar feedback and AGN feedback components and their mutual overlap. This behaviour is expected, as the tagging criteria for both mechanisms are purely geometrical: increasing the characteristic radii naturally enlarges the region of intersection, reducing the fraction uniquely attributed to each driver (and vice versa for more restrictive cuts). By contrast, analogous trends are much weaker for self-gravitating clumps. This reflects the fundamentally different nature of their selection: varying the threshold in $t_{\rm ff}/t_{\rm orb}$ does not simply expand or contract contiguous regions, but can instead select spatially disconnected structures. As a result, changes in the overlap between clumps and the other drivers are less straightforward and do not scale monotonically with the strictness of the cut (in fact, they are close to independent of them). Notably, the differences discussed here remain limited to the $\approx5$-$10$~per~cent level across configurations, indicating that they do not qualitatively affect the inferred disequilibrium budget.

For completeness, and in reference to the integrated percentages discussed in the final paragraphs of Sec.~\ref{sec:Origins_of_diseq:Budget}, the restrictive and permissive configurations respectively account for $\approx20$ and $50$~per~cent of the gas mass in the near-equilibrium regime ($|\varepsilon|\leq0.2$), and $\approx70$ and $90$~per~cent in the extreme disequilibrium regime ($|\varepsilon|\geq1.0$). In the intermediate regime ($0.2<|\varepsilon|<1.0$), they encompass $\approx35$ and $65$~per~cent of the global gas mass. The systematic and differential response across $|\varepsilon|$ regimes further motivates the fiducial configuration as a balanced and optimised choice.

Taken together, these tests demonstrate that while the absolute normalisation of the tagged mass varies with the adopted thresholds, the qualitative association between physical drivers and characteristic disequilibrium regimes -- and their relative partitioning -- remains remarkably stable. This robustness lends confidence to the fiducial tagging strategy and supports the broader conclusion that significant departures from equilibrium in the disc arise from the combined action of stellar feedback, AGN activity, and gravitational instability, rather than from any single dominant mechanism.

\section{Classification Algorithm} \label{app:classAlg}

\begin{figure*}
    \centering
    \includegraphics[width=\textwidth, height = 0.35\textheight]{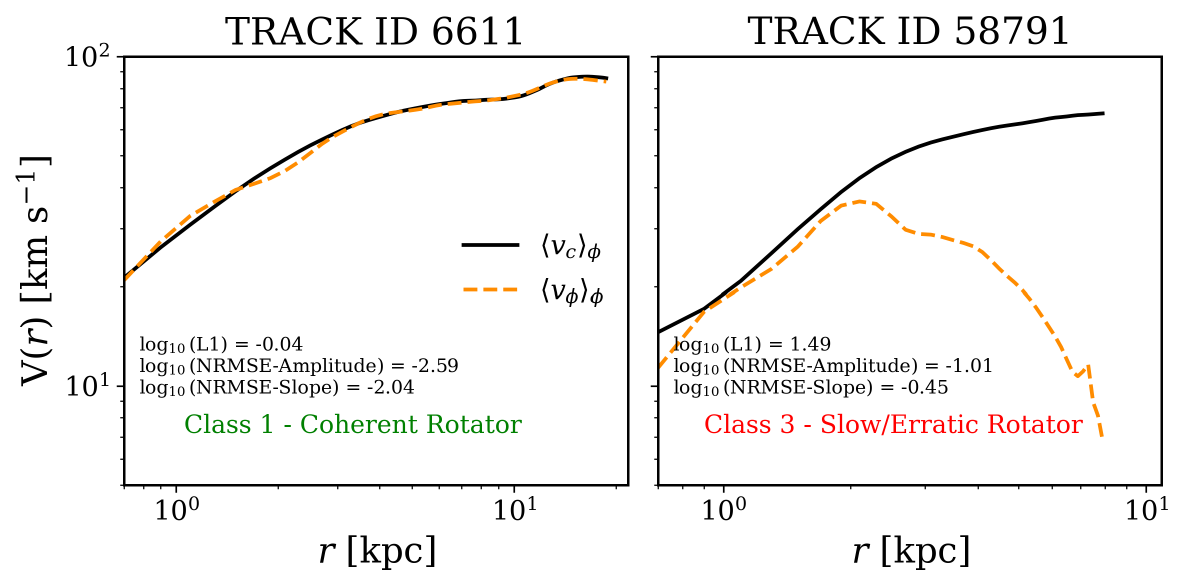} 
    \includegraphics[width=\columnwidth, height = 0.33\textheight]{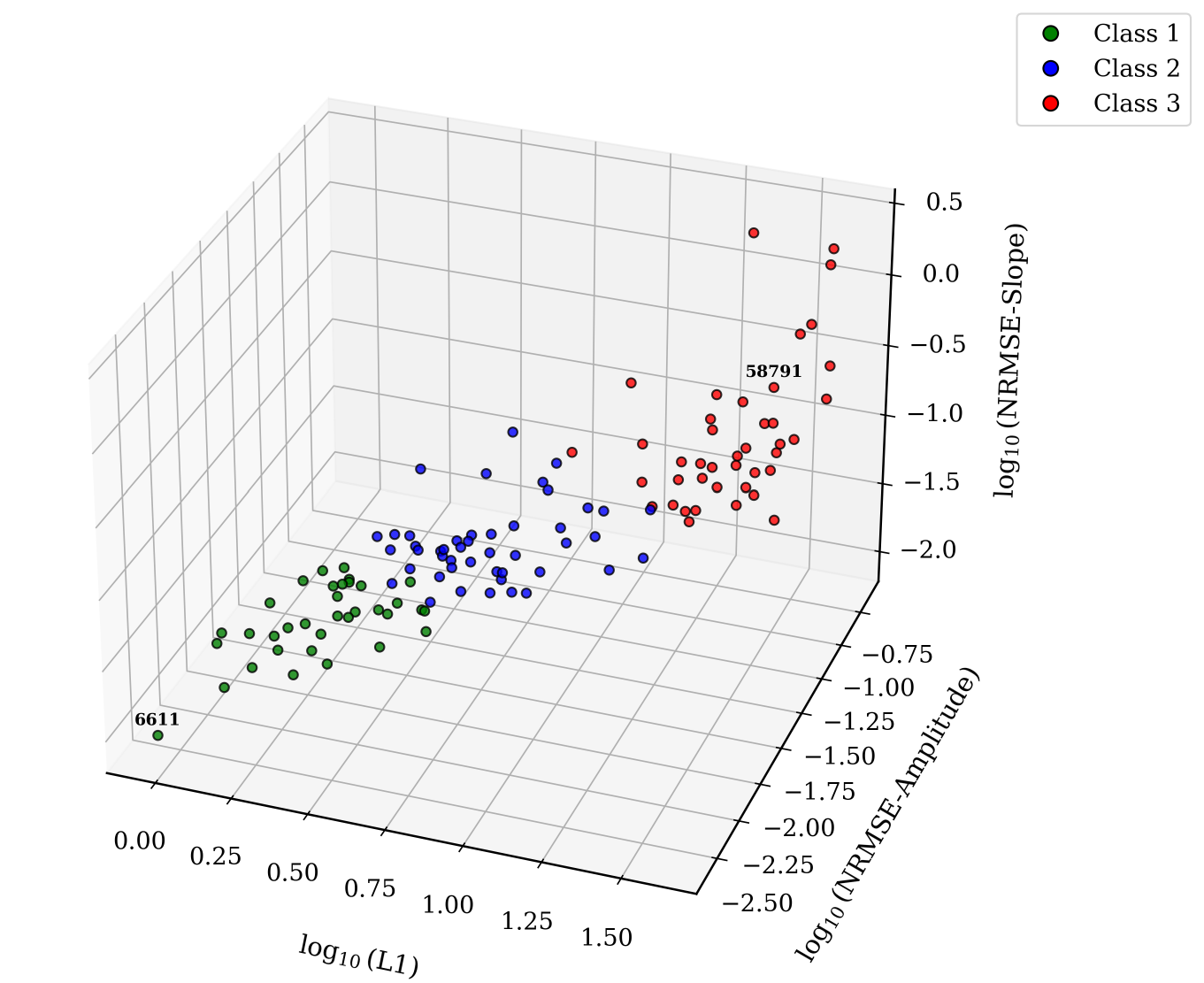}
    \caption{Illustrative examples and clustering result from our dynamical classification algorithm. \textit{Top panels}: Azimuthally averaged rotational velocity (dark orange/dashed lines) and circular velocity (black/solid lines) profiles as a function of radius for two representative galaxies in the sample. \textit{Left}: A dynamically well-behaved system exhibiting excellent agreement between the two profiles across the entire disc extent. \textit{Right}: A strongly perturbed system showing large and systematic residuals. The \textsc{hbt-herons} track IDs of these galaxies, and their associated values for the three classification metrics ($L_1$, NRMSE-Amplitude, and NRMSE-Slope) are indicated in the legends and annotated text. \textit{Bottom panel}: Distribution of the full sample in the three-dimensional metric space, colour-coded by cluster assignment as determined via the $k$-means clustering algorithm. The resulting groups define the dynamical classes discussed in Sec.~\ref{sec:Classification}.}
    \label{fig:Clusters}
\end{figure*}

As outlined in Sec.~\ref{sec:Classification}, our goal is to classify galaxies based on systematic deviations between the azimuthally averaged rotational velocity profile, $\langle v_\phi\rangle_\phi(r)$, and the corresponding circular velocity profile, $\langle v_c\rangle_\phi(r)$. These deviations are used as proxies for assessing the dynamical equilibrium state of each system.

To quantify such deviations in a statistically meaningful and comparative manner, we compute the following three metrics for each galaxy in our sample:
\begin{enumerate}
    \item $L_1$ norm: a global, absolute measure of the discrepancy between the rotational and circular velocity profiles across the radial extent of the gas disc,
    \[ L_1 = \int_{r= r_0}^{r= R_{\rm ext, H\textsc{i}}}|\langle v_c\rangle_\phi(r) - \langle v_\phi\rangle_\phi(r)|dr,\] where $r_0$ is a small inner cutoff to exclude unresolved central regions.
    \item Normalized Root Mean Square Error in Amplitude (NRMSE-Amplitude): a pointwise metric that emphasizes the fractional error between the two profiles,  
    \[\mathrm{NRMSE-Amplitude} = \sum_i{\sqrt{\frac{[\langle v_\phi\rangle_\phi(r_i)-\langle v_c\rangle_\phi(r_i)]^2}{\langle v_c\rangle_\phi(r_i)^2}}},\] where the sum runs over discrete radial bins $r_i$ spanning the disc from $r_0$ to $R_{\rm ext, H\textsc{i}}$ with $\Delta r = 0.2~\mathrm{kpc}$ spacing.
    \item Normalized Root Mean Square Error in Slope (NRMSE-Slope): a scale-invariant measure comparing the local logarithmic slopes of the rotational and circular velocity profiles, \[ \mathrm{NRMSE-Slope} = \sum_i \sqrt{ \frac{\left[ \alpha_{\langle v_\phi\rangle_\phi}(r_i) - \alpha_{\langle v_c\rangle_\phi}(r_i) \right]^2 }{ \left[ 1 + \alpha_{\langle v_c\rangle_\phi}(r_i) \right]^2 } }, \] where \[ \alpha_{A}(r_i) = \left. \frac{\mathrm{d}\ln A(r)}{\mathrm{d}\ln r} \right|_{r=r_i} \] denotes the logarithmic derivative of the quantity $A$ evaluated at $r_i$. The normalization by $1 + \alpha_{\langle v_c\rangle_\phi}(r_i)$ ensures stability and prevents spurious amplification in regions where the slope becomes small or negative. Again, the sum runs over discrete radial bins $r_i$ spanning the disc from $r_0$ to $R_{\rm ext, H\textsc{i}}$ with $\Delta r = 0.2~\mathrm{kpc}$ spacing.
\end{enumerate}

These metrics are designed to capture both amplitude and shape mismatches between the rotational and circular velocity profiles, and they serve as quantitative inputs for the final step of our classification pipeline. Once all three parameters are computed for every galaxy in the sample, we apply a $k$-means clustering algorithm to the resulting feature space. This unsupervised learning method partitions the dataset into distinct clusters -- which we interpret as dynamical classes -- based on similarities in the deviation metrics. The number of clusters is determined using standard internal validation criteria (i.e., the silhouette score), ensuring that the resulting classification reflects genuine structure in the data rather than arbitrary thresholds. These classes form the basis of the perturbed/equilibrium taxonomy discussed in the main text.

Fig.~\ref{fig:Clusters} provides both qualitative and quantitative insight into our classification methodology. The top panels contrast the kinematic behaviour of a stable disc (\textsc{hbt-herons} track ID $= 6611$) with that of a disturbed system (\textsc{hbt-herons} track ID $= 58791$). In the former, the rotational and circular velocity curves track each other almost exactly, indicative of a system close to centrifugal equilibrium. In the latter, systematic deviations in both amplitude and shape highlight significant dynamical disequilibrium. The bottom panel shows the full distribution of galaxies in the space defined by our three residual-based metrics. 

The $k$-means algorithm successfully identifies well-separated groups corresponding to dynamically distinct populations. These groups, or clusters, serve as the basis for our perturbation classification framework. It is worth emphasising that this scheme is designed to formalise the visual impression of how rotation curves depart from their expectation: the clustering provides a reproducible way to carve an otherwise continuous distribution of residual behaviours into classes that clarify the transition from ordered to disordered discs. In this context, limitations such as sample size or the use of a comparatively simple clustering method are not restrictive, since the purpose is organisational rather than predictive -- any reasonable algorithm that captures the same large-scale structure in residual space would yield a broadly equivalent taxonomy.

\bsp	% typesetting comment
\label{lastpage}
\end{document}